%% file: asymptotic-structure-afs.tex
\documentclass[11pt]{article}
\usepackage[english]{babel}
\usepackage[utf8]{inputenc}
\usepackage{lmodern}
\usepackage[T1]{fontenc}
\usepackage{marginnote}
\usepackage{amsmath,amssymb,lmodern,amsthm}
\usepackage{mathtools}
\usepackage{graphicx}
\usepackage{mathrsfs}
\usepackage{bm}
\usepackage[breaklinks]{hyperref}
\usepackage{url}
\usepackage{breakurl}
\usepackage{array}
\usepackage{tensor}
\usepackage{pifont}
\usepackage[toc,page]{appendix}
\usepackage[left=2.5cm,right=2.5cm,top=3cm,bottom=3cm]{geometry}
\usepackage{tikz}
\usetikzlibrary{arrows,positioning}
\usepackage{enumitem}
\usepackage{empheq}
\usepackage{cleveref}
\usepackage{cases}
\usepackage{csquotes}
\usepackage{accents}
\usepackage{setspace}

\theoremstyle{plain}
\usepackage{xargs}
\usepackage{xcolor}
\usepackage{subcaption}
\usepackage[justification=raggedright]{caption}
\usepackage[affil-it]{authblk}
\usepackage{tocloft}

\hypersetup{
	colorlinks=true,
	urlcolor=blue,
	linkcolor={black},
	citecolor={green!40!black}
}
\usepackage{thmtools}
\usepackage[backend=bibtex,style=numeric-comp,maxnames=4,sorting=none,url=false,hyperref=true,date=year,doi=true,eprint=true]{biblatex}

\newlist{properties}{enumerate}{10}
\setlist[properties]{label*=\roman*)}
\crefname{propertiesi}{\text{property}}{\text{properties}}
\Crefname{propertiesi}{\text{Property}}{\text{Properties}}
\newlist{steps}{enumerate}{10}
\setlist[steps]{label*=\arabic*)}
\crefname{stepsi}{\text{step}}{\text{steps}}
\Crefname{stepsi}{\text{Step}}{\text{Steps}}
\newlist{points}{enumerate}{10}
\setlist[points]{label*=\arabic*)}
\crefname{pointsi}{\text{point}}{\text{points}}
\Crefname{pointsi}{\text{Point}}{\text{Points}}

\newcommand{\ubar}[1]{\underaccent{\bar}{#1}}

\numberwithin{equation}{section}
\newtheorem{thm}{Theorem}[]
\theoremstyle{definition}
\newtheorem{prop}{Proposition}[section]
\newtheorem{corollary}{Corollary}[section]
\newtheorem{deff}{Definition}[section]
\newtheorem*{crit*}{Criterion}

\newtheorem{lemma}{Lemma}[section]
\newtheorem{remark}{Remark}[section]
\renewcommand{\iff}{\Longleftrightarrow}
\newcommand{\liff}{\Longleftrightarrow}
\newcommand{\lied}{\pounds}
\newcommand{\defeq}{\vcentcolon=}
\newcommand{\defeqs}{\stackrel{\scri}{\vcentcolon=}}
\newcommand{\defeqc}{\stackrel{\Sc}{\vcentcolon=}}

\newcommand{\scri}{\mathscr{J}}

\newcommand{\pt}[2]{\tensor{\hat{#1}}{#2}}

\newcommand{\ctru}[3]{\tensor[^{{\,}^{\scriptscriptstyle\!#1\!}\!}]{#2}{#3}}
\newcommand{\ptru}[3]{\tensor[^{{\,}^{\scriptscriptstyle\!#1\!}\!}]{\hat{#2}}{#3}}
\newcommand{\ctrd}[3]{\tensor[_{{\ }_{\scriptscriptstyle#1}\!}]{#2}{#3}}
\newcommand{\ctrdu}[4]{\prescript{{\,}^{\scriptscriptstyle\!#2\!}\!}{{\,}{\scriptscriptstyle\!#1\!}}{\tensor{#3}{#4}}}
\newcommand{\ctcrdu}[4]{\prescript{{\,}^{\scriptscriptstyle\!#2\!}\!}{{\,}{\scriptscriptstyle\!#1\!}}{\tensor{\mathring{#3}}{#4}}}
\newcommand{\ctcrd}[3]{\tensor[_{{\ }_{\scriptscriptstyle#1}\!}]{\mathring{#2}}{#3}}
\newcommand{\csru}[2]{\tensor[^{{\,}^{\scriptscriptstyle\!#1\!}\!}]{#2}{}}
\newcommand{\csrd}[2]{\tensor[_{{\ }_{\scriptscriptstyle\!#1\!}\!}]{#2}{}}
\newcommand{\csrddu}[4]{\prescript{{\,}^{\scriptscriptstyle\!#2\!}\!}{{\,}{\scriptscriptstyle\!#1\!}}{\ct{#3}{}}\brkt{#4}}
\newcommand{\csrdu}[3]{\prescript{{\,}^{\scriptscriptstyle\!#2\!}\!}{{\,}{\scriptscriptstyle\!#1\!}}{\cs{#3}}}
\newcommand{\csrdd}[3]{\tensor[_{{\ }_{\scriptscriptstyle\!#1\!}\!}]{#2}{}\brkt{#3}}
\newcommand{\csb}[2]{#1\brkt{#2}}

\newcommand{\ct}[2]{\tensor{{#1}}{#2}}
\newcommand{\pct}[2]{\tensor{\underaccent{*}{#1}}{#2}}
\newcommand{\pctrd}[3]{\tensor[_{{\ }_{\scriptscriptstyle#1}\!}]{\underaccent{*}{#2}}{#3}}
\newcommand{\lct}[2]{\tensor{\accentset{\leftarrow}{#1}}{#2}}
\newcommand{\ppt}[2]{\tensor{\underaccent{*}{\hat{#1}}}{#2}}
\newcommand{\ctn}[2]{\tensor[^{{\,}^{\scriptscriptstyle\!N\!}\!}]{{#1}}{#2}}

\newcommand{\ctc}[2]{\tensor{\mathring{#1}}{#2}}

\newcommand{\cts}[2]{\tensor{\overline{#1}}{#2}}

\newcommand{\ctd}[2]{\tensor{\dot{#1}}{#2}}

\newcommand{\ctcnn}[2]{\tensor[^{{\,}^{\scriptscriptstyle\!N \!}\!}]{\ubar{{#1}}}{#2}}
\newcommand{\ctcnl}[2]{\tensor[^{{\,}^{\scriptscriptstyle\!\ell \!}\!}]{\ubar{{#1}}}{#2}}

\newcommand{\ctcn}[2]{\tensor{\ubar{{#1}}}{#2}}

\newcommand{\ctg}[2]{\tensor{\tilde{#1}}{#2}}
\newcommand{\ctcg}[2]{\tensor{\tilde{\mathring{#1}}}{#2}}
\newcommand{\ctsg}[2]{\tensor{\tilde{\overline{#1}}}{#2}}

\newcommand{\cs}[1]{#1}

\newcommand{\csC}[1]{\mathring{#1}}
\newcommand{\csS}[1]{\overline{#1}}
\newcommand{\csSg}[1]{\tilde{\overline{#1}}}

\newcommand{\ps}[1]{\hat{#1}}

\newcommand{\df}[1]{\text{d}#1}
\newcommand{\td}[2]{\frac{\df#1}{\df#2}}

\newcommand{\prn}[1]{\left(#1\right)}

\newcommand{\brkt}[1]{\left[#1\right]}
\newcommand{\bbrkt}[1]{\Big[#1\Big]}
\newcommand{\cbrkt}[1]{\left\lbrace#1\right\rbrace}
\newcommand{\bcbrkt}[1]{\Big\lbrace#1\Big\rbrace}

\newcommand{\evalat}[1]{\Big|_{#1}}
\newcommand{\bevalat}[1]{\Bigg|_{#1}}
\newcommand{\eqs}{\stackrel{\scri}{=}}

\newcommand{\eqsopen}{\stackrel{\Delta}{=}}
\newcommand{\neqsopen}{\stackrel{\Delta}{\neq}}
\newcommand{\eqc}{\stackrel{\Sc}{=}}

\newcommand{\eqSv}[1]{\stackrel{\Sc_{#1}}{=}}

\newcommand{\cd}[1]{\tensor{\nabla}{#1}}
\newcommand{\cdps}[1]{\tensor{\hat{\nabla}}{#1}}
\newcommand{\cds}[1]{\tensor{\overline{\nabla}}{#1}}
\newcommand{\cdc}[1]{\tensor{\D}{#1}}
\newcommand{\pd}[1]{\tensor{\hat{\nabla}}{#1}}

\newcommand{\cdsg}[1]{\tensor{\tilde{\overline{\nabla}}}{#1}}

\newcommand{\commute}[2]{\left[#1,#2\right]}

\newcommand{\spacef}{\ }

\newcommand{\Q}{\mathcal{Q}}
\newcommand{\T}{\mathcal{T}}
\newcommand{\D}{\mathcal{D}}
\newcommand{\W}{\mathcal{W}}
\newcommand{\Z}{\mathcal{Z}}

\newcommand{\Sc}{\mathcal{S}}

\renewcommand{\P}{\mathcal{P}}
\newcommand{\I}{\mathcal{I}}

\newcommand{\X}{\mathcal{X}}
\newcommand{\M}{\mathcal{M}}
\newcommand{\E}{\mathcal{E}}
\newcommand{\F}{\mathcal{F}}
\newcommand{\G}{\mathcal{G}}
\newcommand{\cH}{\mathcal{H}}
\newcommand{\fX}{\mathfrak{X}}

\newcommand{\fT}{\mathsf{T}}

\newcommand{\ft}{\mathfrak{t}}

\newcommand{\ms}[1]{\cts{g}{#1}}
\newcommand{\msg}[1]{\ct{\tilde{\overline{g}}}{#1}}
\newcommand{\mc}[1]{\ct{q}{#1}}
\newcommand{\mcg}[1]{\ct{\tilde{q}}{#1}}
\newcommand{\mcn}[1]{\ct{\ubar{q}}{#1}}

\renewcommand{\thesubsection}{\thesection.\Alph{subsection}}

\newcolumntype{M}[1]{>{\centering\arraybackslash}m{#1}}
\newcolumntype{N}{@{}m{0pt}@{}}

\def\be{\begin{equation}}
\def\ee{\end{equation}}
\def\bea{\begin{eqnarray}}
\def\eea{\end{eqnarray}}
\def\bean{\begin{eqnarray*}}
	\def\eean{\end{eqnarray*}}

\newcounter{marginnotecount}[section]

\usepackage{subfiles}

	\title{\Large \textbf{Asymptotic Structure}\\ \textbf{with vanishing cosmological constant}}
	\author[]{Francisco Fernández-Álvarez\thanks{francisco.fernandez@ehu.eus}\ }
	\author[]{\ José M. M. Senovilla\thanks{josemm.senovilla@ehu.eus}} 
	\affil[]{Departamento de Física\\ Universidad del País Vasco UPV/EHU\\ Apartado 644, 48080 Bilbao, Spain }
	\date{\today{}}
\begin{document}
	
	\maketitle

	\begin{abstract}
		This is the first of two papers \cite{Fernandez-Alvarez_Senovilla-dS} devoted to the asymptotic structure of space-time in the presence of a non-negative cosmological constant $\Lambda$. This first paper is concerned with the case of $\Lambda =0$. Our approach is fully based on the tidal nature of the gravitational field and therefore on the `tidal energies' built with the Weyl curvature. In particular, we use the (radiant) asymptotic supermomenta computed from the rescaled Weyl tensor at infinity to provide a novel characterisation of radiation escaping from, or entering into, the space-time. Our new criterion is easy to implement and shown to be fully equivalent to the classical one based on the news tensor. One of its virtues is that its formulation can be easily adapted to the case with $\Lambda>0$ covered in the second paper. We derive the {\em general} energy-momentum-loss formulae including the matter terms and all factors associated to the choices of {\em arbitrary} foliation and of super-translation. We also revisit and present a full reformulation of the traditional peeling behaviour with a neat geometrical construction that leads, in particular, to an asymptotic alignment of the supermomenta in accordance with the radiation criterion.
	\end{abstract}

	\tableofcontents
	\newpage
	\section{Introduction}\label{sec:introduction}
			\subfile{./section-1/section-1.tex}

	\section{Asymptotic Structure}\label{sec:structure}
			\subfile{./section-2/section-2.tex}

	\section{News, BMS and asymptotic energy-momentum}\label{sec:news-asymptotic-s-e}
			\subfile{./section-3/section-3.tex}
	\section{Asymptotic propagation of physical fields and the peeling property revisited}\label{ssec:peeling}
			\subfile{./section-4/section-4.tex}

	\section{Asymptotic radiant supermomentum}\label{sec:smomentum}
			\subfile{./section-5/section-5.tex}
	\section{Discussion}
			\subfile{./section-6/section-6.tex}

	\subsubsection*{Acknowledgments}
		Work supported under Grants No. FIS2017-85076-P (Spanish MINECO/AEI/FEDER, EU) and No. IT956-16 (Basque Government). 
		\appendix
		\renewcommand{\thesubsection}{\arabic{subsection}}
			\subfile{./appendices/appendices.tex}

\printbibliography
\end{document}

%% file: section-1/section-1.tex

This is the first of a couple of papers that we present simultaneously and complement each other. Their subject is the asymptotic structure of space-times that admit a conformal completion {\em \`a la} Penrose with a {\em non-negative} cosmological constant $\Lambda \geq 0$. 

In the theory of General Relativity the space-time geometry is dynamical, allowing for changes of the gravitational field that affect causally different points of the space-time. This propagation phenomena, which includes the commonly denominated gravitational waves --or more generically gravitational radiation--, is a fully non-linear effect.

We know for certain that astrophysical objects can lose energy by emission of gravitational waves. With the announcement of the first direct detection in 2016 \cite{LigoVirgoPRL} and the many new detections that followed \cite{wavesRun1and2,wavesRun3}, these waves are now a robust observational fact. Historically, the degree of confidence on the existence of gravitational waves was not always that high. Back in 1918, Einstein published his quadrupole formula \cite{Einstein18} in the weak-field limit establishing the first step towards a theory of gravitational radiation. Nevertheless, the fact that this formula applies only in the linearised theory cast doubts upon the reality of gravitational radiation in the non-linear regime. Later in the 1950-60's, diverse theoretical advances  \cite{Trautman58,Pirani57,Bel1962,Bondi1962,Sachs1962,Penrose62} (see \cite{Zakharov}) showed that gravitational radiation constitutes a characteristic of the full theory and cleared up foregoing uncertainties on the theoretical side. Some landmarks included the so called Bondi news function and Bondi-Trautman momentum and the discovery of the asymptotic group of symmetries BMS (named after Bondi, Metzner and Sachs). 
		The next decade provided the first ever observational evidence of the existence of gravitational waves with the discovery of the binary pulsar PSR B1913+16 and its careful monitoring by Hulse and Taylor \cite{Hulse_Taylor1975} --see e.g. \cite{Damour2015}. 
		
		The theoretical study of isolated systems possibly emitting gravitational radiation is carried out at far distances from the source, what formally one defines as infinity. Penrose \cite{Penrose65} performed great advances in this matter by postulating the geometric representation of infinity as (part of) the boundary of the space-time in a conformal embedding. This boundary is commonly called ``scri'' and represented by $\scri$. 
		The conformal completion depends on a choice of conformal factor that leads to a {\em gauge freedom}: physically relevant results must be gauge independent.
		This approach to the theory of gravitational radiation was further investigated by Geroch, whose covariant formulation 
		constituted a solid geometric characterisation of gravitational radiation at infinity as well as of the fundamental objects involved \cite{Geroch1977}. Subsequently, Ashtekar identified and characterised the radiative degrees of freedom of the gravitational field in terms of classes of equivalence of connections at infinity \cite{Ashtekar81}. All these successful developments, except the conformal completion of Penrose, are valid only when the cosmological constant vanishes, hence they enter in conflict with a second empirical fact: according to observations \cite{Riess1998,Perlmutter1999} our universe is known to undergo a phase of accelerated-expansion which evinces the presence of a (bare or effective) positive cosmological constant. This scenario will be analyzed in the companion paper \cite{Fernandez-Alvarez_Senovilla-dS}.
		
		This first instalment is devoted to the $ \Lambda=0 $ scenario, in which the conformal (called unphysical) space-time has a lightlike conformal boundary. In this context, $ \scri $ is endowed with a conformal class of degenerate metrics and null generators which constitute a \emph{universal structure}. This structure underlies many of the favourable features of the asymptotically flat situation. Nevertheless, one must bear in mind that such universal structure is basically (conformally) stationary and does not contain the dynamical information. Hence, it {\em does not by itself determine all the physical features of the space-time under consideration, in particular, it does not contain information about the radiative sector}. For that, one needs a further ingredient which is not determined from the universal structure, but rather inherited from the conformal ambient space-time: the inherited connection and its corresponding curvature.\footnote{This additional ingredient can also be considered as structure added to the universal one on $\scri$  without referring to the physical space-time, as for instance in \cite{Ashtekar81}, but the properties of the additional structure and its relation to the universal one are chosen in accordance with what one knows from the conformal completion of physical space-times and the objects that thereby infinity {\em inherits}.} Actually, the gauge-invariant part of this curvature is basically the {\em news tensor}, a rank-2 symmetric traceless tensor field on $ \scri $ orthogonal to the generators that fully characterises the presence or absence of gravitational radiation arriving (or departing) from infinity. This can be alternatively understood from the point of view of a {\em characteristic} initial value formulation of the field equations with initial data placed on a null cone ---in this case, a connected component of $\scri$. The initial data on $\scri$ are given by the cone manifold, a family of degenerate metrics on the cone, {\em and a third ingredient: some components of the rescaled Weyl tensor} \cite{Chrusciel_Paetz2013}. The latter actually carries the information about radiation. There are many alternative ways of providing this initial data, and a particularly interesting one is by using the news tensor (see e.g.\cite{Chrusciel_Paetz2013} in a particular gauge), which acts as `potential' of the mentioned part of the rescaled Weyl tensor. This important remark will provide fundamental inspiration for the case with $\Lambda >0$ considered in \cite{Fernandez-Alvarez_Senovilla-dS}.

Asymptotics with $ \Lambda=0 $ can be tackled in the old asymptotic coordinate-based approach \cite{Bondi1962,Sachs1962,Sachs61}, in the NP formalism \cite{Penrose62,Penrose65} or by employing covariant methods and studying the intrinsic structure of $ \scri $ \cite{Geroch1977,Ashtekar81} in a gauge and coordinate-independent way. 
The covariant approach does not fix the conformal gauge, neither it needs of the introduction of coordinates. We incorporate this philosophy into our new method for studying the asymptotic structure and characterising gravitational radiation at $\scri$, which is based on the tidal nature of the gravitational field. We therefore make extensive use of objects defined upon the Bel-Robinson tensor \cite{Bel1958,Senovilla2000} (commonly called `superenergy' quantities) which represent `tidal energies' and suits our approach. We have found a new characterisation of gravitational radiation based on these superenergy quantities that was presented in \cite{Fernandez-Alvarez_Senovilla20} and proved to be equivalent to the standard condition based on the news tensor. We thereby show that gravitational radiation at infinity is of a tidal nature, being mainly associated to {\em cuts} of $\scri$, that is to say, cross sections fully transversal to the null generators of $\scri$.
We establish a geometric relation between quantities at the superenergy level ($ MT^{-2}L^{-3} $) and others at the energy level, being the latter `the source' of the former. 
These techniques are of the kind that we find more appropriate, partly because they are geometrically meaningful and also because they can be compared more easily with the $ \Lambda>0 $ scenario to show why \emph{one cannot simply adapt the known $ \Lambda=0 $ results as shortcuts to the new scenario}. All these ideas are made explicit in the course of this paper and its companion \cite{Fernandez-Alvarez_Senovilla-dS}. For a review of previously known results see e.g. \cite{Ashtekar2015c,Ashtekar2018}.
\\

Although our main aim is the characterisation of gravitational radiation escaping from the physical space-time, there are many other results that are worth remarking. 
Based on Geroch's ideas \cite{Geroch1977}, we provide a complete reformulation of the `peeling' properties in asymptotically flat space-times.
A class of automorphisms at the tangent space of any point in $ \scri $ is built such that they preserve the null cone and provide the asymptotic behaviour of physical fields approaching $ \scri $ along null geodesics. Its application to the physical Weyl tensor provides its peeling behaviour \cite{Sachs1962,Penrose62}, which is presented in the form of a theorem --see \cref{thm:peeling-weyl}. We also present the peeling behaviour of the physical Bel-Robinson tensor --\cref{thm:peeling-br}-- and, as a consequence of this, an alignment of physical supermomenta towards infinity occurs (these supermomenta are the equivalent of the 4-momentum but built with the Bel-Robinson tensor, see section \ref{sec:smomentum} for definitions). Our novel characterisation of gravitational radiation at infinity is presented in section \ref{sec:smomentum} where we introduce our --superenergy-based-- criterion to determine the presence of gravitational radiation at $\scri$ and we compare it with the classical condition. Beautifully, the criterion is in correspondence with the asymptotic alignment of supermomenta and the superenergy at infinity can be understood as sourcing the news tensor field. These features provide a test of our approach towards the characterisation of gravitational radiation by means of the rescaled Bel-Robinson tensor \cite{Fernandez-Alvarez_Senovilla20,Fernandez-Alvarez_Senovilla20b}. The fundamental core of these ideas is applied to the $ \Lambda>0 $ scenario in the companion paper \cite{Fernandez-Alvarez_Senovilla-dS}.

\subsection*{Conventions and notation}\label{ssec:conventions}
			Throughout this paper and its companion \cite{Fernandez-Alvarez_Senovilla-dS} we work in 4 space-time dimensions and distinguish quantities in physical space-time $ \ps{M} $ from those in unphysical space-time $ \cs{M} $ by using hats. Frequently used abbreviations include: KVF (Killing vector field), CKVF (conformal Killing vector field) and PND (principal null direction). 
				\begin{table}[ht!]
					\centering
					\begin{tabular}{ | M{2cm}| M{3cm} | M{3cm} |M{3cm}| M{3cm} | N }
						\hline
						\quad& Physical space-time $ \ps{M} $& Conformal space-time $ \cs{M} $ &  Spacelike hypersurfaces $ \I $  & Surfaces $ \Sc $& \\ \hline 
						\textbf{Metric}& $ \pt{g}{_{\alpha\beta}} $ & $ \ct{g}{_{\alpha\beta}}  $ & $ \ms{_{ab}} $  & $ \mc{_{AB}} $ &\\[1cm] \hline 
						\textbf{Volume form}&	$ \pt{\eta}{_{\alpha\beta\gamma\delta}} $& $ \ct{\eta}{_{\alpha\beta\gamma\delta}} $  & $ \ct{\epsilon}{_{abc}} $  & $ \ctc{\epsilon}{_{AB}} $ & \\[1cm]\hline
						\textbf{Covariant derivative}& $ \cdps{_{\alpha}} $	&$ \cd{_{\alpha}} $	&	$ \cds{_{a}} $& $ \cdc{_{A}} $	& \\[1cm]\hline				
						\textbf{Curvature tensor}& $ \pt{R}{_{\alpha\beta\gamma}^\delta} $ 	& $ \ct{R}{_{\alpha\beta\gamma}^\delta} $ & $ \cts{R}{_{abc}^d} $  & $ \ctc{R}{_{ABC}^D} $ & \\[1cm]\hline
						\textbf{Projector}& --	& -- & $ \ct{P}{^\alpha_\beta} $  & $ \ctc{P}{^a_{b}} $ & \\[1cm]\hline
						\textbf{Bases}&	-- & --	&$ \cbrkt{\ct{e}{^\alpha_a}}\ \cbrkt{\ct{\omega}{_\alpha^a}} $& $ \bcbrkt{\ct{E}{^a_A}}\ \cbrkt{\ct{W}{_a^A}} $ & \\[1cm]\hline	
					\end{tabular}
					\caption[Conventions]{General notation used in the present paper and its companion \cite{Fernandez-Alvarez_Senovilla-dS}. The first two columns correspond to the physical and conformal space-time $ \ps{M} $ and $ \cs{M} $, the third one to any 3-dimensional hypersurface (in particular this notation is used for $ \scri $) and the last one to any 2-dimensional surface $ \Sc $ (like codimension-1 manifolds on $ \scri $).}
					\label{tab:tableconventions}
				\end{table}
			 Part of the notation is summarised in \cref{tab:tableconventions}. \\
			 
			 The following conventions are used:
				\begin{itemize}
					\item Space-time metric signature: $ \prn{-,+,+,+} $.
					\item Space-time indices: $ \alpha,\beta,\gamma,$ etc; three dimensional space-like hypersurfaces indices: $ a,b,c,$ etc; surfaces indices: $ A,B,C,$ etc.
					\item Riemann tensor, Ricci tensor and scalar curvature: $ \ct{R}{_{\alpha\beta\gamma}^\delta}\ct{v}{_\delta}\defeq\prn{\cd{_{\alpha}}\cd{_\beta}-\cd{_{\beta}}\cd{_{\alpha}}}\ct{v}{_\gamma} $, $ \ct{R}{_{\alpha\beta}}\defeq\ct{R}{_{\alpha\mu\beta}^\mu} $, $ \cs{R}\defeq\ct{R}{_{\mu\nu}}\ct{g}{^{\mu\nu}} $.
					\item Choice of orientation: $ \ct{\eta}{_{0123}}=1 $, $ \ct{\epsilon}{_{123}}=1 $, $ \ctc{\epsilon}{_{23}}=1 $.
					\item Symmetrisation and antisymmetrisation: $ 2\ct{T}{_{[\alpha\beta]}}\defeq \prn{\ct{T}{_{\alpha\beta}}-\ct{T}{_{\beta\alpha}}} $,  $ 2\ct{T}{_{(\alpha\beta)}}\defeq \prn{\ct{T}{_{\alpha\beta}}+\ct{T}{_{\beta\alpha}}} $.
					\item Commutator of two vector fields: $ \commute{v}{w}^\alpha\defeq \ct{v}{^\mu}\cd{_{\mu}}\ct{w}{^\alpha}-\ct{w}{^\mu}\cd{_{\mu}}\ct{v}{^{\alpha}} $.
					\item Commutator of endomorphisms and (1,1)-tensors: $ \commute{A}{B}_{\alpha}{}^{\beta}\defeq\prn{\ct{A}{_\alpha^\mu}\ct{B}{_\mu^\beta}-\ct{B}{_\alpha^\mu}\ct{A}{_\mu^\beta}} $.
					\item Hodge dual operation on space-time two-forms: $ 2\ctru{*}{\omega}{_{\alpha\beta}}\defeq \ct{\eta}{^{\mu\nu}_{\alpha\beta}}\ct{\omega}{_{\mu\nu}}$.
					\item  D'Alambert operator: $ \square\defeq\ct{g}{^{\mu\nu}}\cd{_\mu}\cd{_\nu} $.
				\end{itemize}

%% file: section-2/section-2.tex
We assume Einstein field equations (EFE) with a vanishing cosmological constant\footnote{Here $ \varkappa\defeq 8\pi G c^{-4} $, and $ G $ and $ c $ are the gravitational constant and the speed of light, respectively.}
	\begin{equation}\label{eq:efe}
		\pt{R}{_{\alpha\beta}}-\frac{1}{2}\ps{R}\pt{g}{_{\alpha\beta}} =\varkappa \pt{T}{_{\alpha\beta}}\quad .
	\end{equation}
The sort of physical space-times $ \prn{\ps{M},\pt{g}{_{\alpha\beta}}} $ we consider admit a conformal completion, the so called unphysical space-time $ \prn{\cs{M},\ct{g}{_{\alpha\beta}}} $, defined as:
	\begin{properties}
		\item 	There exists an embedding $ \phi: \ps{M} \rightarrow \cs{M} $ such that $ \phi(\ps{M})=\cs{M}\setminus\scri $, where $ \scri $ is the boundary of $ \cs{M} $ and the physical metric is related to the conformal one by
		\begin{equation}
		\ct{g}{_{\alpha\beta}}= \Omega^2\pt{g}{_{\alpha\beta}},
		\end{equation} 
		where, abusing notation, we refer to the pullback of the conformal metric to the physical space-time, $ \prn{\phi^*g}_{\alpha\beta} $, by  $ \ct{g}{_{\alpha\beta}} $.
		\item $\Omega>0$ in $M\setminus\scri$ , $\Omega=0$ on $\scri$ and $\ct{N}{_\alpha}\defeq\cd{_\alpha}\Omega$ (the normal to $ \scri $) is non-vanishing there.\label{it:OmegaAtScri}
		\item The physical metric $ \pt{g}{_{\alpha\beta}} $ is a solution of EFE  \eqref{eq:efe} with $ \Lambda=0 $.\label{it:efesassumption}
		\item The energy-momentum tensor, $ \pt{T}{_{\alpha\beta}} $, vanishes at $ \scri $ and $ \ct{T}{_{\alpha\beta}}\defeq\Omega^{-1}\pt{T}{_{\alpha\beta}} $ is smooth there.\label{it:energytensorassumption}
	\end{properties}
For more details on conformal completions see e.g. \cite{Kroon,Frauendiener2004}. The conformal boundary $\scri$ is disconnected, in general, having a `future' and a `past' component which we denote by $\scri^\pm$, respectively. In our case, since the cosmological constant vanishes, the topology of $ \scri $ is given by $ \scri=\mathbb{R}\times \mathbb{S}^2 $ as long as $ \prn{\ps{M},\pt{g}{_{\alpha\beta}}} $ is weakly asymptotically simple \cite{NewmanRPAC1989}. We will consider more general situations occasionally, though. A typical calculation gives
	\begin{equation}\label{eq:normNscri}
	\ct{N}{_\mu}\ct{N}{^\mu}\eqs 0\spacef,
	\end{equation}
which indicates that $ \scri $ is a lightlike hypersurface (this is a direct consequence of $ \Lambda=0 $). Given any conformal completion $ \prn{\cs{M},\ct{g}{_{\alpha\beta}}} $, it is always possible to rescale the conformal factor
	\begin{equation}\label{eq:conformal-gauge}
		\Omega \rightarrow \ctg{\Omega}{}=\omega\Omega\quad \text{ with } \omega > 0 \text{ on } \cs{M}\spacef,
	\end{equation}
such that associated to a given physical space-time $ \prn{\ps{M},\pt{g}{_{\alpha\beta}}} $ there is a conformal class of completions. One can use this freedom --which we refer to as conformal \emph{gauge freedom}-- conveniently, and we partly restrict this freedom by choosing the gauges that satisfy $\cd{_\alpha}\ct{N}{^\alpha}\eqs 0$ which amounts to choosing functions $ \omega $ that satisfy
	\begin{equation}\label{eq:divergenc-freeEquation}
	4\ct{N}{^\mu}\partial_\mu \omega + \omega \cd{_\alpha}\ct{N}{^\alpha}=4\ct{N}{^\mu}\partial_\mu \omega + \omega \square\Omega \eqs 0\quad .
	\end{equation}
This does not fix completely the gauge ambiguity, and there remains a large freedom since new functions $ \omega $ fulfilling 
	\begin{equation}
	\lied_{\vec{N}}\omega \eqs 0\quad 
	\end{equation}		
are allowed. We call this partial fixing \emph{divergence-free gauge}. Yet another typical calculation yields
	\begin{equation}\label{eq:derivative-N-gauge}
	\cd{_\alpha}\ct{N}{_\beta}\eqs 0\quad 
	\end{equation}
in our gauge. 
	\subsection{Basic geometry of $ \scri $}\label{ssec:basic-geometry}
	Let $ \cbrkt{\ct{e}{^\alpha_a}} $ be a basis of the set of vector fields tangent to $ \scri $, i.e., orthogonal to $ \ct{N}{_\alpha} $,  with $ a=1,2,3 $,  and let $ \cbrkt{\ct{\omega}{_\alpha^a}} $ be a dual basis. In particular, $\ct{N}{^\alpha}=\ct{N}{^a}\ct{e}{^{\alpha}_{a}} $ is collinear with the generators of $ \scri $ and $ \ct{N}{^a} $ is the degeneration vector field of the induced first fundamental form
		\begin{equation}
			\ms{_{ab}}\defeq \ct{e}{^\alpha_{a}}\ct{e}{^\beta_{b}}\ct{g}{_{\alpha\beta}}\spacef,\quad\ct{N}{^a}\ms{_{ab}}=0\spacef.
		\end{equation}
	\Cref{eq:derivative-N-gauge} implies that the second fundamental form of $ \scri $ vanishes
		\begin{equation}\label{eq:Kab=0}
			\ct{K}{_{ab}}:= \ct{e}{^\mu_a}\ct{e}{^\nu_b} \cd{_\mu}\ct{N}{_\nu}=0\spacef,
		\end{equation} 
	hence the intrinsic Lie derivative on $ \scri $ of $ \ms{_{ab}} $ along $ \ct{N}{^a} $ is zero
		\begin{equation}\label{eq:lied-ms}
			\lied_{\vec{N}}\ms{_{ab}}=0\spacef
		\end{equation}
	and the \emph{induced connection} 
		\begin{equation}
			 \forall\spacef \ct{X}{^\alpha}=\ct{X}{^a}\ct{e}{^\alpha_{a}}\spacef,\quad\forall\spacef\ct{Y}{^\alpha}=\ct{Y}{^a}\ct{e}{^\alpha_{a}}\spacef,\quad \ct{X}{^a}\cds{_a} \ct{Y}{^b} \defeq \ct{\omega}{_{\beta}^b}\ct{X}{^\alpha}\cd{_{\alpha}} \ct{Y}{^{\beta}} 
		\end{equation}
	is torsion-free and `metric',
		\begin{equation}
			\cds{_{a}}\ms{_{bc}}=0.
		\end{equation}	
	The induced connection coefficients $ \cts{\Gamma}{^c_{ab}} $ are thus given by
		\begin{equation}
			\ct{e}{^\mu_{a}}\cd{_{\mu}}\ct{e}{^\gamma_{b}}=\cts{\Gamma}{^c_{ab}}\ct{e}{^\gamma_{c}}\spacef.
		\end{equation}
	One can introduce a volume three-form $ \ct{\epsilon}{_{abc}} $ on $ \scri $ by means of the space-time volume four-form $ \ct{\eta}{_{\alpha\beta\gamma\delta}} $,
		\begin{equation}
				-\ct{N}{_\alpha}\ct{\epsilon}{_{abc}}\eqs \ct{\eta}{_{\alpha\mu\nu\sigma}}\ct{e}{^\mu_a}\ct{e}{^\nu_b}\ct{e}{^\sigma_c}\spacef,\label{eq:vol-form-scri}
		\end{equation}
	and a contravariant version determined by $ \ct{\epsilon}{^{abc}}\ct{\epsilon}{_{abc}}=6 $. We fix the corresponding orientations to $ \ct{\epsilon}{_{123}}=\ct{\eta}{_{0123}}=1 $. The choice of gauge also implies that the induced connection is volume preserving,
		\begin{equation}
			\cds{_{a}}\ct{\epsilon}{_{bcd}}=0\spacef.
		\end{equation}
	Although the metric is degenerate, one can define a contravariant object that `raises indices' by
		\begin{equation}\label{eq:inverse-deg-metric}
			\ms{_{ea}}\ms{^{ed}}\ms{_{db}}\defeq \ms{_{ab}}.
		\end{equation}
	There is a freedom in adding to $ \ms{^{ab}} $ any term of the form $ \ct{N}{^a}\ct{v}{^b}+\ct{N}{^b}\ct{v}{^a}  $. One can make a choice, however, by picking out a dual basis $ \cbrkt{\ct{\omega}{_{\alpha}^a}} $ and instead defining $ \ms{^{ab}} $ as
		\begin{equation}
			 \ms{^{ab}}\defeq\ct{\omega}{_\alpha^a}\ct{\omega}{_\beta^b}\ct{g}{^{\alpha\beta}}\spacef,\quad\ms{^{ef}}\ms{_{ef}}=2\spacef,
		\end{equation}
	from where \cref{eq:inverse-deg-metric} follows. $ \scri $ admits a natural definition of \emph{cuts} $ \Sc $, i. e., any closed surface transversal to the generators everywhere. These cross-sections are necessarily spacelike.  In the prominent case when $\scri$ has topology $\mathbb{R}\times \mathbb{S}^2$ cuts $ \Sc $ with topology $ \mathbb{S}^2$ can be chosen. In general, cuts carry a positive-definite metric inherited from --and which essentially is-- $ \ms{_{ab}} $,
		\begin{equation}
			\mc{_{AB}}\defeqc\ct{E}{^a_{A}}\ct{E}{^b_{B}}\ms{_{ab}}\spacef,
		\end{equation}
	where $ \cbrkt{\ct{E}{^a_A}} $ is a basis of the set $ \fX_{\Sc} $ of tangent vector fields on $ \Sc $, with $ A=1,2 $. A basis of tangent vector fields to $ \Sc $ considered within $ M $ is $  \cbrkt{\ct{E}{^\alpha_A}}  $ where $ \ct{E}{^\alpha_A} = \ct{e}{^\alpha_{a}}\ct{E}{^a_A} $. Similarly, we introduce bases $ \cbrkt{\ct{W}{_a^A}} $ and $ \cbrkt{\ct{W}{_{\alpha}^A}} $ of the dual space $ \Lambda_{\Sc} $. At each cut $ \Sc $, there is a unique lightlike vector field $ \ct{\ell}{^\alpha} $ other than $ \ct{N}{^\alpha} $ such that 
		\begin{equation}\label{eq:ell-cut}
			\ct{\ell}{^\mu}\ct{\ell}{_{\mu}}\eqc 0\spacef,\quad\ct{\ell}{_{\alpha}}\eqc\ct{\ell}{_{a}}\ct{\omega}{_{\alpha}^a}\spacef,\quad\ct{\ell}{^\alpha}\ct{\omega}{_{\alpha}^a}\eqc 0\spacef,\quad \ct{\ell}{_{a}}\ct{E}{^a_{A}}\eqc 0,\quad\ct{\ell}{^\alpha}\ct{N}{_{\alpha}}\eqc-1\spacef.
		\end{equation}
	This set of vector fields can be used to complete bases on $ \scri $ \emph{at} $ \Sc $, $ \cbrkt{-\ct{\ell}{_{a}},\ct{W}{_{a}^A}} $, $ \bcbrkt{\ct{N}{^a},\ct{E}{^a_A}} $, and to write the projector to $ \Sc $:
		\begin{equation}
			\ctc{P}{^\alpha_{\beta}}=\ct{E}{^\alpha_{M}}\ct{W}{_{\beta}^M}\eqc\delta^\alpha_{\beta}+\ct{\ell}{_{\beta}}\ct{N}{^\alpha}+\ct{N}{_{\beta}}\ct{\ell}{^\alpha}\spacef.
		\end{equation}
	Also, one has
		\begin{equation}
			\ms{^{mb}}\ct{e}{^\mu_{m}}\ct{g}{_{\mu\alpha}}\eqc\ct{\omega}{_{\alpha}^b}+\ct{\ell}{_{\alpha}}\ct{N}{^b},\quad\ct{\ell}{_{m}}\ms{^{am}}\eqc 0
		\end{equation}
	and the projector to a given cut within $ \scri $ takes the form
		\begin{equation}
				\ctc{P}{^a_{b}}=\delta^a_b+\ct{N}{^a}\ct{\ell}{_{b}}=\ms{^{ac}}\ms{_{bc}}\spacef.
		\end{equation}
	The intrinsic volume two-form $\ctc{\epsilon}{_{AB}}$ of $ \prn{\Sc,\mc{_{AB}}}$ is obtained from
		\begin{align}
		-\ct{\ell}{_{a}}\ctc{\epsilon}{_{AB}} &\eqc \ct{\epsilon}{_{amn}}\ct{E}{^m_A}\ct{E}{^n_B}\spacef,\\
		\ct{N}{^{a}}\ctc{\epsilon}{^{AB}} &\eqc \ct{\epsilon}{^{amn}}\ct{W}{_m^A}\ct{W}{_n^B}\spacef,\label{eq:vol-form-cut}
		\end{align}
	where the orientation is chosen such that $ \ctc{\epsilon}{_{23}}=\cts{\epsilon}{_{123}}=1 $, and the inherited connection
		\begin{equation}
			\forall\spacef \ct{U}{^a}\eqc\ct{E}{^a_A}\ct{U}{^A}\spacef,\quad\forall\spacef \ct{V}{^a}\eqc\ct{E}{^a_A}\ct{V}{^A}\spacef,\quad \ct{V}{^M}\cdc{_M}\ct{U}{^A}\defeqc\ct{W}{_m^A}\ct{V}{^n}\cds{_n}\ct{U}{^m}\spacef
		\end{equation}
	is metric and volume preserving --ergo this is the intrinsic Levi-Civita connection on $ \prn{\Sc,\mc{_{AB}}} $--
		\begin{equation}
			\cdc{_{A}}\mc{_{AB}}=0\spacef,\quad\cdc{_{A}}\ctc{\epsilon}{_{BC}}=0\spacef.
		\end{equation}
	\Cref{eq:derivative-N-gauge} implies that
		\begin{equation}\label{eq:der-N-gauge-afs}
			\cds{_{a}}\ct{N}{^b}=0\spacef.
		\end{equation}
	The relation between the space-time covariant derivative and the induced derivative on $ \scri $ for
	any tensor field $ \ct{T}{^{\alpha_1 ...\alpha_r}_{\beta_1...\beta_q}} $ defined \emph{at least} on $ \mathcal{\scri} $ is
		\begin{align}
			&\ct{\omega}{_{\mu_{1}}^{a_1}}...\ct{\omega}{_{\mu_r}^{a_r}}\ct{e}{^{\nu_1}_{b_q}}...\ct{e}{^{\nu_q}_{b_q}}\ct{e}{^\rho_c}\cd{_\rho}\ct{T}{^{\mu_1...\mu_r}_{\nu_1...\nu_q}}\eqs\cds{_c}\ct{\overline{T}}{^{a_1...a_r}_{b_1...b_q}}\nonumber\\
			&-\sum_{i=1}^{r}\ct{T}{^{a_1...a_{i-1}\sigma a_{i+1}...a_r}_{b_1...b_q}}\ct{N}{_\sigma}\ct{\Psi}{^{a_i}_c}\label{eq:inducedDevscri}\quad
		\end{align}
	where we have defined
		\begin{equation}\label{eq:induced-connection-psi}
			 \cts{T}{^{a_1...a_r}_{b_1...b_q}}\defeqs  \ct{\omega}{_{\mu_{1}}^{a_1}}...\ct{\omega}{_{\mu_r}^{a_r}}\ct{e}{^{\nu_1}_{b_q}}...\ct{e}{^{\nu_q}_{b_q}}\ct{T}{^{\mu_1...\mu_r}_{\nu_1...\nu_q}}\spacef,\quad \ct{\Psi}{^{a}_c}\defeq \ct{\omega}{^a_{\mu}}\ct{e}{^\nu_{c}}\cd{_{\nu}}\cts{\ell}{^\mu}\spacef,
		\end{equation}
	with $ \cts{\ell}{^\alpha} $ any vector field on $ \scri $ satisfying $ \cts{\ell}{^\alpha}\ct{N}{_{\alpha}}=-1 $ and $ \cts{\ell}{^\alpha}\ct{\omega}{_{\alpha}^a}=0 $. We have also
	used \eqref{eq:Kab=0} --for general formulae see \cite{Mars1993b}--, whereas the relation between the induced covariant derivative on $ \scri $ and the intrinsic covariant derivative on $ \Sc $ for a tensor field $ \ct{T}{^{a_1 ...a_r}_{b_1...b_q}} $ defined \emph{at least} on $ \Sc $ reads
		\begin{align}
		&\ct{W}{_{m_{1}}^{A_1}}...\ct{W}{_{m_r}^{A_r}}\ct{E}{^{n_1}_{B_q}}...\ct{E}{^{n_q}_{B_q}}\ct{E}{^r_C}\cds{_r}\ct{T}{^{m_1...m_r}_{n_1...n_q}}\eqc\cdc{_C}\ctc{T}{^{A_1...A_r}_{B_1...B_q}}\nonumber\\&
		-\sum_{i=1}^{q}\ct{T}{^{A_1...A_r}_{B_1...B_{i-1}s B_{i+1}...B_q}}\ct{N}{^{s}}\ct{H}{_{CB_i}}\label{eq:inducedDevcut}\quad 
		\end{align}
	with 
		\begin{equation}\label{eq:afs-HAB}
		 \ctc{T}{^{A_1...A_r}_{B_1...B_q}}\defeqc  \ct{W}{_{m_{1}}^{A_1}}...\ct{W}{_{m_r}^{A_r}}\ct{E}{^{n_1}_{B_q}}...\ct{E}{^{n_q}_{B_q}}\ct{T}{^{m_1...m_r}_{n_1...n_q}}\spacef,\quad\ct{H}{_{AB}}\defeqc\ct{E}{^a_A}\ct{E}{^b_B}\cds{_{a}}\ct{\ell}{_{b}}\spacef.
		\end{equation}
	Observe that under conformal gauge transformations, the following changes apply
		\begin{align}
			\msg{_{ab}}&= \omega^2\ms{_{ab}}\spacef,\label{eq:gauge-ms}\\
			\mcg{_{AB}}&\eqc \omega^2\mc{_{AB}}\spacef.\label{eq:gauge-mc}
		\end{align}
	The  curvature tensor associated to the induced connection satisfies for any vector field $\ct{v}{^a}$
		\begin{equation}\label{eq:ricci-id-afs-scri}
		\prn{\cds{_{a}}\cds{_{b}}-\cds{_{b}}\cds{_{a}}}\ct{v}{^d}=-\cts{R}{_{abc}^d}\ct{v}{^c}\quad 
		\end{equation}
	and it is related to the space-time curvature through the `Gauss equation'
				\begin{equation}\label{eq:gauss-scri}
					\ct{e}{^\alpha_{a}}\ct{e}{^{\beta}_{b}}\ct{e}{^\gamma_{c}}\ct{\omega}{_\delta^{d}}\ct{R}{_{\alpha\beta\gamma}^\delta}\eqs \cts{R}{_{abc}^d}\spacef .
				\end{equation}
	It has the following properties
		\begin{equation}
			\cts{R}{_{abc}^d}=-\cts{R}{_{bac}^d}\spacef,\quad\cts{R}{_{[abc]}^d}=0\spacef,\quad\cts{R}{_{abc}^c}=0\spacef,\quad\cds{_{[e}}\cts{R}{_{ab]c}^d}=0\spacef.
		\end{equation}
	Its non-vanishing trace constitutes a symmetric tensor field
		\begin{equation}
			\cts{R}{_{ab}}\defeq\cts{R}{_{adb}^d}=\cts{R}{_{ba}}\spacef.
		\end{equation}
	The curvature tensor can be expressed as
		\begin{equation}\label{eq:riemann-scri-schouten}
			\cts{R}{_{abc}^d}=2\ms{_{c[a}}\cts{S}{_{b]}^d}-2\delta^d_{[a}\cts{S}{_{b]c}}\spacef,
		\end{equation}
	where the tensor fields $ \cts{S}{_{a}^b} $ and $ \cts{S}{_{ab}}\defeq \ms{_{am}}\ct{S}{_{b}^m} =\cts{S}{_{ba}}$ will be shown to coincide with pullbacks of the space-time Schouten tensor to $ \scri $ --see \cref{ssec:curvature-scri-and-fields}.
	Of course, one can lower the contravariant index of the curvature tensor with the degenerate metric $ \ms{_{ab}} $,
		\begin{equation}
			\cts{R}{_{abcd}}\defeq \ms{_{ed}}\cts{R}{_{abc}^e}\spacef,
		\end{equation}
	however, information is lost in this process and one has to treat the fully covariant version as a different tensor. Using the `metricity' of the induced connection, it follows that $ \cts{R}{_{abcd}} $ has all the symmetries of a Riemann tensor, including
		\begin{equation}
			\cts{R}{_{abcd}}=\cts{R}{_{cdab}}=-\cts{R}{_{abdc}}\spacef.
		\end{equation}
	 In considering the action of $ \cts{R}{_{abc}^d} $ on $ \ct{N}{^a} $, via \cref{eq:ricci-id-afs-scri} and using \cref{eq:der-N-gauge-afs}, one finds
		\begin{equation}
			\cts{R}{_{abc}^d}\ct{N}{^c}=0\spacef.
		\end{equation}
	This implies that
		\begin{equation}\label{eq:riemann-scri-covariant-N}
			\ct{N}{^a}\cts{R}{_{abcd}}=0\spacef,
		\end{equation}
	hence the lower-index version of the curvature tensor is orthogonal to $ \ct{N}{^a} $ in all its indices. This property makes it effectively a two-dimensional tensor field with the symmetries of a Riemann tensor, thus we can write it as
		\begin{equation}\label{eq:riemann-scri-covariant}
			\cts{R}{_{abcd}}=\frac{1}{2}\csS{R}\prn{\ms{_{ac}}\ms{_{db}}-\ms{_{bc}}\ms{_{da}}}
		\end{equation}
	for some scalar field $ \csS{R} $. Using the properties presented so far, it follows that
		\begin{equation}\label{eq:lie-intrinsic-curvature}
			\lied_{\vec{N}}\cts{R}{_{abcd}}=\ct{N}{^e}\cds{_{e}}\cts{R}{_{abcd}}=0\spacef,\quad\ct{N}{^e}\cds{_{e}}\csS{R}=0\spacef.
		\end{equation}
	Using \cref{eq:riemann-scri-schouten}, $ \cts{R}{_{ab}} $ can be expressed as
		\begin{equation}\label{eq:ricci-scri-trace}
			\cts{R}{_{ab}}=\cts{S}{_{ab}}+\ms{_{ab}}\cts{S}{_{m}^m}\spacef,
		\end{equation}
	and $ \cts{R}{_{abcd}} $ as
		\begin{equation}\label{eq:riemann-scri-covariant-Schouten}
			\cts{R}{_{abcd}}=2\ms{_{c[a}}\cts{S}{_{b]d}}-2\ms{_{d[a}}\cts{S}{_{b]c}}\spacef.
		\end{equation}
	Because of \cref{eq:riemann-scri-covariant-N}, one can take the traces of this tensor field with $ \ms{^{ab}} $ without risk of ambiguity. In doing so, if one compares \cref{eq:riemann-scri-covariant-Schouten,eq:riemann-scri-covariant,eq:ricci-scri-trace}, it follows that
		\begin{align}
			\cts{S}{_{mn}}\ms{^{mn}}&=\frac{\csS{R}}{2}\spacef,\label{eq:trace-lower-Schouten-ricci-scri}\\
			2\cts{S}{_{m}^m}+\frac{1}{2}\csS{R}&=\ms{^{rs}}\cts{R}{_{rs}}\spacef.\label{eq:trace-Schouten-ricci-scri}
		\end{align}
	Hence, the following expression holds
		\begin{equation}
			\cts{S}{_{ab}}=\cts{R}{_{ab}}-\frac{1}{2}\ms{_{ab}}\prn{\ms{^{mn}}\cts{R}{_{mn}}-\frac{1}{2}\csS{R}}\spacef.
		\end{equation}
	From \cref{eq:inducedDevcut} it is easily deduced the `Gauss relation' between the intrinsic curvature $ \ctc{R}{_{ABC}^D} $ of any cut $ \Sc $ and the curvature of $ \scri $,
		\begin{equation}
			\ct{E}{^a_A}\ct{E}{^b_{B}}\ct{E}{^c_{C}}\cts{R}{_{abc}^d}\ct{W}{_{d}^D}\ct{v}{_{D}}\eqc\prn{\cdc{_{A}}\cdc{_{B}}-\cdc{_{B}}\cdc{_{A}}}\ct{v}{_{C}}\defeqc\ctc{R}{_{ABC}^D}\ct{v}{_{D}}\spacef,\quad\forall\ct{v}{_{A}}\in \Lambda_\Sc\spacef.
		\end{equation}
	One can readily show that 
		\begin{align}
			\ctc{R}{_{AB}}&\defeq \ctc{R}{_{AMB}^M}\eqc \mc{_{AB}}\cts{S}{_{mn}}\ms{^{mn}}\eqc \frac{1}{2}\csS{R}\mc{_{AB}}\spacef,\label{eq:ricci-cut}\\
			\csC{R}&\defeq \ctc{R}{_{M}^M}\eqc \csS{R}\eqc 2K\spacef,
		\end{align} 
	where $ \cs{K} $ is the Gaussian curvature of $ \prn{\Sc,\mc{_{AB}}} $.
	Instead of single cuts $ \Sc $, one can consider a generic \emph{foliation} where each leaf $ \Sc_{C} $ is defined by a different constant value $ C $ of a function $ F $  such that
		\begin{equation}
			\dot{F}\defeq\ct{N}{^m}\cds{_{m}}F\neq 0\spacef.
		\end{equation}
	Each leaf is a cut, by definition transversal to $ \ct{N}{^a} $. Then, associated to a given foliation there is a one-form
		\begin{equation}
			\cts{\ell}{_{a}}\defeq -\frac{1}{\dot{F}}\cds{_{a}}{F},\quad\ct{N}{^m}\cts{\ell}{_{m}}=-1\spacef.\label{eq:ell-F-foliation}
		\end{equation}
	We set univocally $ \cts{\ell}{_{\alpha}}\defeq\ct{\omega}{_{\alpha}^a}\cts{\ell}{_{a}} $ and require $ \cts{\ell}{_{\mu}}\cts{\ell}{^\mu}=0 $, 
	where $ \ct{\ell}{_{a}} $ is the one-form uniquely defined by \cref{eq:ell-cut}. One can introduce couples of vector fields $ \bcbrkt{\ctcn{E}{^a_{A}}} $ and $ \cbrkt{\ctcn{W}{_{a}^A}} $ --with $ A=2,3 $--  serving as bases for the set of vector fields and forms on $ \scri $ orthogonal to $ \cts{\ell}{_{a}} $ and $ \ct{N}{^{a}} $. Also, on each leaf $ \Sc_{C} $ they constitute bases for the vector fields and forms that are orthogonal to $ \ct{N}{^a} $ and $ \ct{\ell}{_{a}} $ there. Let us introduce the projector
		\begin{equation}
		 \ctcn{P}{^a_{b}}\defeq \delta^a_b+\ct{N}{^a}\cts{\ell}{_b}\spacef,\quad\ctcn{P}{^m_b}\cts{\ell}{_{m}}=0=\ctcn{P}{^a_{m}}\ct{N}{^m}\spacef,\quad\ctcn{P}{^a_{b}}\eqSv{C}\ctc{P}{^a_{b}}\spacef.
		\end{equation}
	We will distinguish quantities projected to a single cut $ \Sc_{C} $ from those projected with $ \ctcn{P}{^a_{b}} $ by using the following notation
		\begin{align}
			\ctcn{v}{_{b}}\defeq\ctcn{P}{^m_{b}}\ct{v}{_{m}}\spacef,\\
			\ctc{v}{_{b}}\stackrel{\Sc_{C}}{\defeq}\ctc{P}{^m_{b}}\ct{v}{_{m}}\spacef,
		\end{align}
	and similarly
		\begin{align}
			\ctcn{v}{_{B}}\defeq\ctcn{E}{^m_{B}}\ct{v}{_{m}}\spacef,\\
			\ctc{v}{_{B}}\stackrel{\Sc_{C}}{\defeq}\ct{E}{^m_{B}}\ct{v}{_{m}}\spacef.
		\end{align}
	Of course, given any one-form field $ \ct{v}{_{a}} $ on $ \scri $,
		\begin{equation}
			\ctcn{v}{_{a}}\eqSv{C}\ctc{v}{_{a}}\spacef.
		\end{equation}
	A simple calculation leads to
		\begin{equation}
			\lied_{\vec{N}}\cts{\ell}{_{a}}=\ct{N}{^e}\cds{_{e}}\cts{\ell}{_{a}}=-\ctcn{P}{^m_{a}}\cds{_{m}}\ln\dot{F}\label{eq:lied-ell-F-foliation}
		\end{equation}
	and the next relations hold
		\begin{align}
				\ctcn{P}{^m_{a}}\cds{_{m}}F&=0\quad\prn{\iff\ctcn{E}{^m_{A}}\cts{\ell}{_{m}}=0}\spacef,\\
				\lied_{\vec{N}}\ctcn{E}{^a_{A}}&=-\ct{N}{^a}\ctcn{E}{^m_{A}}\cds{_{m}}\ln\dot{F}\spacef,\quad\lied_{\vec{N}}\ctcn{W}{_{a}^A}=0\spacef.\label{eq:lied-bases-foliation}
		\end{align}
	In addition, one can define
		\begin{equation}
			\mcn{_{AB}}\defeq\ctcn{E}{^a_{A}}\ctcn{E}{^b_{B}}\ms{_{ab}}\spacef,\quad\mcn{^{AB}}\defeq\ctcn{W}{_{a}^{A}}\ctcn{W}{_{b}^B}\ms{^{ab}}\spacef,
		\end{equation}
	where $ \mcn{_{AB}} $ is such that it coincides with the metric $ \mc{_{AB}} $ of each leaf $ \Sc_{C} $. All cuts are isometric, though, as a quick calculation taking into account the above relations and \cref{eq:lied-ms} yields
		\begin{equation}\label{eq:liemc}
			\lied_{N}\mcn{_{AB}}=0\spacef,
		\end{equation}
	hence $ \mcn{_{AB}} $ and $ \mc{_{AB}} $ are essentially the same object. Hence, the curvature \eqref{eq:ricci-cut} is basically the same for every cut of the foliation, in agreement with \cref{eq:lie-intrinsic-curvature}. In fact, all possible cuts in $\scri$ are locally isometric, even if they do not belong to the same foliation.\\
	
	There is a special sort of foliations that we call \emph{adapted} to $ \ct{N}{^a} $. These are defined by functions $ F $ fulfilling
		\begin{equation}\label{eq:foliation-adapted}
			\lied_{\vec{N}}\dot{F}=0\spacef.
		\end{equation}
	Given any adapted foliation, an appropriate gauge fixing ($ \omega=\dot{F} $) allows, via the transformations of \cref{app:conformal-gauge-transformations}, to set
		\begin{equation}\label{eq:foliation-canonical-adapted}
			\cts{\ell}{_{a}}=-\cds{_{a}}F\ ,\quad\cds{_{[a}}\cts{\ell}{_{b]}}=0\spacef,\quad\lied_{\vec{N}}\cts{\ell}{_{a}}=0\spacef,\quad\lied_{\vec{N}}\ctcn{E}{^a_{A}}=0\spacef.
		\end{equation}
	We refer to this kind of foliations as \emph{canonically adapted to $ \ct{N}{^a} $}. \\
	
	Finally, let us introduce the kinematical quantities
		\begin{align}
			\ctcn{\Theta}{_{ab}}&\defeq \ctcn{P}{^r_{(a}}\ctcn{P}{^s_{b)}}\cds{_{r}}\cts{\ell}{_{s}}\spacef,\label{eq:expansion-tensor-ell}\\
			\ctcn{\sigma}{_{ab}}&\defeq\ctcn{\Theta}{_{ab}}-\frac{1}{2}\ms{_{ab}}\ms{^{rs}}\ctcn{\Theta}{_{rs}}\spacef.\label{eq:shear}
		\end{align}
	Observe that on every cut $ \Sc_{C} $ of the foliation --see definition \eqref{eq:afs-HAB}--
		\begin{equation}
			\ctcn{E}{^a_{A}}\ctcn{E}{^b_{B}}\ctcn{\Theta}{_{ab}}\eqSv{C}\ct{H}{_{AB}}\spacef.
		\end{equation}
	Also, definition \eqref{eq:shear} is nothing but the shear of the one-form $ \cts{\ell}{_{a}} $ orthogonal to each cut $ \Sc_{C} $ of the foliation.
	\\
		
	A deeper characterisation of the curvature and the interplay between the induced connection, the choice of foliation and the space-time fields is given in \cref{ssec:curvature-scri-and-fields}.
		
	\subsection{Fields at infinity}
	 We turn to the study of space-time fields at $ \scri $. Let us introduce 
			\begin{align}
			\ct{S}{_{\alpha\beta}} &\defeq \ct{R}{_{\alpha\beta}}-\frac{1}{6}\cs{R}\ct{g}{_{\alpha\beta}}\quad\label{eq:defSchouten} ,\\
			\pt{S}{_{\alpha\beta}} &\defeq \pt{R}{_{\alpha\beta}}-\frac{1}{6}\ps{R}\pt{g}{_{\alpha\beta}}\quad \label{eq:defSchouten-physical},
			\end{align}
	 	and the physical and conformal Cotton tensors
 			\begin{align}
				\pt{Y}{_{\alpha\beta\gamma}}&\defeq \pd{_{[\alpha}}\pt{S}{_{\beta]\gamma}}\spacef,\label{eq:cottonPhysical}\\
				\ct{Y}{_{\alpha\beta\gamma}}&\defeq \cd{_{[\alpha}}\ct{S}{_{\beta]\gamma}}\spacef.\label{eq:cottonUnphysical}
			\end{align}
		From the contracted Bianchi identity one can show that
			\begin{align}
			\cd{_\mu}\prn{\ct{C}{_{\alpha\beta\gamma}^\mu}}+\ct{Y}{_{\alpha\beta\gamma}}=0\spacef,\label{eq:bianchiIdConformal}\\
			\pd{_\mu}\prn{\pt{C}{_{\alpha\beta\gamma}^\mu}}+\pt{Y}{_{\alpha\beta\gamma}}=0\spacef, \label{eq:bianchiIdPhysical}
			\end{align}
 		where $ \pt{C}{_{\alpha\beta\gamma}^\mu}\stackrel{\ps{M}}{=}\ct{C}{_{\alpha\beta\gamma}^\mu} $ is the Weyl tensor. The relation between the conformal and physical connections gives the relation
 			\begin{equation}
 				\cd{_{[\mu}}\ct{C}{_{\alpha\beta]\gamma}^\delta}\stackrel{\ps{M}}{=}\pd{_{[\mu}}\ct{C}{_{\alpha\beta]\gamma}^\delta}+\Omega^{-1}\prn{\ct{g}{_{\gamma[\mu}}\ct{C}{_{\alpha\beta]\rho}}^\delta\ct{N}{^\rho}-\delta^\delta_{[\mu}\ct{C}{_{\alpha\beta]\rho\gamma}}\ct{N}{^\rho}}
 			\end{equation}
 		 which, after taking the trace, yields 
 			\begin{equation}\label{eq:property-weyl-conformal-weyl-physical}
			\cd{_\mu}\prn{\Omega^{-1}\ct{C}{_{\alpha\beta\gamma}^\mu}}\stackrel{\ps{M}}{=} \Omega^{-1}\pd{_\mu}\ct{C}{_{\alpha\beta\gamma}^\mu}\spacef.
			\end{equation}
		Regarding the matter content, in \cite{Fernandez-Alvarez_Senovilla-dS} the following equation is derived
			\begin{align}
				2\varkappa\ct{N}{_{[\alpha}}\ct{T}{_{\beta]\gamma}}-\varkappa\ct{N}{_{[\alpha}}\ct{g}{_{\beta]\gamma}}\cs{T}+\Omega\varkappa\cd{_{[\alpha}}\ct{T}{_{\beta]\gamma}}-\frac{1}{3}\Omega\varkappa\ct{g}{_{\gamma[\beta}}\cd{_{\alpha]}}\cs{T}\nonumber\\ =\cd{_{[\alpha}}\ct{S}{_{\beta]\gamma}}+\ct{d}{_{\alpha\beta\gamma}^\mu}\ct{N}{_\mu}+\varkappa\ct{g}{_{\gamma[\alpha}}\ct{T}{_{\beta]\mu}}\ct{N}{^\mu}\quad				\label{eq:aux9}	
			\end{align}
 from where one can deduce that at $ \scri $
			\begin{equation}\label{eq:matter-scri}
				\ct{T}{_{\alpha\beta}} \eqs \mu \ct{N}{_{\alpha}}\ct{N}{_{\beta}}
			\end{equation}
		for some function $ \mu $ on $\scri$. Also, using the field equations into definition \eqref{eq:defSchouten-physical},
			\begin{equation}\label{eq:physical-cotton-matter}
				\pt{Y}{_{\alpha\beta\gamma}}=\kappa\pd{_{[\alpha}}\pt{T}{_{\beta]\gamma}}-\frac{1}{3}\varkappa\pt{g}{_{\gamma[\beta}}\cd{_{\alpha]}}\ps{T}\spacef,
			\end{equation}
		which in terms of $ \ct{T}{_{\alpha\beta}} $ and the unphysical connection reads
			\begin{equation}\label{eq:cottonPhysicalMatter}
			\frac{1}{\varkappa}\pt{Y}{_{\alpha\beta\gamma}}=\Omega\cd{_{[\alpha}}\ct{T}{_{\beta]\gamma}}-\ct{N}{^\lambda}\ct{T}{_{\lambda[\beta}}\ct{g}{_{\alpha]\gamma}}+2\ct{N}{_{[\alpha}}\ct{T}{_{\beta]\gamma}}-\ct{N}{_{[\alpha}}\ct{g}{_{\beta]\gamma}}\cs{T}-\frac{1}{3}\Omega\ct{g}{_{\gamma[\beta}}\cd{_{\alpha]}}\cs{T}\spacef.
			\end{equation}
		Taking into account \cref{eq:matter-scri}, the physical Cotton tensor vanishes at $ \scri $
			\begin{equation}\label{eq:physical-cotton-vanishing}
			\pt{Y}{_{\alpha\beta\gamma}}\eqs 0\spacef.
			\end{equation}
		This allows to define the \emph{rescaled Cotton tensor}
			\begin{equation}
			\ct{y}{_{\alpha\beta\gamma}} \defeq \Omega^{-1}\pt{Y}{_{\alpha\beta\gamma}}\spacef.
			\end{equation}
		In fact, \cref{eq:physical-cotton-vanishing} is one of the conditions involved in the vanishing of the Weyl tensor at $ \scri $ (see \cite{Kroon})
			\begin{lemma}[Vanishing of the Weyl tensor at $ \scri $]
				Assume that $ \scri $ has $ \mathbb{R}\times\mathbb{S}^2 $ topology and that
					\begin{enumerate}
					\item $ \pt{Y}{_{\alpha\beta\gamma}}\eqs 0 $ and $ \Omega\cd{_\sigma}\pt{Y}{_{\alpha\beta\gamma}}\eqs 0 $,
					\item $ \ct{C}{_{\alpha\beta\gamma}^\delta} $ is regular at $ \scri $ and $  \Omega\cd{_{\mu}}\ct{C}{_{\alpha\beta\gamma}^\mu}\eqs 0 $.
					\end{enumerate}
				Then,
					\begin{equation}
						\ct{C}{_{\alpha\beta\gamma}^\delta}\eqs 0\spacef.
					\end{equation}
			\end{lemma}
			\begin{proof}
			 We will use the lightlike decomposition of a Weyl candidate presented in section II of the companion paper \cite{Fernandez-Alvarez_Senovilla-dS}. First, from \cref{eq:bianchiIdPhysical,eq:property-weyl-conformal-weyl-physical} one has
					\begin{equation}\label{eq:aux1}
						\Omega\cd{_{\mu}}\ct{C}{_{\alpha\beta\gamma}^\mu}-\ct{N}{_{\mu}}\ct{C}{_{\alpha\beta\gamma}^\mu}+\Omega\pt{Y}{_{\alpha\beta\gamma}}= 0.
					\end{equation}
				This equation evaluated at $ \scri $ gives
					\begin{equation}\label{eq:aux2}
						\ct{N}{_{\mu}}\ct{C}{_{\alpha\beta\gamma}^\mu}\eqs 0,
					\end{equation}
				immediately implying that $ \ct{C}{_{\alpha\beta\gamma}^\mu} $ has Petrov type N at $ \scri $ with $ \ct{N}{^\alpha} $ the repeated PND. But this condition shows that the only components that survive at $ \scri $ are those of the symmetric traceless rank-2 tensor field
					\begin{equation}
						\ctcnl{E}{_{\alpha\beta}}\defeq\cts{\ell}{^\mu}\cts{\ell}{^\nu}\ctcn{P}{^\rho_\alpha}\ctcn{P}{^\sigma_\beta}\ct{C}{_{\rho\mu\sigma\nu}}\spacef,
					\end{equation}
				where $ \cts{\ell}{^\alpha} $ is any lightlike vector field on $ \scri $, \emph{which we choose to be orthogonal to $\mathbb{S}^2$-cuts}, such that $ \cts{\ell}{^\mu}\ct{N}{_{\mu}}\eqs-1 $ and
					\begin{equation}\label{eq:aux5}
						 \ctcn{P}{^\alpha_{\beta}}\defeq\delta^\alpha_{\beta}+\ct{N}{^\alpha}\cts{\ell}{_{\beta}} +\cts{\ell}{^\alpha}\ct{N}{_{\beta}}\spacef
					\end{equation}
				 is the projector to the two dimensional space orthogonal to $ \ct{N}{^\alpha} $ and $ \cts{\ell}{^\beta} $. Now, take the derivative of \cref{eq:aux1} and evaluate it at $ \scri $
					\begin{equation}
						\ct{N}{_{\sigma}}\cd{_{\mu}}\ct{C}{_{\alpha\beta\gamma}^\mu}-\ct{N}{_{\mu}}\cd{_{\sigma}}\ct{C}{_{\alpha\beta\gamma}^\mu}\eqs 0.
					\end{equation}
				Contract this equation with $ \cts{\ell}{^\sigma}\cts{\ell}{^\alpha}\cts{\ell}{^\gamma} $ to obtain
					\begin{equation}
						\ctcn{P}{^{\mu\rho}}\cd{_{\mu}}\prn{\cts{\ell}{^\sigma}\cts{\ell}{^\nu}\ct{C}{_{\sigma\beta\nu\rho}}}-\ct{C}{_{\alpha\beta\gamma\rho}}\ctcn{P}{^{\rho\mu}}\prn{\ctcn{P}{^\alpha_{\tau}}\cts{\ell}{^\gamma}\cd{_{\mu}}\cts{\ell}{^\tau}+\ctcn{P}{^\gamma_{\tau}}\cts{\ell}{^\alpha}\cd{_\mu}\cts{\ell}{^\tau}}\eqs 0,
					\end{equation}
				where we have used \cref{eq:aux2,eq:aux5}. If we contract now with $ \ctcn{P}{^\beta_{\delta}} $, we find
					\begin{equation}
						\ctcn{P}{^{\mu\rho}}\cd{_{\mu}}\prn{\ctcnl{E}{_{\beta\rho}}}+\prn{\ctcnl{s}{_{\gamma\delta}^\mu}+\ctcnl{s}{_\gamma^\mu_\delta}}\cd{_{\mu}}\cts{\ell}{^\alpha}\eqs 0,
					\end{equation}
				where $ \ctcnl{s}{_{\alpha\beta\gamma}}\defeq\cts{\ell}{^\mu}\ctcn{P}{^\rho_{\alpha}}\ctcn{P}{^\sigma_{\beta}}\ctcn{P}{^\nu_{\gamma}}\ct{C}{_{\rho\sigma\nu\mu}} $. But by the properties in appendix B of \cite{Fernandez-Alvarez_Senovilla-dS} and \cref{eq:aux2} this tensor field vanishes at $ \scri $, and therefore
					\begin{equation}
							\ctcn{P}{^{\mu\rho}}\cd{_{\mu}}\prn{\ctcnl{E}{_{\beta\rho}}}\eqs 0\spacef.
					\end{equation}
				This equation is equivalently written by means of the intrinsic connection on each cut $ \Sc $ defined by $ \cts{\ell}{^\alpha} $ as
					\begin{equation}
						\cdc{_{M}}\prn{\ctcnl{E}{_{A}^M}}\eqc 0\spacef.
					\end{equation}
				By assumption, the topology of the cuts is $ \mathbb{S}^2 $, and then $ \ctcnl{E}{_{AB}} $ is a traceless divergence-free symmetric tensor on $ \mathbb{S}^2 $ and must vanish \cite{Liu1998},
					\begin{equation}
						\ctcnl{E}{_{AB}}\eqc 0\spacef.
					\end{equation}
				Since this happens on any cut $\Sc\subset \scri $ with $\mathbb{S}^2$ topology and transversal to $ \ct{N}{^\alpha} $ --and $ \scri $ can be foliated by such cuts--, it holds everywhere on $ \scri $, implying
					\begin{equation}
						\ct{C}{_{\alpha\beta\gamma}^\delta}\eqs 0\spacef.
					\end{equation}
			\end{proof}
 		Accordingly, the \emph{rescaled Weyl tensor}
					\begin{equation}\label{eq:rescaledWeyl-def}
					\ct{d}{_{\alpha\beta\gamma}^\delta}\defeq\Omega^{-1} \ct{C}{_{\alpha\beta\gamma}^\delta}\spacef
					\end{equation}
		 is therefore well defined at $ \scri $. Let us introduce the scalar \cite{Friedrich2002}
			\begin{equation}\label{eq:friedrichscalar}
			\cs{f}\defeq \frac{1}{4}\cd{_\mu}\ct{N}{^\mu}+\frac{\Omega}{24}\cs{R}\quad 
			\end{equation}	
		which vanishes at $ \scri $ in our gauge.
		The fields $ \prn{\Omega,\ct{d}{_{\alpha\beta\gamma}^\mu}, \cs{f},\ct{g}{_{\alpha\beta}},\ct{S}{_{\alpha\beta}}} $ satisfy \cite{Friedrich81,Friedrich81b} (see, e.g., \cite{Paetz2013,Kroon} for a review and detailed derivation of the equations, respectively)
			\begin{align}
				&\cd{_\alpha}\ct{N}{_\beta} = -\frac{1}{2}\Omega\ct{S}{_{\alpha\beta}}+ \cs{f}\ct{g}{_{\alpha\beta}}+ \frac{1}{2}\Omega^2\varkappa\ct{\underline{T}}{_{\alpha\beta}}\quad ,\label{eq:cefesDerN}\\
				&\ct{N}{_\mu}\ct{N}{^\mu}=\frac{\Omega^3}{12}\varkappa \cs{T}+ 2\Omega f\quad ,\label{eq:cefesNormN}\\
				&\cd{_\alpha}f= -\frac{1}{2}\ct{S}{_{\alpha\mu}}\ct{N}{^\mu}+\frac{1}{2}\Omega\varkappa\ct{N}{^\mu}\ct{\underline{T}}{_{\alpha\mu}}-\frac{1}{24}\Omega^2\varkappa\cd{_\alpha}\cs{T}- \frac{1}{8}\Omega\varkappa\ct{N}{_\alpha}\cs{T}\spacef,\label{eq:cefesDerF}\\
				&\ct{d}{_{\alpha\beta\gamma}^\mu}\ct{N}{_\mu}+\cd{_{[\alpha}}\prn{\ct{S}{_{\beta]\gamma}}}-\Omega\ct{y}{_{\alpha\beta\gamma}}=0\spacef,\label{eq:cefesDerSchouten}\\
				&\ct{y}{_{\alpha\beta\gamma}}+\cd{_{\mu}}\ct{d}{_{\alpha\beta\gamma}^\mu}= 0\spacef,\label{eq:cefesDerWeyl}\\
				&\ct{R}{_{\alpha\beta\gamma\delta}} =\Omega\ct{d}{_{\alpha\beta\gamma\delta}}+ \ct{g}{_{\alpha[\gamma}}\ct{S}{_{\delta]\beta}}-\ct{g}{_{\beta[\gamma}}\ct{S}{_{\delta]\alpha}}\spacef.	\label{eq:cefesRiemann}						
			\end{align}
			They constitute the conformal EFE --see also \cite{Paetz2013}.
		Evaluation at $ \scri $ yields 
			\begin{align}
				&\cd{_\alpha}\ct{N}{_\beta} \eqs 0\quad ,\label{eq:cefesScriDerN}\\
				&\ct{N}{_\mu}\ct{N}{^\mu}\eqs 0 \quad ,\label{eq:cefesScriNormN}\\
				&\cd{_\alpha}f\eqs -\frac{1}{2}\ct{S}{_{\alpha\mu}}\ct{N}{^\mu}\spacef,\label{eq:cefesScriDerF}\\
				&\ct{d}{_{\alpha\beta\gamma}^\mu}\ct{N}{_\mu}+\cd{_{[\alpha}}\prn{\ct{S}{_{\beta]\gamma}}}\eqs 0\spacef,\label{eq:cefesScriDerSchouten}\\
				&\ct{y}{_{\alpha\beta\gamma}}+\cd{_{\mu}}\ct{d}{_{\alpha\beta\gamma}^\mu}\eqs 0\spacef,\label{eq:cefesScriDerWeyl}\\
				&\ct{R}{_{\alpha\beta\gamma\delta}} \eqs\ct{g}{_{\alpha[\gamma}}\ct{S}{_{\delta]\beta}}-\ct{g}{_{\beta[\gamma}}\ct{S}{_{\delta]\alpha}}\spacef.	\label{eq:cefesScriRiemann}	
			\end{align}
		Relevant properties of the fields at infinity can be deduced from \cref{eq:cefesScriDerF,eq:cefesScriDerN,eq:cefesScriDerSchouten,eq:cefesScriDerWeyl,eq:cefesScriNormN,eq:cefesScriRiemann}. In particular, from \cref{eq:physical-cotton-matter,eq:cefesScriDerWeyl}
			\begin{equation}\label{eq:matter-decay-cotton}
			\pt{T}{_{\alpha\beta}}\lvert_{\scri}\sim \mathcal{O}\prn{\Omega^p} \text{ with } p>2  \implies \ct{y}{_{\alpha\beta\gamma}}\eqs 0\eqs\cd{_{\mu}}\ct{d}{_{\alpha\beta\gamma}^\mu} \spacef.
			\end{equation}
		A more detailed description of the components of $ \ct{S}{_{\alpha\beta}} $, $ \ct{d}{_{\alpha\beta\gamma}^\delta} $ and their relation with fields within $ \scri $ is presented next in \cref{ssec:curvature-scri-and-fields}.
		
	\subsection{Curvature on $ \scri $ and its relation to space-time fields}\label{ssec:curvature-scri-and-fields}
		If one considers the Gauss relation \eqref{eq:gauss-scri} and uses \cref{eq:cefesScriRiemann}, \cref{eq:riemann-scri-schouten} is obtained with
			\begin{align}
				\cts{S}{_{ab}}&\defeqs\frac{1}{2}\ct{e}{^\alpha_{a}}\ct{e}{^\beta_{b}}\ct{S}{_{\alpha\beta}}\spacef,\\
				\cts{S}{_{a}^b}&\defeqs\frac{1}{2}\ct{e}{^\alpha_{a}}\ct{\omega}{_\beta^{b}}\ct{S}{_{\alpha}^\beta}.
			\end{align}
		For simplicity, consider a foliation given by $ F $ with $ \cts{\ell}{_{a}} $ as in \cref{eq:ell-F-foliation} and $ \cts{\ell}{_{\alpha}}\defeq\ct{\omega}{_{\alpha}^a}\cts{\ell}{_{a}} $ determined by $ \cts{\ell}{_{\mu}}\cts{\ell}{^\mu}=0 $ --see \cref{ssec:basic-geometry}--. Now, since $ \cs{f}\eqs 0 $, \cref{eq:cefesScriDerF} gives
			\begin{align}
				\ct{N}{^r}\cts{S}{_{r}^a}&=\ct{N}{^a}\lied_{\vec{\ell}}f\spacef,\label{eq:aux3}\\
				\ct{N}{^r}\cts{S}{_{ra}}&=0\spacef.\label{eq:aux4}
			\end{align}
		Provided \cref{eq:cefesScriDerN}, a general formula \cite{Yano1957} gives
			\begin{equation}
				\lied_{\vec{N}}\cts{\Gamma}{^a_{bc}}=\cts{R}{_{cdb}^a}\ct{N}{^d}
			\end{equation}
		which in conjunction with \cref{eq:riemann-scri-schouten,eq:trace-Schouten-ricci-scri,eq:trace-lower-Schouten-ricci-scri} and \cref{eq:aux3,eq:aux4} leads to
			\begin{align}
				\lied_{\vec{N}}\cts{\Gamma}{^a_{bc}}&=\ct{N}{^a}\cts{S}{_{bc}}+\ms{_{bc}}\ct{N}{^m}\cts{S}{_{m}^a}=\ct{N}{^a}\prn{\cts{S}{_{bc}}+\ms{_{bc}}\lied_{\vec{\ell}}f}\spacef,\\
				\lied_{\vec{\ell}}f&=\cts{S}{_{m}^m}-\cts{S}{_{mn}}\ms{^{mn}}=\cts{S}{_{m}^m}-\frac{\csS{R}}{2}\spacef.
			\end{align}
		In a middle step in deriving the second formula, we have contracted the second and fourth indices of \cref{eq:riemann-scri-schouten} with $ \delta^a_b=-\ct{N}{^a}\cts{\ell}{_{b}}+\ctcn{P}{^a_b} $. Also, a direct calculation using \cref{eq:riemann-scri-covariant-Schouten,eq:aux3} yields
		 	\begin{equation}
		 		\ct{N}{^c}\cts{R}{_{cab}^d}\cts{\ell}{_{d}}=\ms{_{ab}}\lied_{\vec{\ell}}f+\cts{S}{_{ab}}\spacef,
		 	\end{equation}
		 whereas application of the `Ricci identity' leads to
		 	\begin{equation}
		 		\ct{N}{^c}\cds{_{c}}\ctcn{\Theta}{_{ab}}-\lied_{\vec{N}}\cts{\ell}{_{a}}\lied_{\vec{N}}\cts{\ell}{_{b}}-\ctcn{P}{^m_{(a}}\ctcn{P}{^n_{b)}}\cds{_{m}}\lied_{\vec{N}}\cts{\ell}{_{n}}=\ms{_{ab}}\lied_{\vec{\ell}}f+\cts{S}{_{ab}}\spacef,\label{eq:expansion-tensor-schouten}
		 	\end{equation}
		 where we have symmetrised the free indices, introduced \eqref{eq:expansion-tensor-ell} and taken into account that
		 	\begin{equation}
		 		\ct{N}{^d}\cds{_{d}}\ctcn{P}{^a_{b}}=\ct{N}{^a}\ct{N}{^d}\cds{_{d}}\cts{\ell}{_{b}}\spacef.
		 	\end{equation}
		 One can take the trace-free part of \cref{eq:expansion-tensor-schouten},
		 	\begin{align}
	 			\ct{N}{^c}\cds{_{c}}\ctcn{\sigma}{_{ab}}&=\cts{S}{_{ab}}-\frac{1}{2}\ms{_{ab}}\ms{^{mn}}\cts{S}{_{mn}}+\lied_{\vec{N}}\cts{\ell}{_{a}}\lied_{\vec{N}}\cts{\ell}{_{b}}+\ctcn{P}{^m_{(a}}\ctcn{P}{^n_{b)}}\cds{_{m}}\lied_{\vec{N}}\cts{\ell}{_{n}}\nonumber\\
	 			&-\frac{1}{2}\ms{_{ab}}\ms{^{mn}}\prn{\lied_{\vec{N}}\cts{\ell}{_{m}}\lied_{\vec{N}}\cts{\ell}{_{n}}+\ctcn{P}{^e_{n}}\ctcn{P}{^f_{m}}\cds{_{f}}\lied_{\vec{N}}\cts{\ell}{_{e}}}\spacef,\label{eq:shear-schouten}
		 	\end{align}
		 which in terms of the function $ F $ giving the foliation --see \cref{eq:ell-F-foliation,eq:lied-ell-F-foliation}-- reads
		 	\begin{align}
		 		\ct{N}{^c}\cds{_{c}}\ctcn{\sigma}{_{ab}}&=\cts{S}{_{ab}}-\frac{1}{2}\ms{_{ab}}\ms{^{mn}}\cts{S}{_{mn}}+\ctcn{P}{^m_{a}}\cds{_{m}}\prn{\ln\dot{F}}\ctcn{P}{^n_{b}}\cds{_{n}}\prn{\ln\dot{F}}-\ctcn{P}{^m_{(a}}\ctcn{P}{^n_{b)}}\cds{_{n}}\prn{\ctcn{P}{^r_{m}}\cds{_{r}}\ln\dot{F}}\nonumber\\
		 		&-\frac{1}{2}\ms{_{ab}}\ms{^{ef}}\brkt{\ctcn{P}{^m_{e}}\cds{_{m}}\prn{\ln\dot{F}}\ctcn{P}{^n_{f}}\cds{_{n}}\prn{\ln\dot{F}}-\ctcn{P}{^m_{e}}\ctcn{P}{^n_{f}}\cds{_{n}}\prn{\ctcn{P}{^r_{m}}\cds{_{r}}\ln\dot{F}}}\spacef.\label{eq:shear-schouten-foliation}
		 	\end{align}
		 On each cut, one can take the pullback with $ \cbrkt{\ct{E}{^a_{A}}} $ to find
		 	\begin{align}
		 		\ct{N}{^c}\cds{_{c}}\ctcn{\sigma}{_{AB}}&\eqc\ctc{S}{_{AB}}-\frac{1}{2}\mc{_{AB}}\ctc{S}{_{M}^M}+\cdc{_{A}}\prn{\ln\dot{F}}\cdc{_{B}}\prn{\ln\dot{F}}+\cdc{_{A}}\cdc{_{B}}\ln\dot{F}\nonumber\\
		 		&-\frac{1}{2}\mc{_{AB}}\brkt{\cdc{_{M}}\cdc{^{M}}\ln\dot{F}+\cdc{_{M}}\prn{\ln\dot{F}}\cdc{^{M}}\prn{\ln\dot{F}}}\spacef,
		 	\end{align}
		 where \cref{eq:lied-bases-foliation} has been used and we have introduced
		 	\begin{equation}\label{eq:schouten-cuts-traces}
		 		\ctc{S}{_{AB}}\defeqc\ct{E}{^a_{A}}\ct{E}{^b_{B}}\cts{S}{_{ab}},\hspace{1cm} \ctc{S}{_{M}^M}\eqc\cts{S}{_{mn}}\ms{^{mn}}\eqc\frac{\csS{R}}{2}\eqc\frac{\csC{R}}{2}\spacef.
		 	\end{equation}
		\\
		
		Now, let us consider the following `lightlike projections' of the rescaled Weyl tensor,
			\begin{align}
				\ctn{D}{^{\alpha\beta}}&\defeqs\ct{N}{^\mu}\ct{N}{^\nu}\ct{d}{_\mu^\alpha_\nu^\beta}=\ctn{D}{^{ab}}\ct{e}{^\alpha_{a}}\ct{e}{^\beta_{b}}\spacef,\label{eq:DpDef}\\					\ctn{C}{^{\alpha\beta}}&\defeqs\ct{N}{^\mu}\ct{N}{^\nu}\ctru{*}{d}{_\mu^\alpha_\nu^\beta}\eqs\ctn{C}{^{ab}}\ct{e}{^\alpha_{a}}\ct{e}{^\beta_{b}}\spacef,\label{eq:CpDef}
			\end{align}
		where
			\begin{equation}
				\ctru{*}{d}{_{\alpha\beta\gamma}^\delta}\defeq\frac{1}{2}\ct{\eta}{_{\alpha\beta}^{\mu\nu}}\ct{d}{_{\mu\nu\gamma}^\delta}.
			\end{equation}	
		see \cite{Fernandez-Alvarez_Senovilla-dS} for this kind of decomposition in general; some of the properties listed therein will be used too. Contract \cref{eq:cefesScriDerSchouten} with $ \ct{N}{^{\beta}} $, raise the index $ \gamma $ and contract with $ \ct{e}{^\alpha_{a}}\ct{\omega}{_{\gamma}^b}$ to get
			\begin{equation}
				\ctn{D}{_{a}^b}\defeq\ms{_{ma}}\ctn{D}{^{mb}}=\ct{N}{^m}\cds{_{m}}\cts{S}{_{a}^b}-\ct{N}{^b}\cds{_{a}}\prn{\lied_{\vec{N}}f}\spacef.
			\end{equation}
		One may lower the contravariant index with $ \ms{_{ab}} $ so that
			\begin{equation}\label{eq:Dcovariant-schouten}
				\ctn{D}{_{ab}}\defeq\ms{_{mb}}\ctn{D}{_{a}^m}=\ct{N}{^m}\cds{_{m}}\cts{S}{_{ab}} = \lied_{\vec{N}}\cts{S}{_{ab}} \spacef.
			\end{equation}
		Notice that $ \ctn{D}{_{ab}} $ is symmetric and effectively two-dimensional $ \ct{N}{^m}\ctn{D}{_{am}}=0 $. In addition, if firstly one takes the Hodge dual of \cref{eq:cefesScriDerSchouten} with $ \ct{\eta}{^{\alpha\beta\gamma\delta}} $ and contract once with $ \ct{N}{^\alpha} $ and the remaining two free indices with $ \ct{\omega}{_\alpha^{a}} $, then
			\begin{equation}\label{eq:Ccontravariant-schouten}
				\ctn{C}{^{ab}}=\ct{\epsilon}{^{rsa}}\cds{_{r}}\cts{S}{_{s}^b}
			\end{equation}
		follows. Also, lowering an index,
			\begin{equation}
				\ctn{C}{^{a}_b}\defeq\ms{_{mb}}\ctn{C}{^{am}}=\ct{\epsilon}{^{rsa}}\cds{_{r}}\cts{S}{_{sb}}\spacef.
			\end{equation}
		It will become useful to consider the component
			\begin{equation}\label{eq:CA-schouten}
				-\sqrt{2}\ctcnn{C}{_{a}}\defeqs\cts{\ell}{_{r}}\ctn{C}{^r_{a}}=\cts{\ell}{_r}\ct{\epsilon}{^{mpr}}\cds{_{m}}\cts{S}{_{pa}}\spacef.
			\end{equation}
		On each cut, projecting with $ \ct{E}{^a_{A}} $, one has
			\begin{equation}
				-\sqrt{2}\ctcnn{C}{_{A}}\eqc\ctc{\epsilon}{^{MP}} \cdc{_{M}}\ctc{S}{_{PA}}\spacef.
			\end{equation}
		By general properties presented in \cite{Fernandez-Alvarez_Senovilla-dS}, one also has $ \ctcnn{D}{_{A}}=\ctc{\epsilon}{_{AB}}\ctcnn{C}{^{B}} $, hence
			\begin{equation}
				-\sqrt{2}\ctcnn{D}{_{A}}\defeqc \ct{E}{^a_A}\cts{\ell}{_{m}}\ctn{D}{^m_{a}}\eqc 2\cdc{_{[M}}\ctc{S}{_{A]}^M}.
			\end{equation}
		From \cref{eq:cefesScriDerWeyl}, contracting twice with $ \ct{N}{^\alpha} $, one arrives at
			\begin{equation}\label{eq:divDscri}
				\cds{_{m}}\ctcnn{D}{^{bm}}=-\ct{y}{_{m}^b_{f}}\ct{N}{^{m}}\ct{N}{^{f}},
			\end{equation}
		which by means of the rescaled energy momentum tensor $ \ct{T}{_{\alpha\beta}} $ reads --see \cref{eq:cottonPhysicalMatter}--
			\begin{equation}
				\cds{_{m}}\ctcnn{D}{^{bm}}\eqs\varkappa\ct{N}{^\mu}\ct{N}{^\nu}\ctrd{0}{T}{_{\mu\nu}}\ct{N}{^b}\spacef,
			\end{equation}
		where  
			\begin{equation}\label{eq:cotton-matter-contribution}
				\ct{N}{^\mu}\ct{N}{^\nu}\ctrd{0}{T}{_{\mu\nu}}\defeqs\Omega^{-1}\varkappa\ct{N}{^\mu}\ct{N}{^\nu}\ct{T}{_{\mu\nu}}
			\end{equation}
		is regular at $ \scri $ because $ \ct{T}{_{\mu\nu}}\ct{N}{^\mu}\ct{N}{^\nu}\eqs0 $ due to \cref{eq:matter-scri}. \Cref{eq:divDscri} may be expanded and contracted with $ \cts{\ell}{_{b}} $ to get on any cut
			\begin{equation}\label{eq:matter-term}
					\cts{\ell}{_{m}}\cts{\ell}{_{n}}\ct{N}{^p}\cds{_{p}}\ctcnn{D}{^{nm}}\eqc -\sqrt{2}\cdc{_M}\ctcnn{D}{^{M}}-\ctcn{\sigma}{_{AB}}\ct{N}{^m}\cds{_{m}}\ctc{S}{^{AB}}+\ct{y}{_{m}^b_{f}}\ct{N}{^{m}}\cts{\ell}{_{b}}\ct{N}{^{f}}\spacef.
			\end{equation}

%% file: section-3/section-3.tex
	This section is devoted to the study of the asymptotic group of symmetries at $ \scri $, the isolation of the radiative degrees of freedom of the gravitational field and the definition of an asymptotic energy-momentum, which are closely related tasks. 
	\subsection{Geroch's tensor rho and news tensor}
		A result by Geroch \cite{Geroch1977} gives the existence and uniqueness of a symmetric tensor field $ \ct{\rho}{_{ab}} $ on $ \scri $ whose gauge behaviour and differential properties play a fundamental role in finding the so called `news' tensor, $ \ct{N}{_{ab}} $ -- in the classical characterisation, the tensor field which determines the presence of outgoing gravitation radiation at $ \scri $. In \cite{Fernandez-Alvarez_Senovilla-dS}, related general results for two dimensional Riemannian manifolds are proven --see. Those results can be particularised for the present case, leading to Geroch's tensor. However, we take a different approach here due to the particular structure of the three-dimensional manifold $ \scri $.
			\begin{lemma}\label{thm:general-gauge-trace}
				Let $ \ct{t}{_{ab}} $ be any symmetric tensor field on $ \scri $, orthogonal to $ \ct{N}{^a} $, whose behaviour under conformal rescalings \eqref{eq:gauge-ms} is
					\begin{equation}\label{eq:gauge-behaviour-appropriate}
							\ctg{t}{_{ab}}=\ct{t}{_{ab}}-a\frac{1}{\omega}\cds{_a}\cts{\omega}{_b}+\frac{2a}{\omega^2}\cts{\omega}{_a}\cts{\omega}{_b}-\frac{a}{2\omega^2}\cts{\omega}{_c}\cts{\omega}{^c}\ms{_{ab}}
					\end{equation}
				for some fixed constant $ a\in\mathbb{R} $, where 
				\begin{equation}\label{eq:omegaN}
				 \cts{\omega}{_{a}}\defeqc \cds{_{a}}\omega , \hspace{1cm} \ct{N}{^a}\cts{\omega}{_{a}}=0 \, \, .
				 \end{equation}
				 
				 Then,
					\begin{equation}\label{eq:gauge-derivative-appropiate-behaviour}
						\cdsg{_{[c}}\ctg{t}{_{a]b}}=\cds{_{[c}}\ct{t}{_{a]b}}+\frac{1}{\omega}\prn{a\frac{\csS{R}}{2}-\ms{^{ed}}\ct{t}{_{ed}}}\cts{\omega}{_{[c}}\ms{_{a]b}}\quad .
					\end{equation}						
				In particular, for any symmetric gauge-invariant tensor field $ \ct{B}{_{ab}} $ on $ \scri $ orthogonal to $ \ct{N}{^a} $,
					\begin{equation}\label{eq:aux23}
						\cdsg{_{[c}}\ctg{B}{_{a]b}}=\cds{_{[c}}\ct{B}{_{a]b}}-\frac{1}{\omega}\ct{B}{_{ed}}\ms{^{ed}}\cts{\omega}{_{[c}}\ms{_{a]b}}\quad\spacef.
					\end{equation}
			\end{lemma}
				\begin{proof}
			A direct calculation yields
					\begin{equation}
						\cdsg{_{[c}}\ctg{t}{_{a]b}}=\cds{_{[c}}\ct{t}{_{a]b}}+\frac{1}{\omega}\ct{t}{_{b[c}}\cts{\omega}{_{a]}}+\frac{1}{\omega}\ms{_{b[c}}\ct{t}{_{a]e}}\ms{^{de}}\cts{\omega}{_d}+\frac{1}{\omega} \frac{a\csS{R}}{2}\cts{\omega}{_{[c}}\ms{_{a]b}}\spacef.
					\end{equation}
				Observe that the term
					\begin{equation}
					\frac{1}{\omega}\ct{t}{_{b[c}}\cts{\omega}{_{a]}}+\frac{1}{\omega}\ms{_{b[c}}\ct{t}{_{a]e}}\ms{^{de}}\cts{\omega}{_d}
					\end{equation}
					
				is effectively two-dimensional (it is orthogonal to $ \ct{N}{^a} $,due to \eqref{eq:omegaN}). Hence, one can use the two-dimensional identity \cite{Edgar2002}
					\begin{equation}\label{eq:dimensional-identity}
						\ct{A}{_{CAE}}=2\mc{_{E[A}}\ct{A}{_{C]DM}}\mc{^{DM}},\text{ for any tensor  such that }\ct{A}{_{CAE}}=-\ct{A}{_{ACE}}\quad
					\end{equation}
				in order to write
					\begin{equation}
						\frac{1}{\omega}\ct{t}{_{b[c}}\cts{\omega}{_{a]}}+\frac{1}{\omega}\ms{_{b[c}}\ct{t}{_{a]e}}\ms{^{de}}\cts{\omega}{_d}=-\frac{1}{\omega}\ct{t}{_{ed}}\ms{^{ed}}\cts{\omega}{_{[c}}\mc{_{a]b}}\quad,
					\end{equation}
				arriving at the final result. For a gauge invariant tensor $ a=0 $ in \cref{eq:gauge-behaviour-appropriate}, therefore one only has to set this value in \cref{eq:gauge-derivative-appropiate-behaviour} to obtain \cref{eq:aux23}.
			\end{proof}
			\begin{corollary}\label{thm:gauge-invariant-diff}
				A symmetric gauge-invariant tensor field $ \ct{m}{_{ab}} $ on $ \scri $, orthogonal to $ \ct{N}{^a} $, satisfies
					\begin{equation}
						\cds{_{[c}}\ctg{m}{_{b]a}}=\cds{_{[c}}\ct{m}{_{b]a}}
					\end{equation}
				if and only if $ \ct{m}{_{ed}}\ms{^{ed}}=0 $.
			\end{corollary}
			\begin{corollary}[The tensor $ \rho $]\label{thm:rho-tensor}
				There is a unique symmetric tensor field $ \ct{\rho}{_{ab}} $ on $ \scri $ orthogonal to $ \ct{N}{^a} $ whose behaviour under conformal rescalings \eqref{eq:gauge-ms} is as in \eqref{eq:gauge-behaviour-appropriate} and satisfies the equation
					\begin{equation}\label{eq:rho-diff-eq}
						\cds{_{[c}}\ct{\rho}{_{a]b}}=0
					\end{equation}
				in any conformal frame. This tensor field must have a trace $ \ct{\rho}{_{ed}}\ms{^{ed}}=a\csS{R}/2 $
				and is given in the gauge where the cuts of $ \scri $ are endowed with the round metric by $ \ct{\rho}{_{ab}}=\ms{_{ab}}a\csS{R}/4 $.
			\end{corollary}	
				\begin{proof}
					Existence is proved by noticing that $ \ct{\rho}{_{ab}}=\ms{_{ab}}a\csS{R}/4 $ fulfils $ \cds{_a}\ct{\rho}{_{bc}} =0$ in the round metric sphere. Concerning uniqueness, notice that \cref{thm:general-gauge-trace} fixes the trace of $ \ct{\rho}{_{ab}} $ to $ \ct{\rho}{_{ed}}\ms{^{ed}}=a\csS{R}/2 $, and recall the assumption that \cref{eq:rho-diff-eq} holds in any gauge. Then, if two different solutions $ \ctrd{1}{\rho}{_{ab}} $ and $ \ctrd{2}{\rho}{_{ab}} $ exist, $ \cds{_{[c}}\prn{\ctrd{1}{\rho}{_{a]b}}-\ctrd{2}{\rho}{_{a]b}}}=0 $. However, in that case and since $ \ct{\rho}{_{ad}}\ct{N}{^d}=0 $, the difference $ \ctrd{1}{\rho}{_{ab}}-\ctrd{2}{\rho}{_{ab}} $ is a traceless, Codazzi tensor on $ \mathbb{S}^2 $ and, as a consequence of the uniqueness of this kind of tensors \cite{Liu1998}, $ \ctrd{1}{\rho}{_{ab}}-\ctrd{2}{\rho}{_{ab}}=0 $.
			\end{proof}
			\begin{remark}
				Geroch's original tensor corresponds to $ a=1 $.
			\end{remark}
			\begin{remark}
				Since these results are in essence two-dimensional, one could have taken a different path for the proof. Namely, use the general results for tensors $ \rho $ on two-dimensional Riemannian manifolds presented in \cite{Fernandez-Alvarez_Senovilla-dS}.
			\end{remark}
			\begin{remark}
				Applying the results of \cite{Fernandez-Alvarez_Senovilla-dS}, one also finds that the Lie derivative on a cut of  the projection to that cut  of $ \ct{\rho}{_{ab}} $ along any conformal Killing vector field (CKVF) $ \ct{\chi}{^A} $ is proportional to $ \cdc{_{A}}\cdc{_{B}}\cdc{_{C}}\ct{\chi}{^C} $, and in particular vanishes for Killing vector fields (KVF).
			\end{remark}
			\begin{remark}\label{rem:rhoinvariant}
				Contraction of \cref{eq:rho-diff-eq} with $ \ct{N}{^a} $ gives
					\begin{equation}
						\lied_{\vec{N}}\ct{\rho}{_{ab}}=\ct{N}{^e}\cds{_{e}}\ct{\rho}{_{ab}}=0\spacef.
					\end{equation}
				Therefore, $ \ct{\rho}{_{ab}} $ is `constant' along the generators of $ \scri $. 
			\end{remark}
		A direct calculation of the gauge behaviour of $ \cts{S}{_{ab}} $ shows that
			\begin{equation}\label{eq:schouten-gauge}
				\ctsg{S}{_{ab}}=\cts{S}{_{ab}}-\frac{1}{\omega}\cds{_a}\cts{\omega}{_b}+\frac{2}{\omega^2}\cts{\omega}{_a}\cts{\omega}{_b}-\frac{1}{2\omega^2}\cts{\omega}{_c}\cts{\omega}{^c}\ms{_{ab}}\spacef,
			\end{equation}
		with $ \cts{\omega}{_{a}}\defeqs\cds{_{a}}\omega $. Also, this can be projected to any cut $ \Sc $
			\begin{equation}
				\ctcg{S}{_{AB}}\eqc\ctc{S}{_{AB}}-\frac{1}{\omega}\cdc{_A}\ctc{\omega}{_B}+\frac{2}{\omega^2}\ctc{\omega}{_A}\ctc{\omega}{_B}-\frac{1}{2\omega^2}\ctc{\omega}{_C}\ctc{\omega}{^C}\mc{_{AB}}\spacef,
			\end{equation}
		where we have used the second in \eqref{eq:omegaN}. 
		That is, $ \cts{S}{_{ab}} $ has the adequate gauge behaviour and trace \eqref{eq:schouten-cuts-traces} that imply by \cref{thm:general-gauge-trace}
			\begin{equation}
				\cds{_{[c}}\ctcg{S}{_{a]b}}=\cds{_{[c}}\ctc{S}{_{a]b}}\spacef
			\end{equation}
		and allows to write the following result
			\begin{prop}[News tensor]\label{thm:news}
				The tensor field on $ \scri $
					\begin{equation}\label{eq:schouten-news-rho}
					\ct{N}{_{ab}}\defeq \cts{S}{_{ab}}-\ct{\rho}{_{ab}}\spacef,
					\end{equation}
				is symmetric, traceless, gauge invariant, orthogonal to $ \ct{N}{^a} $ and satisfies the gauge-invariant equation
					\begin{equation}\label{eq:diffNab}
					\cds{_{[a}}\cts{S}{_{b]c}}=\cds{_{[a}}\ct{N}{_{b]c}}\spacef,
					\end{equation}
				where $ \ct{\rho}{_{ab}} $ is the tensor field of  \cref{thm:rho-tensor} (for $ a=1 $). Besides, $ \ct{N}{_{ab}} $ is unique with these properties.
			\end{prop}	
			\begin{proof}
					The tensor field $ \ct{N}{_{ab}} $ is symmetric, traceless, gauge invariant and orthogonal to $ \ct{N}{^a} $ as a consequence of \cref{eq:aux4,eq:schouten-gauge,thm:rho-tensor}. That \cref{eq:diffNab} is gauge invariant follows from \cref{thm:gauge-invariant-diff}. The uniqueness of $ \ct{N}{_{ab}} $ is a consequence of \cref{thm:rho-tensor} too and \cref{eq:diffNab}.
			\end{proof}	
		The	tensor field $ \ct{N}{_{ab}} $ is the \emph{news tensor}, and can be projected on any cut $ \Sc $
			\begin{equation}\label{eq:news-tensor}
				\ct{N}{_{AB}}\defeq \ctcn{E}{^{a}_A}\ctcn{E}{^{b}_B}\ct{N}{_{ab}}\spacef,\quad\mc{^{MN}}\ct{N}{_{MN}}=0\spacef,
			\end{equation}
		and the same can be done with $ \ct{\rho}{_{ab}}$
			\begin{equation}
				\ct{\rho}{_{AB}}\defeq \ctcn{E}{^{a}_A}\ctcn{E}{^{b}_B}\ct{\rho}{_{ab}} \, \, .
			\end{equation}
		
	Note that contraction of \cref{eq:diffNab} with $ \ct{N}{^c} $ and $ \ct{E}{^c_{C}}\ct{E}{^a_{A}}\ct{E}{^b_{B}} $, respectively, yields
			\begin{equation}\label{eq:news-schouten-derivatives}
				\lied_{\vec{N}}\cts{S}{_{ab}}=\ct{N}{^c}\cds{_{c}}\cts{S}{_{ab}}=\ct{N}{^c}\cds{_{c}}\ct{N}{_{ab}}=\lied_{\vec{N}}\ct{N}{_{ab}}\spacef,\quad\cdc{_{[C}}\ctc{S}{_{A]B}}\eqc\cdc{_{[C}}\ct{N}{_{A]B}}\spacef
			\end{equation}
	and also observe that in general, one has
			\begin{equation}
				\lied_{\vec{N}}\ct{N}{_{ab}}\neq 0
			\end{equation}
	and the notation 
			\begin{equation}
				\ctd{N}{_{AB}}\defeq \ctcn{E}{^a_{A}}\ctcn{E}{^b_{B}}\lied_{\vec{N}}\ct{N}{_{ab}}
			\end{equation}
	will be used. From \cref{eq:Ccontravariant-schouten,eq:Dcovariant-schouten} and \cref{eq:news-schouten-derivatives}, one gets
			\begin{align}
				\ctn{C}{^a_{b}}&=\ct{\epsilon}{^{rsa}}\cds{_{r}}\ct{N}{_{sb}}\spacef,\label{eq:Ccontravariant-news}\\
				\ctn{D}{_{ab}}&=\ct{N}{^m}\cds{_{m}}\ct{N}{_{ab}}\spacef\label{eq:Dcovariant-news}
			\end{align}
		and from \cref{eq:CA-schouten}
			\begin{equation}
				-\sqrt{2}\ctcnn{C}{_{a}}=\cts{\ell}{_r}\ct{\epsilon}{^{mpr}}\cds{_{m}}\ct{N}{_{pa}}\spacef.
			\end{equation}
		On each cut
			\begin{align}
				\ctcnn{C}{^{A}_B}&\defeqc\ct{W}{_{a}^A}\ct{E}{^b_{B}}\ctn{C}{^a_{b}}\eqc\ctd{N}{_{BM}}\ctc{\epsilon}{^{AM}}\spacef,\label{eq:CAB-news}\\
				\ctcnn{C}{_{A}}&\defeqc\ct{E}{^a_{A}}\ctcnn{C}{_{a}}=-\frac{1}{\sqrt{2}}\ctc{\epsilon}{^{RP}}\cdc{_{R}}\ct{N}{_{PA}}\spacef\label{eq:CA-news},\\
				\ctcnn{D}{_{AB}}&\defeqc \ct{E}{^a_{A}}\ct{E}{^b_{B}}\ctn{D}{_{ab}}\eqc\ct{\dot{N}}{_{AB}},\label{eq:DAB-news}\\
				\ctcnn{D}{_{A}} &\eqc -\frac{1}{\sqrt{2}}\cdc{_{M}}\ctc{N}{_{A}^M}\spacef,\label{eq:DA-news}
			\end{align}
		where  we have used \cref{eq:vol-form-cut}, \cref{eq:lied-bases-foliation}.
		Some of these formulae will be used in \cref{sec:smomentum}.\\
		
		Looking back to \cref{eq:shear-schouten-foliation}, inserting decomposition \eqref{eq:schouten-news-rho}, it follows that
			\begin{align}
		 		\ct{\dot{\sigma}}{_{ab}}\defeq\ct{N}{^c}\cds{_{c}}\ctcn{\sigma}{_{ab}}&=\ct{N}{_{ab}}+\ct{\rho}{_{ab}}-\frac{\csS{R}}{4}\ms{_{ab}}+\ctcn{P}{^m_{a}}\cds{_{m}}\prn{\ln\dot{F}}\ctcn{P}{^n_{b}}\cds{_{n}}\prn{\ln\dot{F}}-\ctcn{P}{^m_{(a}}\ctcn{P}{^n_{b)}}\cds{_{n}}\prn{\ctcn{P}{^r_{m}}\cds{_{r}}\ln\dot{F}}\nonumber\\
		 		&-\frac{1}{2}\ms{_{ab}}\ms{^{ef}}\brkt{\ctcn{P}{^m_{e}}\cds{_{m}}\prn{\ln\dot{F}}\ctcn{P}{^n_{f}}\cds{_{n}}\prn{\ln\dot{F}}-\ctcn{P}{^m_{e}}\ctcn{P}{^n_{f}}\cds{_{n}}\prn{\ctcn{P}{^r_{m}}\cds{_{r}}\ln\dot{F}}}\spacef,
		 	\end{align}
	 or on each cut
		 	\begin{align}
		 		\ct{\dot{\sigma}}{_{AB}}\defeq \ctcn{E}{^a_A} \ctcn{E}{^b_B} \ct{N}{^c}\cds{_{c}}\ctcn{\sigma}{_{ab}}
				&\eqc\ct{N}{_{AB}}+\ct{\rho}{_{AB}}-\frac{\csS{R}}{4}\mc{_{AB}}+\cdc{_{A}}\prn{\ln\dot{F}}\cdc{_{B}}\prn{\ln\dot{F}}-\cdc{_{A}}\prn{\cdc{_{B}}\ln\dot{F}}\nonumber\\
		 		&-\frac{1}{2}\mc{_{AB}}\mc{^{EF}}\brkt{\cdc{_{E}}\prn{\ln\dot{F}}\cdc{_{F}}\prn{\ln\dot{F}}-\cdc{_{E}}\prn{\cdc{_{F}}\ln\dot{F}}}\spacef.\label{eq:shear-news}
		 	\end{align}
	 This equation relates the news tensor to the `time' derivative of the shear tensor $ \ctcn{\sigma}{_{ab}} $. Observe that in general \emph{only for  canonically adapted foliations --\cref{eq:foliation-canonical-adapted}-- in which the gauge fixing\footnote{The so called `Bondi gauge' usually refers to this kind of gauge-fixing.} gives the round metric on the cuts} one obtains --recalling \cref{eq:lied-bases-foliation}--
		 	\begin{align}
		 		\ct{\dot{\sigma}}{_{ab}}&=\ct{N}{_{ab}}\label{eq:shear-news-adapted}\spacef.
		 	\end{align}
			 
	\subsection{Symmetries and universal structure}
		The conformal boundary for vanishing cosmological constant presents a universal structure \cite{Geroch1977} which gives rise to an asymptotic symmetry group known as the $ \mathsf{BMS} $ group --named after Bondi, Metzner and Sachs \cite{Bondi1962,Sachs1962,Penrose1966}--, which has been widely studied \cite{Geroch81,Wald1984,Ashtekar81,Geroch1977,Stewart1991} and recently has attracted renewed attention with proposals of generalisations and extensions \cite{Campiglia2014,Barnich2010,Flanagan17,Speziale2021,Bonga2020}. The $ \mathsf{BMS} $ group admits different characterisations, from coordinate-based methods, to covariant ones. We focus on the latter and, particularly, in Geroch's approach --see also \cite{Ashtekar87} for more details. The asymptotic infinitesimal symmetries are those vector fields preserving the \emph{universal structure}
			\begin{deff}[Universal structure]
			Let $ \ct{N}{^a} $ be the tangent vector field to the generators of $ \scri $ and $ \ms{_{ab}} $ its degenerate metric. Then, the universal structure of $ \scri $ consists of the conformal family of pairs
					\begin{equation*}
						\prn{\ms{_{ab}},\ct{N}{^a}}\spacef.
					\end{equation*}
			 Two pairs belong to the same conformal family if and only if $ \prn{\cts{g'}{_{ab}},\ct{N'}{^a}}=\prn{\Psi^2\ms{_{ab}},\Psi^{-1}\ct{N}{^a}} $, where $ \Psi $ is a positive function on $ \scri $.
			\end{deff}
			\begin{remark}
				An equivalent formulation is to consider the gauge-invariant object \cite{Geroch1977}
					\begin{equation}
						\ms{_{ab}}\ct{N}{^c}\ct{N}{^d}\spacef
					\end{equation}
				and the infinitesimal symmetries are those that leave it invariant, that is
					\begin{equation}
						\lied_{\vec{\xi}}\prn{\ms{_{ab}}\ct{N}{^c}\ct{N}{^d}}=0\spacef.
					\end{equation}
			\end{remark}
		The algebra $ \mathfrak{bms} $ is characterised by the infinitesimal symmetries $ \ct{\xi}{^a} $
		defined by
			\begin{align}
				\lied_{\vec{\xi}}\ct{N}{^a}&=-\psi\ct{N}{^a}\spacef,\\
				\lied_{\vec{\xi}}\ms{_{ab}}&=2\psi\ms{_{ab}}.
			\end{align}
		The  infinitesimal symmetries $ \ct{\tau}{^a} $ proportional to $ \ct{N}{^a} $,
			\begin{equation}
				\ct{\tau}{^a}=\alpha\ct{N}{^a}
			\end{equation}
		are called \emph{supertranslations}. They form an infinite-dimensional subalgebra $ \ft $ of $ \mathfrak{bms} $ and the group of supertranslations $ \T $ is a Lie ideal of $ \mathsf{BMS} $. One has
			\begin{align}
				\lied_{\vec{\tau}}\ct{N}{^a}&=0\spacef,\\
				\lied_{\vec{\tau}}\ms{_{ab}}&=0
			\end{align}
		and
			\begin{equation}
				\lied_{\vec{N}}\alpha=0\spacef.
			\end{equation}
		The resulting symmetry group $ \mathsf{BMS} $ consists of the semidirect product of the Lorentz group $ \mathsf{SO(1,3)} $ with the  normal subgroup of supertranslations $ \fT $,
			\begin{equation}
				\mathsf{BMS}=\fT\ltimes\mathsf{SO(1,3)}\spacef.
			\end{equation}
			\begin{remark}
			\Cref{rem:rhoinvariant} together with $\ct{N}{^a} \ct{\rho}{_{ab}}=0$ makes $ \ct{\rho}{_{ab}}$ invariant under supertranslations. 
			\end{remark}
		Geroch identified a 4-dimensional subspace of infinitesimal translations, given by those elements of $ \ft $ satisfying 
			\begin{equation}\label{eq:infinitesimal-translations}
			 	\cds{_{a}}\cds{_{b}}\alpha+\alpha\ct{\rho}{_{ab}}=\frac{1}{2}\prn{\ms{^{mn}}\cds{_{m}}\cds{_{n}}\alpha+\alpha\frac{\csS{R}}{4}}\ms{_{ab}}\spacef.
			\end{equation}
		For an interpretation of this equation, see section in \cite{Fernandez-Alvarez_Senovilla-dS}.
		These infinitesimal symmetries enter into the definition of the Bondi-Trautman energy-momentum.

	\subsection{Asymptotic energy-momentum of the gravitational field}
		Any weakly asymptotically simple $ \Lambda=0 $ space-time \cite{Kroon} features the existence of a total energy-momentum at $ \scri $ --the so called Bondi-Trautman 4-momentum \cite{Trautman58,Bondi1962,Stewart1991}. It includes the energy of the gravitational field and quantifies the energy-loss due to outgoing gravitational waves. Geroch \cite{Geroch1977} presented it as a particular case of a generalised momentum built upon a vector field associated to any supertranslation $ \ct{\tau}{^a}=\alpha\ct{N}{^a} $ and given by --see \cite{Dray84} for the spin-coefficient version--
			\begin{equation}\label{eq:geroch-field}
			 	\ctrd{\tau}{\M}{^a}\defeq \alpha\ctn{D}{^{am}}\cts{\ell}{_{m}}-2\prn{\alpha\cds{_{m}}\cts{\ell}{_{n}}+\cts{\ell}{_{m}}\cds{_{n}}\alpha}\ms{^{np}}\ct{N}{_{pq}}\ms{^{q[m}}\ct{N}{^{a]}}\spacef,
			\end{equation}
		where $ \cts{\ell}{_{a}} $ is any one-form satisfying $ \cts{\ell}{_{m}}\ct{N}{^m}=-1 $. It is possible to include Geroch's approach into our formalism of foliations --\cref{ssec:basic-geometry}. Let $ \cts{\ell'}{_a} $ be any field associated to a foliation, as in \cref{eq:ell-F-foliation}, with the defining function $ F $ giving the cuts $ \Sc_{C} $ at constant values $ F=C=\text{constant} $ and lightlike extension $ \cts{\ell'}{_{\alpha}} $. The  two-form  $ \csrd{\tau}{\mathbf{M}}\defeq\ctrd{\tau}{\M}{^a}\ct{\epsilon}{_{abc}}\df{x}^b\wedge\df{x}^c $  integrated over any cut $ \Sc_{C} $ gives a charge associated to that cut and the supertranslation $ \ct{\tau}{^a} $
			\begin{equation}
			  \csrdd{_{\tau}}{\E}{\Sc_{C}}\defeq-\frac{1}{8\pi} \int_{\Sc_{C}}\csrd{\tau}{\mathbf{M}}=-\frac{1}{8\pi}\int_{\Sc_{C}}\ctrd{\tau}{\M}{^a}\cts{\ell'}{_{a}}\csC{\epsilon}'
			\end{equation}
		This formula can be shown to be independent of the choice of $ \cts{\ell}{_a} $ in \cref{eq:geroch-field}, thus without loss of generality, let us write $ \cts{\ell'}{_{a}}=\cts{\ell}{_a}  $. Using \cref{eq:geroch-field} and introducing \cref{eq:shear}, the charge can be rewritten as
			\begin{equation}\label{eq:geroch-charge}
			 	\csrdd{_{\tau}}{\E}{\Sc_{C}}=-\frac{1}{8\pi}\int_{\Sc_{C}}\prn{\alpha\ctn{D}{^{rs}}\cts{\ell}{_{r}}\cts{\ell}{_{s}}+\alpha\ctcn{\sigma}{_{rs}}\ct{N}{^{rs}}}\csC{\epsilon}\spacef.
			\end{equation}
		The first term contains essentially a Coulomb contribution from the gravitational field --see \cite{Fernandez-Alvarez_Senovilla-dS}. The difference of the quantity \eqref{eq:geroch-charge} for any two cuts $ \Sc_{2} $ and $ \Sc_{1} $, with the former to the future of the latter, is derived by computing the divergence of \cref{eq:geroch-field} and integrating over the three-dimensional portion $ \Delta \in \scri $ bounded by the two cuts:
			\begin{equation}\label{eq:two-cuts-energy-loss}
			 	\csrdd{_{\tau}}{\E}{\Sc_{2}}-\csrdd{_{\tau}}{\E}{\Sc_{1}}=-\frac{1}{8\pi}\int_{\Delta}\prn{\alpha\ct{y}{_m^f_p}\ct{N}{^m}\cts{\ell}{_{f}}\ct{N}{^p}+\alpha\cts{S}{_{rs}}\ct{N}{^{rs}}+\ct{N}{^{rs}}\cds{_{r}}\cds{_{s}}\alpha}\cs{\epsilon}\spacef.
			\end{equation}
		Indeed, it is possible to differentiate along the foliation to obtain the infinitesimal change in the charge \eqref{eq:geroch-charge} on a cut $ \Sc_{C} $,
			\begin{align}
				\frac{\df \csrdd{_{\tau}}{\E}{\Sc_{C}}}{\df C}=&-\frac{1}{8\pi}\int_{\Sc_C}\frac{\alpha}{\dot{F}}\bbrkt{\ct{N}{^{PQ}}\ctc{S}{_{PQ}}+\ct{N}{^{PQ}}\cdc{_{P}}\cdc{_{Q}}\ln\dot{F}-2\cdc{_{P}}\prn{\ct{N}{^{QP}}}\cdc{_{Q}}\ln\dot{F}\nonumber\\
				&+\cdc{_{Q}}\cdc{_{P}}\ct{N}{^{QP}}+\ct{y}{_m^f_p}\ct{N}{^m}\cts{\ell}{_{f}}\ct{N}{^p}+\ct{N}{^{AB}}\cdc{_{A}}\prn{\ln\dot{F}}\cdc{_{B}}\prn{\ln\dot{F}}				}\csC{\epsilon}
			\end{align}
		where \cref{eq:shear-news,eq:DA-news,eq:DAB-news,eq:matter-term} have been used. Since the cuts are topological spheres, total divergences integrate out and one can simplify the expression above to reach the nice formula
			\begin{equation}\label{eq:charge-loss}
					\frac{\df \csrdd{_{\tau}}{\E}{\Sc_{C}}}{\df C}=-\frac{1}{8\pi}\int_{\Sc_C}\frac{1}{\dot{F}}\brkt{\alpha\ct{N}{^{AB}}\ctc{S}{_{AB}}+\ct{N}{^{AB}}\cdc{_{A}}\cdc{_{B}}\alpha+\alpha\varkappa\ct{N}{^\mu}\ct{N}{^\nu}\ctrd{0}{T}{_{\mu\nu}}}\csC{\epsilon}\spacef,
			\end{equation}
		where the matter fields enter the integral through (see \cref{eq:cottonPhysicalMatter,eq:cotton-matter-contribution})
			\begin{equation}
				\ct{y}{_m^f_p}\ct{N}{^m}\cts{\ell}{_{f}}\ct{N}{^p}\eqs\varkappa\ct{N}{^\mu}\ct{N}{^\nu}\ctrd{0}{T}{_{\mu\nu}}\spacef.
			\end{equation}
		 If one chooses $ \ct{\tau}{^a} $ as an infinitesimal translation --i.e., $ \alpha $ satisfying \cref{eq:infinitesimal-translations}--, \eqref{eq:geroch-charge} gives the \emph{Bondi-Trautman energy-momentum}. In that case, \cref{eq:two-cuts-energy-loss}  and \cref{eq:charge-loss} yield
		\begin{align}
				\csrdd{{\tau}}{\E}{\Sc_{2}}-\csrdd{{\tau}}{\E}{\Sc_{1}}&=-\frac{1}{8\pi}\int_{\Delta}\alpha\prn{\ct{N}{^{qp}}\ct{N}{_{qp}}+\varkappa\ct{N}{^\mu}\ct{N}{^\nu}\ctrd{0}{T}{_{\mu\nu}}}\cs{\epsilon}\spacef.\label{eq:energy-difference}\\
			  	\frac{\df \csrdd{{\tau}}{\E}{\Sc_{C}}}{\df C}&=-\frac{1}{8\pi}\int_{\Sc_{C}}\frac{\alpha}{\dot{F}}\prn{\ct{N}{^{PQ}}\ct{N}{_{PQ}}+\varkappa\ct{N}{^\mu}\ct{N}{^\nu}\ctrd{0}{T}{_{\mu\nu}}}\csC{\epsilon}\spacef.\label{eq:energy-loss}
			\end{align}
		The second one is the \emph{Bondi-Trautman energy-momentum-loss formula}. Both \cref{eq:two-cuts-energy-loss} and  \cref{eq:energy-loss} feature the same property: if the energy-momentum of the matter fields decay such that $\ct{N}{^\mu}\ct{N}{^\nu}\ctrd{0}{T}{_{\mu\nu}}=0$ --in particular in the absence of matter fields around infinity-- 
		a non-vanishing news tensor still diminishes the total energy-momentum. To our knowledge, it is the first time \cref{eq:energy-loss} is presented including the matter term and the factors associated to the choice of foliation $ \dot{F} $ and translation $ \alpha $; in the literature, either just \cref{eq:energy-difference} is given \cite{Geroch1977}, even without the matter contribution (see \cite{Frauendiener2021} for a recent derivation), or just \cref{eq:energy-loss} is considered, typically without the matter term nor the factor $ \dot{F} $ corresponding to the choice of foliation \cite{Penrose63,Newman68,Kroon}. Moreover, when $ \alpha $ is set to a constant --equivalently, a `time' translation is selected-- to get the total energy-loss, the dimensional analysis of \cref{eq:energy-loss} becomes obscure --see \cite{Fernandez-Alvarez_Senovilla20} for the discussion of the units including $ \alpha $ and $ \dot{F} $. For later convenience, let us define the energy-momentum loss associated to gravitational waves only (i.e., excluding the matter term)
			\begin{equation}
				\frac{\df \csrddu{{\tau}}{G}{\E}{\Sc_{C}}}{\df C}=-\frac{1}{8\pi}\int_{\Sc_{C}}\frac{\alpha}{\dot{F}}\ct{N}{^{PQ}}\ct{N}{_{PQ}}\csC{\epsilon}\spacef.\label{eq:energy-loss-gw}
			\end{equation}
		Observe that \cref{eq:energy-loss} is the general energy-momentum loss, whereas the commonly presented energy-loss formulae involving the square of the Lie derivative along $ \ct{N}{^a} $ \eqref{eq:shear-news} of the shear tensor \eqref{eq:shear} arise by making all or some of the following elections: $ \alpha=1 $, a canonical foliation \eqref{eq:foliation-canonical-adapted}, a round-metric gauge and the absence of the asymptotic matter term \eqref{eq:cotton-matter-contribution}.

%% file: section-4/section-4.tex
		 We deal now with the behaviour of physical fields when they are parallel transported along null geodesics. The outcome of this process when applied to the physical Weyl tensor, typically receives the name of \emph{peeling} property or behaviour \cite{Sachs1962,Geroch1977,Kroon,Wald1984,Stewart1991}. We adopt Geroch's approach and refine it to construct endomorphisms at the tangent space of every point of $ \scri^+ $, considered as the endpoint in $ \prn{M,\ct{g}{_{\alpha\beta}}} $ of future-pointing null geodesics arriving there. \\
		 
		 Let $ \gamma\prn{\lambda} $ be a curve parametrised by $ \lambda\in[-1,0] $, with one endpoint at $ p_{0}\in\scri^+ $ (corresponding to $ \lambda_{0}\defeq\lambda\evalat{p_{0}}=0 $) and the past endpoint at $ p_{1}\in\ps{M} $ (with $ \lambda_{1}\defeq\lambda\evalat{p_{1}}=-1 $). Points belonging to $ \gamma $ corresponding to fixed values $ \lambda=\lambda_{i} $ will be labelled by $ p_{i} $,	
				\begin{center}
				\begin{tabular}{  M{1cm} M{1cm}  M{1cm}  M{1cm}  N }
					 $ \gamma\prn{\lambda}: $& $ \brkt{-1,0} $&$  \longrightarrow  $ &  $\cs{M} $  & \\ 
					 \spacef & $ \lambda $&$ \longrightarrow  $ &  $ p $\spacef.  &\\
				\end{tabular}
				\end{center}
		Denote the tangent vector field to the curve by $ \ct{\ell}{^\alpha} $ and choose the parametrisation such that
			\begin{equation}\label{eq:null-geodesic-normalisation}
				\ct{\ell}{^\mu}\ct{N}{_{\mu}}= \frac{\df\Omega}{\df\lambda}\eqs^+ -1 \spacef.
			\end{equation}
		At first order near $ \lambda_{0}=0 $, $ \Omega\approx-\lambda $. Observe that we do not require at this stage $ \ct{\ell}{^\alpha} $ to be lightlike, though we have chosen it to be future-pointing. Next, denote by
			\begin{align*}
				\ct{t}{_{\alpha}^\beta}\prn{\lambda_{i},\lambda_{j}}\spacef,\text{ the parallel propagator w.r.t. }\ct{\Gamma}{^a_{bc}}\spacef,\\
				\pt{t}{_{\alpha}^\beta}\prn{\lambda_{i},\lambda_{j}}\spacef,\text{ the parallel propagator w.r.t. }\pt{\Gamma}{^a_{bc}}\spacef,
			\end{align*}
		such that given any one-form $ \ct{v}{_{\alpha}} $ defined at $ p_{j} $, the result of parallel-transporting it along $  \gamma\prn{\lambda} $  from $ p_{j} $ to $ p_{i} $ results on the new one-form $ \pct{v}{_{\alpha}}\prn{\lambda_{i}} $ at $ p_{i} $ given by
			\begin{equation}
				\pct{v}{_{\alpha}}\prn{\lambda_{i}}=\ct{t}{_{\alpha}^\mu}\prn{\lambda_{i},\lambda_{j}}\ct{v}{_{\mu}}\prn{\lambda_{j}}\spacef.
			\end{equation}
		Observe that indices $ \alpha $ and $ \mu  $ in this relation belong to different tangent spaces. The propagator $ \ct{t}{_{\alpha}^\beta} $ is a `bi-tensor' \cite{Stephani2004} which is defined by the differential equation
			\begin{align}
				\td{\ct{t}{_{\alpha}^\beta}\prn{\lambda,\lambda_{j}}}{\lambda}=\ct{\ell}{^\rho}\ct{\Gamma}{^\mu_{\alpha\rho}}\prn{\lambda}\ct{t}{_{\mu}^\beta}\prn{\lambda,\lambda_{j}}\spacef,\label{eq:propagator-diff}\\
				\td{\pt{t}{_{\alpha}^\beta}\prn{\lambda,\lambda_{j}}}{\lambda}=\ct{\ell}{^\rho}\pt{\Gamma}{^\mu_{\alpha\rho}}\prn{\lambda}\pt{t}{_{\mu}^\beta}\prn{\lambda,\lambda_{j}}\spacef,
			\end{align}
		with `initial' condition
			\begin{equation}
				\ct{t}{_{\alpha}^\beta}\prn{\lambda_{j},\lambda_{j}}=\delta^\alpha_{\beta}\spacef,\quad \pt{t}{_{\alpha}^\beta}\prn{\lambda_{j},\lambda_{j}}=\delta^\alpha_{\beta}\spacef,
			\end{equation}
		and satisfies
			\begin{equation}\label{eq:propagator-properties}
			\ct{t}{_{\alpha}^\mu}\prn{\lambda,\lambda_{i}}\ct{t}{_{\mu}^\beta}\prn{\lambda_{i},\lambda}=\delta^\beta_{\alpha}\spacef,\quad \ct{t}{_{\alpha}^\mu}\prn{\lambda_{j},\lambda_{i}}=\ct{t}{^\mu_{\alpha}}\prn{\lambda_{i},\lambda_{j}}
			\end{equation}
		where
			\begin{equation}
				\ct{t}{^\mu_{\alpha}}\prn{\lambda_{i},\lambda_{j}}\defeq\ct{g}{^{\mu\sigma}}\prn{\lambda_{i}}\ct{g}{_{\nu\alpha}}\prn{\lambda_{j}}\ct{t}{_{\sigma}^\nu}\prn{\lambda_{i},\lambda_{j}}\spacef.
			\end{equation}
		 The main idea \cite{Geroch1977} is to perform 3 different parallel transports of any covariant tensor field:
		 	\begin{enumerate}
		 	\item From $ p_{0} $ to $ p\prn{\lambda} $ with $ \ct{t}{_{\alpha}^\beta}\prn{\lambda,\lambda_{0}} $,
		 	\item From $ p(\lambda) $ to $ p_{1} $ with $ \pt{t}{_{\alpha}^\beta}\prn{\lambda_{1},\lambda} $,
		 	\item From $ p_{1} $ to $ p_{0} $ with $ \ct{t}{_{\alpha}^\beta}\prn{\lambda_{0},\lambda_{1}} $.
		 	\end{enumerate}
		 That is, a transport along $ \gamma $ back and forth, departing from $ \scri^+ $ and interchanging the conformal connection by the physical one for a stretch of $ \gamma $. If one chains one operation after another, the result is an endomorphism on the co-tangent space at $ p_{0} $:
		 	\begin{equation}
		 		\ctrdu{\lambda}{\gamma}{L}{_{\alpha}^\beta}\defeq\ct{t}{_{\alpha}^\mu}\prn{\lambda_{0},\lambda_{1}}\pt{t}{_{\mu}^\rho}\prn{\lambda_{1},\lambda}\ct{t}{_{\rho}^\beta}\prn{\lambda,\lambda_{0}}.
		 	\end{equation}
		 The upper and lower indices on the left-hand side of $\ctrdu{\lambda}{\gamma}{L}{_{\beta}^\alpha} $ indicate its dependence on the curve $ \gamma $ and the point $ p(\lambda) $. Since the notation may become cumbersome, we drop this two labels in most of the calculations and recover them only when  doing so happens to be convenient. Since $ \ct{L}{_{\beta}^\alpha} $ is a tensor at $ p_0 $, acting on covariant objects, we introduce the notation
		 	\begin{equation}
		 		\lct{T}{_{{\alpha_1}...{\alpha_r}}}\defeq \ct{L}{_{\alpha_1}^{\mu_{1}}}...\ct{L}{_{\alpha_r}^{\mu_{r}}}\ct{T}{_{{\mu_1}...{\mu_r}}}.
		 	\end{equation}
		 The action on the metric at $ p_{0} $ gives
		 	\begin{equation}\label{eq:operatorL-metric}
		 		\lct{g}{_{\alpha\beta}}=\csrd{\lambda}{\Xi}^2 \ct{g}{_{\alpha\beta}},
		 	\end{equation}
		 where one uses that the connections are metric-compatible with respect to $ \ct{g}{_{\alpha\beta}} $ and $ \pt{g}{_{\alpha\beta}} $, respectively, and introduces the definition
		 	\begin{equation}\label{eq:operatorL-xi}
		 		\csrd{\lambda}{\Xi}\defeq \frac{\Omega(\lambda)}{\Omega(\lambda_1)}\spacef
		 	\end{equation}
		 --we will drop the label on the left-hand side. \Cref{eq:operatorL-metric} implies that the endomorphism $ \ct{L}{_{\alpha}^\beta} $ preserves the null cone (and obviously also the future orientation) and therefore it is proportional to a Lorentz transformation at $ p_0 $.  Recalling the first of \cref{eq:propagator-properties}, it is easy to verify that
		 	\begin{equation}
		 		\ctru{-1}{L}{_{\beta}^\alpha}\defeq \ct{t}{_{\beta}^\mu}\prn{\lambda_{0},\lambda}\pt{t}{_{\mu}^\rho}\prn{\lambda,\lambda_{1}}\ct{t}{_{\rho}^\alpha}\prn{\lambda_{1},\lambda_{0}}
		 	\end{equation}
		 is the inverse operator, that is,
		 	\begin{equation}
		 		\ct{L}{_{\alpha}^\rho}\ctru{-1}{L}{_\rho^{\beta}}=\delta^\beta_{\alpha}\spacef.
		 	\end{equation}	 
		 The version of $ \ct{L}{_{\beta}^\alpha} $ that acts on contravariant fields is defined as
		 	\begin{equation}
		 		\ctrdu{\lambda}{\gamma}{\tilde{L}}{^{\beta}_\alpha}\defeq\ct{t}{_{\alpha}^\mu}\prn{\lambda_{0},\lambda}\pt{t}{_{\mu}^\rho}\prn{\lambda,\lambda_{1}}\ct{t}{_{\rho}^\beta}\prn{\lambda_{1},\lambda_{0}},
		 	\end{equation}
		 and a simple calculation using the second of \cref{eq:propagator-properties} shows that 
		 	\begin{equation}\label{eq:operatorL-relation-inverse}
		 		\ctru{-1}{L}{_\alpha^{\beta}}=\ct{\tilde{L}}{^{\beta}_\alpha}=\frac{1}{\Xi^2}\ct{L}{^\beta_{\alpha}}\spacef,
		 	\end{equation}
		 where $ \ct{L}{^\beta_{\alpha}}=\ct{g}{^{\beta\mu}}\ct{L}{_{\mu}^\nu}\ct{g}{_{\nu\alpha}} $. Therefore, taking into account \cref{eq:operatorL-relation-inverse} one can work only with $ \ct{L}{_{\beta}^\alpha} $. Some useful relations are
		 	\begin{align}
		 		\lct{\eta}{_{\alpha\beta\gamma\delta}}&=\Xi^4\ct{\eta}{_{\alpha\beta\gamma\delta}}\spacef,\label{eq:operatorL-eta}\\
		 		\ctru{-1}{L}{_{\rho}^\alpha}\ctru{-1}{L}{_{\sigma}^\beta}\ct{g}{^{\rho\sigma}}&=\frac{1}{\Xi^2}\ct{g}{^{\alpha\beta}}\spacef,\\
		 		\lct{v}{_{\mu}}\lct{w}{^\mu}&=\Xi^2\ct{v}{_{\mu}}\ct{w}{^\mu}\spacef\spacef\forall\spacef\ct{v}{^\alpha},\spacef\ct{w}{^\alpha}\label{eq:operatorL-scalar}\spacef,
		 	\end{align}
		 where
		 	\begin{equation}
		 		\lct{w}{^\alpha}\defeq \ct{g}{^{\alpha\mu}}\lct{w}{_{\mu}}\spacef.
		 	\end{equation}
		 
		 The next task to be addressed is to find the explicit form of the operator $ \ct{L}{_{\alpha}^\beta} $. We believe that this can be done for more general curves, however the most relevant case -- and easiest to deal with-- is when $ \ct{\ell}{^\alpha} $ is geodesic and lightlike with $ \lambda $ an affine parameter,
		 	\begin{equation}
		 		\ct{\ell}{_{\mu}}\ct{\ell}{^\mu}=0\spacef,\quad\ct{\ell}{^\rho}\cd{_{\rho}}\ct{\ell}{_{\alpha}}=0.
		 	\end{equation}
		 We assume this restriction from now on. Observe that for null geodesics (e.g. \cite{Wald1984}) one can always write
			\begin{equation}
				\pt{\ell}{_{\alpha}}=\ct{\ell}{_{\alpha}}
			\end{equation}
		where $ \pt{\ell}{_{\alpha}} $ is lightlike and geodesic with respect to the physical metric. This fact allows to deduce the action of $ \ct{L}{_{\alpha}^\beta} $ on $ \ct{\ell}{_{\alpha}} $ 
			\begin{equation}\label{eq:operatorL-ell}
				\lct{\ell}{_{\alpha}}=\ct{L}{_{\alpha}^\beta}\ct{\ell}{_\beta}= \ct{\ell}{_{\alpha}}\spacef.
			\end{equation}
			
		 Observe that $ \ct{L}{_{\alpha}^\beta} $ has at most 16 independent components.  It can be expressed in the bases $ \cbrkt{-\ct{N}{_{\alpha}},-\ct{\ell}{_{\alpha}},\ct{q}{_{\alpha}},\ct{r}{_{\alpha}}} $ and $ \cbrkt{\ct{\ell}{^{\alpha}},\ct{N}{^{\alpha}},\ct{q}{^{\alpha}},\ct{r}{^{\alpha}}} $, with $ \ct{q}{_{\alpha}} $ and $ \ct{r}{_{\alpha}} $ arbitrary unit one-forms orthogonal to $ \ct{N}{^\alpha} $ and $ \ct{\ell}{^\alpha} $ at $ p_{0} $, as
		 	\begin{align}
		 		\ctrdu{\lambda}{\gamma}{L}{_{\beta}^\alpha}=&\csrdu{\lambda}{\gamma}{A}\ct{N}{^\alpha}\ct{N}{_{\beta}}+\csrdu{\lambda}{\gamma}{B}\ct{N}{^\alpha}\ct{\ell}{_{\beta}}+\csrdu{\lambda}{\gamma}{C}\ct{\ell}{^\alpha}\ct{\ell}{_{\beta}}+ \csrdu{\lambda}{\gamma}{D}\ct{\ell}{^\alpha}\ct{N}{_{\beta}}+\csrdu{\lambda}{\gamma}{F}\ct{q}{^\alpha}\ct{r}{_{\beta}}+\csrdu{\lambda}{\gamma}{G}\ct{r}{^\alpha}\ct{q}{_{\beta}}+\csrdu{\lambda}{\gamma}{H}\ct{q}{^\alpha}\ct{q}{_{\beta}}+\csrdu{\lambda}{\gamma}{I}\ct{r}{^\alpha}\ct{r}{_{\beta}}\nonumber\\
		 		&+\ct{N}{^\alpha}\ctcrdu{\lambda}{\gamma}{v}{_{\beta}}+\ct{\ell}{^{\alpha}}\ctcrdu{\lambda}{\gamma}{w}{_{\beta}}+\ctcrdu{\lambda}{\gamma}{x}{^{\alpha}}\ct{N}{_{\beta}}+\ctcrdu{\lambda}{\gamma}{y}{^{\alpha}}\ct{\ell}{_{\beta}}\spacef.
		 	\end{align}
		 The dependence on the curve $ \gamma $ and the point $ \lambda $ is contained in the 8 scalars and 4 2-dimensional vector fields correspondingly labelled in the formula above. By \cref{eq:operatorL-ell,eq:operatorL-relation-inverse} direct simplifications take place:
		 	\begin{equation}
		 		A=0\spacef,\quad B=-1\spacef,\quad D=-\Xi^2\spacef,\quad\ctc{v}{_{\beta}}=0\spacef,\quad\ctc{x}{^\alpha}=0\spacef.
		 	\end{equation}
		 Projecting \cref{eq:operatorL-metric} with the elements of the bases, one arrives at the expression
		 	\begin{equation}
		 		\ct{L}{_{\beta}^\alpha}=-\ct{N}{^\alpha}\ct{\ell}{_{\beta}}+C\ct{\ell}{^\alpha}\ct{\ell}{_{\beta}}-\Xi^2\ct{\ell}{^\alpha}\ct{N}{_{\beta}}+F\prn{\ct{q}{^\alpha}\ct{r}{_{\beta}}-\ct{r}{^\alpha}\ct{q}{_{\beta}}}+H\ctc{P}{^\alpha_{\beta}}+\ct{\ell}{^\alpha}\ctc{w}{_{\beta}}+\ctc{y}{^\alpha}\ct{\ell}{_{\beta}},
		 	\end{equation}
		 with
		 	\begin{equation}\label{eq:aux12}
		 		2\Xi^2C=-\Xi^2\ctc{y}{_{\mu}}\ctc{y}{^\mu}=-\ctc{w}{_{\mu}}\ctc{w}{^\mu}\spacef,\quad F^2+H^2=\Xi^2
		 	\end{equation}
		 and
		 	\begin{align}
		 		0&=-\ctc{w}{_{r}}-F\ctc{y}{_{q}}-H\ctc{y}{_{r}}\spacef,\label{eq:aux11} \\
		 		0&=-\ctc{w}{_{q}}+F\ctc{y}{_{r}}-H\ctc{y}{_{q}}\spacef,\label{eq:aux10} \\
		 		0&=-\Xi^2\ctc{y}{_{q}}-F\ctc{w}{_{r}}-H\ctc{w}{_{q}}\spacef, \\
		 		0&=-\Xi^2\ctc{y}{_{r}}+F\ctc{w}{_{q}}-H\ctc{w}{_{r}}\spacef.
		 	\end{align}
		 By construction, one has 
		 	\begin{equation}\label{eq:operatorL-initial-condition}
		 		\ctrd{\lambda_1}{L}{_{\beta}^\alpha}=\delta^\beta_{\alpha},
		 	\end{equation}
		 which implies
		 	\begin{equation}\label{eq:operatorL-initial-condition-terms}
		 		\csrd{\lambda_{1}}{C}=\csrd{\lambda_{1}}{F}=0\spacef,\quad\csrd{\lambda_{1}}{H}=1\spacef,\quad\ctcrd{\lambda_{1}}{w}{_{\beta}}=0\spacef,\quad\ctcrd{\lambda_{1}}{y}{_{\beta}}=0\spacef\spacef.
		 	\end{equation}
		 Since $ \ct{L}{_{\alpha}^\beta} $ depends on $ \lambda $, it makes sense to search for a differential equation for it. To that purpose, notice that another version of \cref{eq:propagator-diff} can be written for $ \ct{t}{_{\alpha}^\beta}\prn{\lambda,\lambda_{j}} $ by using \cref{eq:propagator-properties},
		 	\begin{equation}
		 		\td{\ct{t}{_{\alpha}^\beta}\prn{\lambda_{j},\lambda}}{\lambda}=-\ct{k}{^\rho}\ct{\Gamma}{^\beta_{\rho\mu}}\ct{t}{_{\alpha}^\mu}\prn{\lambda_{j},\lambda}\spacef.
		 	\end{equation}
		The final differential formula for $ \ct{L}{_{\alpha}^\beta} $ reads
			\begin{equation}\label{eq:operatorL-diff-eq}
				\td{\ct{L}{_{\beta}^\alpha}}{\lambda}=\frac{1}{\Omega\prn{\lambda}}\ct{L}{_{\beta}^\mu}\ct{\Lambda}{_{\mu}^\alpha}
			\end{equation}
		where 
			\begin{align}
				\ct{\Lambda}{_{\beta}^\alpha}=\ctrd{\lambda}{\Lambda}{_{\beta}^\alpha}&\defeq \Omega\prn{\lambda}\ct{\ell}{^\sigma}\prn{\lambda}\ct{\gamma}{^\nu_{\mu\sigma}}\prn{\lambda}\ct{t}{_{\nu}^\alpha}\prn{\lambda,\lambda_{0}}\ct{t}{_{\beta}^\mu}\prn{\lambda_{0},\lambda}\\
				&=\td{\Omega}{\lambda}\delta^\alpha_{\beta}+\ct{\ell}{^\alpha}\pct{N}{_{\beta}}-\ct{\ell}{_{\beta}}\pct{N}{^\alpha}\spacef,
			\end{align}
		with
			\begin{equation}\label{eq:operatorL-N-transported}
				\pct{N}{_{\beta}}=\pctrd{\lambda}{N}{_{\beta}}\defeq\ct{t}{_{\beta}^\mu}\prn{\lambda_{0},\lambda}\ct{N}{_{\mu}}\prn{\lambda}\spacef,\quad\pctrd{\lambda_0}{N}{_{\beta}}=\ct{N}{_{\beta}}\spacef,
			\end{equation}
		and 
		$$ \ct{\gamma}{^\alpha_{\beta\gamma}} =\frac{1}{\Omega} \prn{\delta^\alpha_\beta \ct{N}{_\gamma} +\delta^\alpha_\gamma \ct{N}{_\beta} -\ct{N}{^\alpha} \ct{g}{_{\beta\gamma}}}
		$$
		is the tensor field defined in (III.3) of the companion paper \cite{Fernandez-Alvarez_Senovilla-dS}
		 giving the difference between the unphysical and physical connections. 
Observe that $ \ct{\Lambda}{_{\beta}^\alpha} $ is a tensor at $ p_{0} $ that depends on $ \gamma\prn{\lambda} $. \Cref{eq:operatorL-diff-eq} is a Fuchsian system with a regular singular point at $ \lambda=\lambda_{0}=0 $ --recall that $ \Omega(\lambda_{0})=0 $. In components, one has the following non-trivial equations
			\begin{align}
				\td{F}{\lambda}&=\frac{\cs{F}}{\Omega\prn{\lambda}}\td{\Omega}{\lambda}\spacef,\label{eq:aux6}\\
				\td{H}{\lambda}&=\frac{\cs{H}}{\Omega\prn{\lambda}}\td{\Omega}{\lambda}\spacef,\label{eq:aux7}\\
				\td{C}{\lambda}&=\frac{1}{\Omega\prn{\lambda}}\brkt{2C\td{\Omega}{\lambda}+\pct{N}{_{\rho}}\ctc{y}{^\rho}}\spacef,\label{eq:aux13}\\
				\td{\ctc{y}{_{r}}}{\lambda}&=\frac{1}{\Omega\prn{\lambda}}\brkt{\ctc{y}{_{r}}\td{\Omega}{\lambda}-\ctc{r}{_{\rho}}\pct{N}{^\rho}}\spacef,\label{eq:operatorL-diff-y-r}\\
				\td{\ctc{y}{_{q}}}{\lambda}&=\frac{1}{\Omega\prn{\lambda}}\brkt{\ctc{y}{_{q}}\td{\Omega}{\lambda}-\ctc{q}{_{\rho}}\pct{N}{^\rho}}\spacef,\label{eq:operatorL-diff-y-q}\\
				\td{\ctc{w}{_{r}}}{\lambda}&=\frac{1}{\Omega\prn{\lambda}}\brkt{F\pct{N}{_{\mu}}\ct{q}{^\mu}+H\pct{N}{_{\mu}}\ct{r}{^\mu}+2\ctc{w}{_{r}}\td{\Omega}{\lambda}}\spacef,\\
				\td{\ctc{w}{_{q}}}{\lambda}&=\frac{1}{\Omega\prn{\lambda}}\brkt{H\pct{N}{_{\mu}}\ct{q}{^\mu}-F\pct{N}{_{\mu}}\ct{r}{^\mu}+2\ctc{w}{_{q}}\td{\Omega}{\lambda}}\spacef.
			\end{align}
		Using \cref{eq:operatorL-initial-condition} as initial condition, \cref{eq:aux6,eq:aux7} yield
			\begin{equation}
				\csrdu{\lambda}{\gamma}{F}=0\spacef,\quad\csrdu{\lambda}{\gamma}{H}=\csrd{\lambda}{\Xi}\spacef,
			\end{equation}		
		and then, from \cref{eq:aux10,eq:aux11},
			\begin{equation}
				\ctc{w}{_{\alpha}}=-\Xi\ctc{y}{_{\alpha}}\spacef.
			\end{equation}
		Since $ C $ is determined by $ \ctc{y}{_{\alpha}} $ through \cref{eq:aux12}, it only remains to solve for $ \ctc{y}{_{\alpha}} $. \Cref{eq:operatorL-diff-y-q,eq:operatorL-diff-y-r} are two uncoupled linear ODEs, whose solution with the initial condition \eqref{eq:operatorL-initial-condition-terms} reads
			\begin{equation}\label{eq:operatorL-y-solution}
				\ctc{y}{^\alpha}=-\Omega\prn{\lambda}\int_{\lambda_{1}}^{\lambda}\frac{1}{\Omega^2\prn{\lambda'}}\pctrd{\lambda'}{N}{^{\alpha}}\df{\lambda'}\spacef.
			\end{equation}
			This solution is smooth in the limit $ \lambda=\lambda_{0} $. Taking this into account, if one multiplies \cref{eq:aux13} by $ \Omega $ and evaluates at $ \lambda_{0} $, it follows that
				\begin{equation}
					-2\csrd{\lambda_{0}}{C}=\ctcrd{\lambda_{0}}{y}{_\mu}\ctcrd{\lambda_{0}}{y}{^\mu}=0\spacef.
				\end{equation}
			All in all, the final expression of $ \ct{L}{_{\beta}^\alpha} $ is
				\begin{equation}\label{eq:operatorL-final-form}
					\ct{L}{_{\beta}^\alpha}=-\ct{N}{^\alpha}\ct{\ell}{_{\beta}}-\frac{1}{2}\ctc{y}{_{\mu}}\ctc{y}{^{\mu}}\ct{\ell}{^\alpha}\ct{\ell}{_{\beta}}-\Xi^2\ct{\ell}{^\alpha}\ct{N}{_{\beta}}+\Xi\ctc{P}{^\alpha_{\beta}}-\Xi\ct{\ell}{^\alpha}\ctc{y}{_{\beta}}+\ctc{y}{^\alpha}\ctc{\ell}{_{\beta}},
				\end{equation}
			with 
			$ \ctc{y}{_{\alpha}} $ determined by \cref{eq:operatorL-y-solution}, depending on the choice of curve through $ \pct{N}{^\alpha} $ --given in \cref{eq:operatorL-N-transported}-- and on $ \lambda $. Consider the decomposition 
				\begin{align}
					\ctrdu{\lambda}{\gamma}{L}{_{\beta}^\alpha}&=\ctrdu{\lambda}{\gamma}{p}{_{\beta}^\mu}\ctrdu{\lambda}{\gamma}{K}{_{\mu}^\alpha}\spacef,\\
					\ctrdu{\lambda}{\gamma}{p}{_{\beta}^\mu}&\defeq -\ct{N}{^\mu}\ct{\ell}{_{\beta}}-\csrdu{\lambda}{\gamma}{\Xi}^2\ct{\ell}{^\mu}\ct{N}{_{\beta}}+\csrdu{\lambda}{\gamma}{\Xi}\ctc{P}{^\mu_{\beta}}\spacef,\\
					\ctrdu{\lambda}{\gamma}{K}{_{\mu}^\alpha}&\defeq \delta^\alpha_{\mu}-\frac{1}{2}\ctc{y}{^2}\ct{\ell}{^\alpha}\ct{\ell}{_{\mu}}-\ct{\ell}{^\alpha}\ctc{y}{_{\mu}}+\ctc{y}{^\alpha}\ct{\ell}{_{\mu}}\spacef.
				\end{align}
			The interest of this decomposition is that $ \ctrdu{\lambda}{\gamma}{K}{_{\mu}^\alpha}$ carries mostly details of the curve $ \gamma $, whereas $ \ctrdu{\lambda}{\gamma}{p}{_{\beta}^\mu} $ contains essentially powers of $ \Omega $ and no information about the curve $ \gamma $: just the value of $ \Omega $ at the chosen point $ p_{1} $ --see \eqref{eq:operatorL-xi}. \\
			\begin{table}[hb!]
				\centering
				\begin{tabular}{ |M{3cm}| M{3cm} | M{3cm}| M{3cm}| N }
					\hline
						\quad Weyl-tensor candidate & Non-vanishing $ \csru{(a)}{\Psi}_{i} $ &  $ \csru{(a)}{\psi}_{i} $ when $ \lambda=\lambda_{0} $  & PND &\\ \hline 
					$ \ctru{(4)}{C}{_{\alpha\beta\gamma\delta}} $& $ \csru{(4)}{\psi}_{4} $ & $ \Omega^{-2}(\lambda_{1}) \cs{\phi}_{4} $& $ \prn{\ct{\ell}{^\alpha},\ct{\ell}{^\alpha},\ct{\ell}{^\alpha},\ct{\ell}{^\alpha}} $&\\[1cm] \hline 
					$ \ctru{(3)}{C}{_{\alpha\beta\gamma\delta}} $& $ \csru{(3)}{\psi}_{3} $ & $ \Omega^{-3}(\lambda_{1}) \cs{\phi}_{3} $&$ \prn{\ct{\ell}{^\alpha},\ct{\ell}{^\alpha},\ct{\ell}{^\alpha},\ct{N}{^\alpha}} $&\\[1cm] \hline
					$ \ctru{(2)}{C}{_{\alpha\beta\gamma\delta}} $& $ \csru{(2)}{\psi}_{2} $ & $ \Omega^{-4}(\lambda_{1}) \cs{\phi}_{2} $&$ \prn{\ct{\ell}{^\alpha},\ct{\ell}{^\alpha},\ct{N}{^\alpha},\ct{N}{^\alpha}} $&\\[1cm] \hline				
					$ \ctru{(1)}{C}{_{\alpha\beta\gamma\delta}} $& $ \csru{(1)}{\psi}_{1} $ & $ \Omega^{-5}(\lambda_{1}) \cs{\phi}_{1} $&$ \prn{\ct{\ell}{^\alpha},\ct{N}{^\alpha},\ct{N}{^\alpha},\ct{N}{^\alpha}} $&\\[1cm] \hline
					$ \ctru{(0)}{C}{_{\alpha\beta\gamma\delta}} $& $ \csru{(0)}{\psi}_{0} $ & $ \Omega^{-6}(\lambda_{1}) \cs{\phi}_{0} $&$ \prn{\ct{N}{^\alpha},\ct{N}{^\alpha},\ct{N}{^\alpha},\ct{N}{^\alpha}} $&\\[1cm] \hline	
				\end{tabular}
				\caption[Asymptotic propagation of the Weyl tensor]{The asymptotic propagation of the physical Weyl tensor \eqref{eq:asymptotic-propagation-weyl-tensor} is composed by the five terms listed above. Each one has the symmetries of a Weyl tensor and one non-vanishing Weyl scalar which in the limit $ \lambda\rightarrow\lambda_{0}=0 $ coincides up to a multiplicative constant with one of the scalars of the rescaled Weyl tensor $ \ct{d}{_{\alpha\beta\gamma}^\delta} $. The repeated PNDs are listed in the last column.}
				\label{tab:weyl-candidates}
			\end{table}
			We are mainly interested in the \emph{asymptotic behaviour} of $ \ct{L}{_{\beta}^\alpha} $, i.e. when $ \lambda\rightarrow\lambda_{0} $. It is very interesting the fact that details on the choice of $ \gamma $ become irrelevant at zeroth order in this regime because
				\begin{equation}
					\ctrdu{\lambda_{0}}{^\gamma}{K}{_{\beta}^\alpha}=\delta_{\beta}^{\alpha}\spacef.
				\end{equation}		
			In other words, \emph{the asymptotic behaviour is ruled by}	$ \ctrdu{\lambda}{\gamma}{p}{_{\beta}^\mu} $ which we come to call \emph{the asymptotic propagator}. In order to derive this behaviour for \emph{any} physical field, one has to follow the next steps:
				\begin{steps}
					\item Propagate the physical field from $ p\prn{\lambda} $ to $ p_{0} $ using $ \ct{t}{_{\beta}^\alpha}(\lambda_{0},\lambda) $ --hence defining a new tensor of the same type at $ p_{0} $ on $ \scri^+ $.\label{it:asymptotic-propagation-first-step}
					\item Apply to the covariant version of the new tensor at $ p_{0} $ as many copies of $ \ct{L}{_{\beta}^\alpha} $ as free indices has the field.\label{it:asymptotic-propagation-second-step}
					\item Expand the expression obtained previously in terms of $ \lambda $ near $ \lambda_{0}=0 $. \label{it:asymptotic-propagation-third-step}
				\end{steps}
			Note that this program `compares' the parallel propagation of the physical tensor field in $ \ps{M} $ from the point $ p_1 $ to $ p(\lambda) $, with the propagation in $ \cs{M} $ between these two points. Expanding $ \lambda $ around the limit value $ \lambda_0=0  $, one takes this comparison towards infinity of $ \ps{M} $.\\ 
	
			The canonical example is the application to the physical Weyl tensor. Consider $ \pt{C}{_{\alpha\beta\gamma\delta}}(\lambda) $, i.e. the physical Weyl tensor at $ p(\lambda) $. Now, take \cref{it:asymptotic-propagation-first-step} to define a tensor at $ p_{0} $
				\begin{equation}
					\pct{\hat{C}}{_{\alpha\beta\gamma\delta}}=\frac{1}{\Omega^2\prn{\lambda}}\pct{C}{_{\alpha\beta\gamma\delta}}=\frac{1}{\Omega\prn{\lambda}}\pct{d}{_{\alpha\beta\gamma\delta}}\spacef.
				\end{equation}  
			Notice that
				\begin{equation}
					\pct{d}{_{\alpha\beta\gamma\delta}}\evalat{\lambda=\lambda_{0}=0}=\ct{d}{_{\alpha\beta\gamma\delta}}\evalat{p_{0}}\spacef,
				\end{equation}
			thus $ \pct{\hat{C}}{_{\alpha\beta\gamma\delta}}$ contains a pole of order $ 1 $ in the limit $\lambda\rightarrow \lambda_{0} =0$. Nevertheless, this divergence is overcome in \cref{it:asymptotic-propagation-second-step},
				\begin{equation}\label{eq:asymptotic-propagation-weyl-tensor}
					 \lct{\underaccent{*}{\hat{C}}}{_{\alpha\beta\gamma\delta}}=\Omega \ctru{(4)}{C}{_{\alpha\beta\gamma\delta}}+\Omega^2\ctru{(3)}{C}{_{\alpha\beta\gamma\delta}}+\Omega^3\ctru{(2)}{C}{_{\alpha\beta\gamma\delta}}+\Omega^4\ctru{(1)}{C}{_{\alpha\beta\gamma\delta}}+\Omega^5\ctru{(0)}{C}{_{\alpha\beta\gamma\delta}},
				\end{equation}
			where $ \ctru{(a)}{C}{_{\alpha\beta\gamma\delta}} $ with $ a=0,1,2,3,4 $ are Weyl-tensor candidates, regular in the limit to $ \lambda_{0}=0 $ and with algebraic properties listed in \cref{tab:weyl-candidates}. They depend on $ \lambda $ and we assume that they can be expanded around $ \lambda_{0}=0 $ as
				\begin{equation}\label{eq:weyl-terms-expanded}
					\ctru{(a)}{C}{_{\alpha\beta\gamma\delta}}=\ctru{(a,0)}{C}{_{\alpha\beta\gamma\delta}}+\sum_{i=1}^{\infty}\ctru{(a,i)}{C}{_{\alpha\beta\gamma\delta}}\lambda^i\spacef.
				\end{equation}
			Their explicit expressions are written as
				\begin{align}
					\ctru{(4)}{C}{_{\alpha\beta\gamma\delta}}&=\frac{4}{\Omega^2\prn{\lambda_{1}}}\pct{d}{_{\tau\omega\chi\eta}}\ct{K}{_{\mu}^\tau}\ct{K}{_{\nu}^\omega}\ct{K}{_{\rho}^\chi}\ct{K}{_{\sigma}^\eta}\ct{N}{^\nu}\ct{N}{^\sigma}\ctc{P}{_{[\alpha}^\mu}\ct{\ell}{_{\beta]}}\ctc{P}{_{[\gamma}^\rho}\ct{\ell}{_{\delta]}}\spacef\label{eq:peeling-C-4},\\
					\ctru{(3)}{C}{_{\alpha\beta\gamma\delta}}&=\frac{1}{\Omega^3\prn{\lambda_{1}}}\pct{d}{_{\tau\omega\chi\eta}}\ct{K}{_{\mu}^\tau}\ct{K}{_{\nu}^\omega}\ct{K}{_{\rho}^\chi}\ct{K}{_{\sigma}^\eta}\bbrkt{-4\ctc{P}{^\mu_{[\alpha}}\ct{N}{^\nu}\ct{\ell}{_{\beta]}}\ct{N}{^\rho}\ct{\ell}{_{[\gamma}}\ct{N}{_{\delta]}}\ct{\ell}{^\sigma}-4\ct{N}{^\mu}\ct{\ell}{^\nu}\ct{\ell}{_{[\alpha}}\ct{N}{_{\beta]}}\ctc{P}{^\rho_{[\gamma}}\ct{\ell}{_{\rho]}}\ct{N}{^\sigma}\nonumber\\
					&-4\ctc{P}{^\mu_{[\alpha}}\ctc{P}{^\nu_{\beta]}}\ct{N}{^\sigma}\ct{\ell}{_{[\delta}}\ctc{P}{^\rho_{\gamma]}}-4\ct{N}{^\mu}\ct{\ell}{_{[\alpha}}\ctc{P}{^\nu_{\beta]}}\ctc{P}{^\rho_{[\gamma}}\ctc{P}{^\sigma_{\delta]}}}\spacef,\\
					\ctru{(2)}{C}{_{\alpha\beta\gamma\delta}}&=\frac{1}{\Omega^4\prn{\lambda_{1}}}\pct{d}{_{\tau\omega\chi\eta}}\ct{K}{_{\mu}^\tau}\ct{K}{_{\nu}^\omega}\ct{K}{_{\rho}^\chi}\ct{K}{_{\sigma}^\eta}\bbrkt{\ctc{P}{^\mu_{\alpha}}\ctc{P}{^\nu_{\beta}}\ctc{P}{^\rho_{\gamma}}\ctc{P}{^\sigma_{\delta}}-\ctc{P}{^\mu_{\alpha}}\ctc{P}{^\nu_{\beta}}\prn{\ctc{P}{^\rho_{\gamma}}\ct{N}{_{\delta}}+\ctc{P}{^\sigma_{\delta}}\ct{N}{_{\gamma}}}-\nonumber\\
					&-\ctc{P}{^\rho_{\gamma}}\ctc{P}{^\sigma_{\delta}}\prn{\ctc{P}{^\mu_{\alpha}}\ct{\ell}{^\nu}\ctc{N}{_{\beta}}+\ctc{P}{^\nu_{\beta}}\ct{N}{^\mu}\ct{\ell}{_{\alpha}}}+4\ct{\ell}{^\mu}\ct{N}{^\nu}\ct{\ell}{^\rho}\ct{N}{^\sigma}\ct{N}{_{[\alpha}}\ct{\ell}{_{\beta]}}\ct{N}{_{[\gamma}}\ct{\ell}{_{\delta]}}+4\ctc{P}{^\mu_{[\alpha}}\ct{N}{_{\beta]}}\ctc{P}{^\rho_{[\gamma}}\ct{\ell}{_{\delta]}}\ct{N}{^\sigma}\ct{\ell}{^\nu}\nonumber\\
					&+4\ctc{P}{^\mu_{[\alpha}}\ct{\ell}{_{\beta]}}\ctc{P}{^\rho_{[\gamma}}\ct{N}{_{\rho]}}\ct{\ell}{^\sigma}\ct{N}{^\nu}+4\ctc{P}{^\mu_{[\alpha}}\ctc{P}{^\nu_{\beta]}}\ct{N}{_{[\gamma}}\ct{\ell}{_{\delta]}}\ct{\ell}{^\rho}\ct{N}{^\sigma}+4\ctc{P}{^\rho_{[\gamma}}\ctc{P}{^\sigma_{\delta]}}\ct{N}{_{[\alpha}}\ct{\ell}{_{\beta]}}\ct{\ell}{^\mu}\ct{N}{^\nu}}\spacef,\\
					\ctru{(1)}{C}{_{\alpha\beta\gamma\delta}}&=\frac{1}{\Omega^5\prn{\lambda_{1}}}\pct{d}{_{\tau\omega\chi\eta}}\ct{K}{_{\mu}^\tau}\ct{K}{_{\nu}^\omega}\ct{K}{_{\rho}^\chi}\ct{K}{_{\sigma}^\eta}\bbrkt{-4\ctc{P}{^\mu_{[\alpha}}\ct{\ell}{^\nu}\ct{N}{_{\beta]}}\ct{\ell}{^\rho}\ct{N}{_{[\gamma}}\ct{\ell}{_{\delta]}}\ct{N}{^\sigma}-4\ct{\ell}{^\mu}\ct{N}{^\nu}\ct{N}{_{[\alpha}}\ct{\ell}{_{\beta]}}\ctc{P}{^\rho_{[\gamma}}\ct{N}{_{\rho]}}\ct{\ell}{^\sigma}\nonumber\\
					&-4\ctc{P}{^\mu_{[\alpha}}\ctc{P}{^\nu_{\beta]}}\ct{\ell}{^\sigma}\ct{N}{_{[\delta}}\ctc{P}{^\rho_{\gamma]}}-4\ct{\ell}{^\nu}\ct{N}{_{[\beta}}\ctc{P}{^\mu_{\alpha]}}\ctc{P}{^\rho_{[\gamma}}\ctc{P}{^\sigma_{\delta]}}}\spacef,\\
					\ctru{(0)}{C}{_{\alpha\beta\gamma\delta}}&=\frac{4}{\Omega^6\prn{\lambda_{1}}}\pct{d}{_{\tau\omega\chi\eta}}\ct{K}{_{\mu}^\tau}\ct{K}{_{\nu}^\omega}\ct{K}{_{\rho}^\chi}\ct{K}{_{\sigma}^\eta}\ct{\ell}{^\nu}\ct{\ell}{^\sigma}\ctc{P}{_{[\alpha}^\mu}\ct{N}{_{\beta]}}\ctc{P}{_{[\gamma}^\rho}\ct{N}{_{\delta]}}\spacef.\\
				\end{align}
			Observe that the leading-order term of \cref{eq:peeling-C-4} reads
			 	\begin{equation}\label{eq:asymptotic-leading-term-weyl}
					\ctru{(4,0)}{C}{_{\alpha\beta\gamma\delta}}=\frac{4}{\Omega^2\prn{\lambda_{1}}}\ct{d}{_{\mu\nu\rho\sigma}}\ct{N}{^\nu}\ct{N}{^\sigma}\ctc{P}{_{[\alpha}^\mu}\ct{\ell}{_{\beta]}}\ctc{P}{_{[\gamma}^\rho}\ct{\ell}{_{\delta]}}
				\end{equation}
			and is determined by the rescaled Weyl tensor $ \ct{d}{_{\alpha\beta\gamma}^\delta} $ projected to a Petrov-type N Weyl-candidate tensor. It should be observed that this leading term has always the structure 
			\begin{equation}
			\ct{\ell}{_{[\alpha}}\ct{d}{_{\beta][\gamma}}\ct{\ell}{_{\delta]}}
			\end{equation}
			where
			\begin{equation}\label{eq:d-projected-N}
			\ct{d}{_{\beta\gamma}}\defeq \ct{d}{_{\mu\rho\nu\sigma}}\ct{N}{^\rho}\ct{N}{^\sigma}\ctc{P}{_\beta^\mu}\ctc{P}{_\gamma^\nu}
			\end{equation}
			is a traceless symmetric tensor orthogonal to both $\ct{N}{^\alpha}$ and $\ct{\ell}{^\alpha}$. Hence, the leading term vanishes if and only if $\ct{d}{_{\beta\gamma}}$ vanishes, which is equivalent to saying that $\ct{N}{^\alpha}$ is a PND of the rescaled Weyl tensor. This will serve as further inspiration for our radiation condition in the next section.\\
			
			Now one can perform \cref{it:asymptotic-propagation-third-step}, finally arriving at the next result:
				\begin{thm}[Peeling of the Weyl tensor]\label{thm:peeling-weyl}
					Let $ \prn{\cs{M},\ct{g}{_{\alpha\beta}}} $ be a conformal completion of a physical space-time with $ \Lambda=0 $ as presented on page  \pageref{it:energytensorassumption} 
					and let $ \gamma $ be a lightlike geodesic with affine parameter $ \lambda $ and tangent vector field $ \ct{\ell}{^\alpha} $ as in \cref{eq:null-geodesic-normalisation}. Also, let one end point $ p_{0} $ ($ \lambda=\lambda_{0}=0 $) of $ \gamma $ be at $ \scri^+ $ and the other one, $ p_{1} $ ($ \lambda=\lambda_{1}=-1 $), in $ \ps{M} $. Then, the asymptotic behaviour of the physical Weyl tensor $ \pt{C}{_{\alpha\beta\gamma\delta}} $ along $ \gamma $ follows by application of \cref{it:asymptotic-propagation-first-step,it:asymptotic-propagation-second-step,it:asymptotic-propagation-third-step} on page \pageref{it:asymptotic-propagation-first-step} and reads
						\begin{equation}\label{eq:peeling-weyl}
						  \lct{\underaccent{*}{\hat{C}}}{_{\alpha\beta\gamma\delta}}=	\lambda \ctru{(N)}{d}{_{\alpha\beta\gamma\delta}}+\lambda^2\ctru{(III)}{e}{_{\alpha\beta\gamma\delta}}+\lambda^3\ctru{(II/D)}{f}{_{\alpha\beta\gamma\delta}}+\lambda^4\ctru{(I)}{g}{_{\alpha\beta\gamma\delta}}+\lambda^5\ctru{(I)}{h}{_{\alpha\beta\gamma\delta}}+\mathcal{O}\prn{\lambda^6}\spacef,
						\end{equation}
					near $ \lambda=\lambda_{0}=0 $, where the tensors
						\begin{align}
							\ctru{(N)}{d}{_{\alpha\beta\gamma\delta}}\defeq&- \ctru{(4,0)}{C}{_{\alpha\beta\gamma\delta}}\spacef,\label{eq:rrw-d}\\
							\ctru{(III)}{e}{_{\alpha\beta\gamma\delta}}\defeq& \ctru{(3,0)}{C}{_{\alpha\beta\gamma\delta}}+\frac{\Omega_{2}}{2}\ctru{(4,0)}{C}{_{\alpha\beta\gamma\delta}}-\ctru{(4,1)}{C}{_{\alpha\beta\gamma\delta}}\spacef,\label{eq:rrw-e}\\
							\ctru{(II/D)}{f}{_{\alpha\beta\gamma\delta}}\defeq& -\ctru{(2,0)}{C}{_{\alpha\beta\gamma\delta}}-\Omega_{2}\ctru{(3,0)}{C}{_{\alpha\beta\gamma\delta}}+\frac{\Omega_{3}}{6}\ctru{(4,0)}{C}{_{\alpha\beta\gamma\delta}}+\ctru{(3,1)}{C}{_{\alpha\beta\gamma\delta}}-\ctru{(4,2)}{C}{_{\alpha\beta\gamma\delta}}\nonumber\\
							&+\frac{\Omega_{2}}{2}\ctru{(4,1)}{C}{_{\alpha\beta\gamma\delta}}\spacef,\label{eq:rrw-f}\\
							\ctru{(I)}{g}{_{\alpha\beta\gamma\delta}}\defeq& \ctru{(1,0)}{C}{_{\alpha\beta\gamma\delta}}+\frac{3\Omega_{2}}{2}\ctru{(2,0)}{C}{_{\alpha\beta\gamma\delta}}+\prn{\frac{\Omega_{2}^2}{4}-\frac{\Omega_{3}}{3}}\ctru{(3,0)}{C}{_{\alpha\beta\gamma\delta}}+\frac{\Omega_{4}}{4!}\ctru{(4,0)}{C}{_{\alpha\beta\gamma\delta}}\nonumber\\
							&-\ctru{(2,1)}{C}{_{\alpha\beta\gamma\delta}}+\ctru{(3,2)}{C}{_{\alpha\beta\gamma\delta}}-\ctru{(4,3)}{C}{_{\alpha\beta\gamma\delta}}+\frac{\Omega_{2}}{2}\ctru{(4,2)}{C}{_{\alpha\beta\gamma\delta}}-\Omega_{2}\ctru{(3,1)}{C}{_{\alpha\beta\gamma\delta}}\nonumber\\
							&+\frac{\Omega_{3}}{6}\ctru{(4,1)}{C}{_{\alpha\beta\gamma\delta}}\spacef,\label{eq:rrw-g}\\
							\ctru{(I)}{h}{_{\alpha\beta\gamma\delta}}\defeq& -\ctru{(0,0)}{C}{_{\alpha\beta\gamma\delta}}-2\Omega_{2}\ctru{(1,0)}{C}{_{\alpha\beta\gamma\delta}}+\prn{\frac{\Omega_{3}}{2}-\frac{3}{4}\Omega_{2}^2}\ctru{(2,0)}{C}{_{\alpha\beta\gamma\delta}}+\prn{\frac{1}{6}\Omega_{2}\Omega_{3}-\frac{\Omega_{4}}{12}}\ctru{(3,0)}{C}{_{\alpha\beta\gamma\delta}}\nonumber\\
							&+\frac{\Omega_{5}}{5!}\ctru{(4,0)}{C}{_{\alpha\beta\gamma\delta}}+\ctru{(1,1)}{C}{_{\alpha\beta\gamma\delta}}-\ctru{(2,2)}{C}{_{\alpha\beta\gamma\delta}}+\ctru{(3,3)}{C}{_{\alpha\beta\gamma\delta}}-\ctru{(4,4)}{C}{_{\alpha\beta\gamma\delta}}\nonumber\\
							&+\frac{3}{2}\Omega_{2}\ctru{(2,1)}{C}{_{\alpha\beta\gamma\delta}}-\Omega_{2}\ctru{(3,2)}{C}{_{\alpha\beta\gamma\delta}}+\frac{\Omega_{2}}{2}\ctru{(4,3)}{C}{_{\alpha\beta\gamma\delta}}+\frac{\Omega_{2}^2}{4}\ctru{(3,1)}{C}{_{\alpha\beta\gamma\delta}}\nonumber\\
							&-\frac{\Omega_{3}}{3}\ctru{(3,1)}{C}{_{\alpha\beta\gamma\delta}}+\frac{\Omega_{3}}{6}\ctru{(4,2)}{C}{_{\alpha\beta\gamma\delta}}+\frac{\Omega_{4}}{4!}\ctru{(4,1)}{C}{_{\alpha\beta\gamma\delta}}\spacef.\label{eq:rrw-h}
						\end{align}
					are Weyl-tensor candidates labelled with their Petrov type, respectively; $ \Omega_{i} $, with $ i=1,2,3,4,5 $, is the $ i$-th derivative of $ \Omega $ w.r.t. $ \lambda $ evaluated at $ \lambda=\lambda_{0}=0 $, and $ \ctru{(a)}{C}{_{\alpha\beta\gamma\delta}} $, with $ a=0,1,2,3,4 $, are the Weyl-tensor candidates of \cref{tab:weyl-candidates} each one having one non-vanishing Weyl scalar $ \csru{(a)}{\Psi}_{a} $ in the tetrad containing $ \ct{\ell}{^\alpha} $ and $ \ct{N}{^\alpha} $.
				\end{thm}
			 	\begin{proof}
				 	The asymptotic propagation along $ \gamma $ of the physical Weyl tensor is given in \cref{eq:asymptotic-propagation-weyl-tensor}. Then, one expands around $ \lambda_{0}=0 $ 
				 	and rearranges the terms by powers of $ \lambda $. The algebraic structure of the first 5 terms of \cref{eq:rrw-e,eq:rrw-f,eq:rrw-h,eq:rrw-g,eq:rrw-d} follows from the properties listed in \cref{tab:weyl-candidates}.
			 	\end{proof}
			 	\begin{remark}
				 	The Weyl-tensor candidates of \cref{eq:rrw-e,eq:rrw-f,eq:rrw-h,eq:rrw-g,eq:rrw-d} have the algebraic structure specified in \cref{tab:rrw-tensors}. Notice that though this constitutes the so called \emph{peeling} property, the present derivation is purely geometric, showing neatly that we derive the behaviour of the physical field (the Weyl tensor in this case) as it approaches $ \scri^+ $ along null geodesics, thereby providing a solid foundation for the so-called peeling behaviour. Notice, further, that once this construction has been performed, it can be applied to any physical field whatsoever by just following the \cref{it:asymptotic-propagation-first-step,it:asymptotic-propagation-second-step,it:asymptotic-propagation-third-step} on page \pageref{it:asymptotic-propagation-first-step} and using the explicit form of $ \ct{L}{_{\beta}^\alpha} $ in \cref{eq:operatorL-final-form}.
			 	\end{remark}
			 	\begin{remark}
				 	 The Weyl scalars of the first three elements in \cref{eq:peeling-weyl} have the following expressions:
				 	 	\begin{align}
				 	 		\eta_{4}&=-\frac{1}{\Omega^{2}(\lambda_{1})}\phi_{4}\spacef,\quad&\tau_{4}&=-\frac{\Omega_{3}}{6\Omega^2\prn{\lambda_{1}}}\phi_{4}-\csru{(4,2)}{\psi}_{4}+\frac{\Omega_{2}}{2}\csru{(4,1)}{\psi}_{4}\spacef,\\
				 	 		\chi_{4}&=\frac{\Omega_{2}}{2\Omega^{2}(\lambda_{1})}\phi_{4}-\csru{(4,1)}{\psi}_{4}\spacef,\quad&\tau_{3}&=-\frac{\Omega_{2}}{\Omega^3\prn{\lambda_{1}}}\phi_{3}+\csru{(3,1)}{\psi}_{3}\spacef,\\
				 	 		\chi_{3}&=\frac{1}{\Omega^3\prn{\lambda_{1}}}\phi_{3}\spacef,\quad&\tau_{2}&=-\frac{1}{\Omega^4\prn{\lambda_{1}}}\phi_{2}\spacef,\label{eq:xi3}
				 	 	\end{align}
				 	 where $ \phi_{i} $ with $ i=2,3,4 $ are the scalars of the rescaled Weyl tensor $ \ct{d}{_{\alpha\beta\gamma}^\delta} $ and $ \csru{(a,i)}{\psi}_{i} $ are the scalars corresponding to the tensors $ \ctru{(a,i)}{C}{_{\alpha\beta\gamma\delta}} $ of \cref{eq:weyl-terms-expanded}.
			 	\end{remark}
		 		\begin{table}[h!]
					\centering
					\begin{tabular}{ |M{3cm}| M{3cm} | M{3cm}| N }
						\hline
							\quad Weyl-tensor candidate & Non-vanishing  scalars in general  & degeneracy of $ \ct{\ell}{^\alpha} $ as PND in general &\\ \hline 
						$\ctru{(N)}{d}{_{\alpha\beta\gamma\delta}} $& $ \cs{\eta}_{4} $ &  $ 4 $&\\[1cm] \hline 
						$ \ctru{(III)}{e}{_{\alpha\beta\gamma\delta}} $ &$ \chi_{3} $ $ \chi_{4} $ &$3 $&\\[1cm] \hline
						$ \ctru{(II/D)}{f}{_{\alpha\beta\gamma\delta}} $&   $ \cs{\tau}_{4} $ $ \cs{\tau}_{3} $ $ \cs{\tau}_{2} $ &$ 2 $&\\[1cm] \hline				
						$ \ctru{(I)}{g}{_{\alpha\beta\gamma\delta}} $& $ \cs{\nu}_{4} $ $ \cs{\nu}_{3} $ $ \cs{\nu}_{2} $ $ \cs{\nu}_{1} $ &$ 1 $&\\[1cm] \hline
						$ \ctru{(I)}{h}{_{\alpha\beta\gamma\delta}} $& $ \cs{\mu}_{4} $ $ \cs{\mu}_{3} $ $ \cs{\mu}_{2} $ $ \cs{\mu}_{1} $ $ \cs{\mu}_{0} $ & $ 0 $&\\[1cm] \hline	
					\end{tabular}
					\caption[Asymptotic structure of the Weyl tensor]{The vector $ \ct{\ell}{^\alpha} $, tangent to $ \gamma $,  is a PND of the first four terms in the asymptotic propagation of the physical Weyl tensor. The degree of degeneracy decreases towards higher order terms; this effect is commonly referred to as the \emph{peeling} property of the Weyl tensor. Observe that for particular situations these Weyl-tensor candidates can be more degenerate; e.g., $ \ctru{(III)}{e}{_{\alpha\beta\gamma}^\delta} $ in general has Petrov type III but it can have Petrov type N ($ \chi_{4}\neq 0 $, $ \chi_{3}=0 $) or zero ($ \chi_{3}=0=\chi_{4} $).}
					\label{tab:rrw-tensors}
				\end{table}

%% file: section-5/section-5.tex
As it has been expounded in the introduction, we give a characterisation of the gravitational radiation grounded on the rescaled version \cite{Fernandez-Alvarez_Senovilla-dS} of the Bel-Robinson tensor $ \ct{\T}{_{\alpha\beta\gamma\delta}} $ \cite{Bel1958}
	\begin{equation}
			\ct{\D}{_{\alpha\beta\gamma\delta}} \defeq \Omega^{-2}\ct{\T}{_{\alpha\beta\gamma\delta}} =  \ct{d}{_{\alpha\mu\gamma}^\nu}\ct{d}{_{\delta\nu\beta}^\mu} + \ctru{*}{d}{_{\alpha\mu\gamma}^\nu}\ctru{*}{d}{_{\delta\nu\beta}^\mu}\spacef.
	\end{equation}
One constructs the \emph{asymptotic radiant supermomentum} as
	\begin{equation}\label{eq:radiant-super-momentum}
		\ct{\Q}{^\alpha}\defeq -\ct{N}{^\mu}\ct{N}{^\nu}\ct{N}{^\rho}\ct{\D}{^\alpha_{\mu\nu\rho}}\spacef.
	\end{equation}
The definition and description of general \emph{radiant supermomenta} are presented in the companion paper \cite{Fernandez-Alvarez_Senovilla-dS}; the following fundamental properties were also presented in  \cite{Fernandez-Alvarez_Senovilla20}:
	\begin{properties}
		\item \label{it:Q-future-property} $  \ct{\Q}{^\mu}$ is lightlike $  \ct{\Q}{^\mu}\ct{\Q}{_\mu} \eqs 0  $ and future pointing at $\scri$, which follows from the causal character of $ \ct{N}{^{\alpha}} $ and known properties of superenergy tensors \cite{Bergqvist2004,Senovilla2000}.
		\item Under gauge transformations it changes as 
		 	\begin{equation}\label{eq:Qgauge}
		 		\ct{\Q}{^\alpha} \rightarrow \omega^{-7} \prn{\ct{\Q}{^\alpha} -3\frac{\Omega}{\omega} \ct{\D}{^\alpha_{\beta\rho\tau}}\ct{N}{^\beta}\ct{N}{^\rho}\cd{^{\tau}}\omega} +\mathcal{O}(\Omega^2).
		 	\end{equation}
		
		\item \label{it:div-free-Q}It is divergence-free at $ \scri $, independently of the matter content,
			\begin{equation}\label{eq:Q-divergence}
				\cd{_\mu}\ct{\Q}{^\mu}\eqs 0\spacef. 
			\end{equation}
	\end{properties}
The last property is easily verified by noting first that in general \cite{Fernandez-Alvarez_Senovilla-dS}
	\begin{equation}\label{eq:div-superenergy-tensor}
			\cd{_\mu}\ct{\D}{_{\alpha\beta\gamma}^\mu}= 2\ct{d}{_{\mu\gamma\nu\alpha}}\ct{y}{_{\beta}^{\nu\mu}}+2\ct{d}{_{\mu\gamma\nu\beta}}\ct{y}{_{\alpha}^{\nu\mu}}+ \ct{g}{_{\alpha\beta}}\ct{d}{^{\mu\nu\rho}_\gamma}\ct{y}{_{\mu\nu\rho}}\quad, 
	\end{equation}
so that, recalling \cref{eq:cefesScriDerN}, one can write
	\begin{equation}
		\cd{_{\mu}}\ct{\Q}{^\mu}\eqs 4 \ct{N}{^\alpha}\ctn{D}{^{\beta\gamma}}\ct{y}{_{\alpha\beta\gamma}}\eqs \ctcnn{D}{^{BC}}\ctcn{E}{^\beta_{B}}\ctcn{E}{^{\gamma}_{C}}\ct{N}{^{\alpha}}\ct{y}{_{\alpha\beta\gamma}}+\sqrt{2}\ctcnn{D}{^B}\ctcn{E}{^{\beta}_{B}}\ct{N}{^\alpha}\ct{N}{^{\gamma}}\ct{y}{_{\alpha\beta\gamma}}\spacef,
	\end{equation}
where in the last equality we have exploited the fact that $ \ctn{D}{^{\alpha\beta}}=\ctn{D}{^{ab}}\ct{e}{^\alpha_{a}}\ct{e}{^\beta_{b}} $ and expanded in the bases $ \cbrkt{\ct{N}{^a},\ctcn{E}{^\alpha_{A}}} $, $ \cbrkt{-\cts{\ell}{_{\alpha}},\ctcn{W}{_{\alpha}}^A} $ --also, $ \ctcnn{D}{^{AB}}\defeq\ctcn{W}{_{\alpha}^A}\ctcn{W}{_{\beta}^B} \ctn{D}{^{\alpha\beta}}$. From \cref{eq:cottonPhysicalMatter}, taking into account \cref{eq:matter-scri}, it follows that
	\begin{align}
		\ctcn{E}{^\beta_{B}}\ctcn{E}{^{\gamma}_{C}}\ct{N}{^{\alpha}}\ct{y}{_{\alpha\beta\gamma}}&\eqs \frac{1}{2}\varkappa\Omega^{-1}\mcn{_{BC}}\ct{N}{^{\mu}}\ct{N}{^\nu}\ct{T}{_{\mu\nu}}\spacef,\\
		\ctcn{E}{^{\beta}_{B}}\ct{N}{^\alpha}\ct{N}{^{\gamma}}\ct{y}{_{\alpha\beta\gamma}}&\eqs 0\spacef.
	\end{align}
By properties described in  \cite{Fernandez-Alvarez_Senovilla-dS} , $ \ctcnn{D}{^{BC}}\mcn{_{BC}}=0 $ and then \cref{it:div-free-Q} follows\footnote{In \cite{Fernandez-Alvarez_Senovilla20}, \cref{it:div-free-Q} was presented in a less general situation. As we have shown, $ \ct{\Q}{^\alpha} $ is divergence-free at $ \scri $ independently of the matter content.}.\\

The asymptotic radiant supermomentum $ \ct{\Q}{^\alpha} $ is geometrically well defined, as it is built only with the generators of $ \scri $ and the rescaled Weyl tensor $ \ct{d}{_{\alpha\beta\gamma}^\delta} $. Moreover, it has a good gauge behaviour at $ \scri $, $ \ctg{\Q}{^\alpha}=\omega^{-7}\ct{\Q}{^\alpha} $. These facts, together with the close relation with the intrinsic fields on $ \scri $ exhibited by the rescaled Weyl tensor --see \cref{ssec:curvature-scri-and-fields}--, suggests a link between $ \ct{\Q}{^\alpha} $ and the news tensor $ \ct{N}{_{ab}} $ of \cref{eq:news-tensor}. To show that this is the case, first decompose the asymptotic radiant supermomentum as	
	\begin{equation}
	\ct{\Q}{^\alpha}\eqs \cs{\W}\cts{\ell}{^\alpha} + \cts{\Q}{^\alpha}=\cs{\W}\cts{\ell}{^\alpha} + \cts{\Q}{^a}\ct{e}{^\alpha_a}\quad ,
	\end{equation}
where $ \cts{\ell}{_\alpha}=\ct{\omega}{_{\alpha}^a}\cts{\ell}{_{a}} $ is a lightlike field at $ \scri $ associated to a foliation as in \cref{eq:ell-F-foliation} whose restriction on each cut gives the $ \ct{\ell}{_\alpha} $ of \cref{eq:ell-cut}. The quantity 
	\begin{equation}\label{eq:radiant-superenergy}
	\cs{\W}\defeq -\ct{N}{_\mu}\ct{\Q}{^\mu}\geq 0
	\end{equation} 
is the \emph{asymptotic radiant superenergy} and the vector field
	\begin{equation}
		\cts{\Q}{^a}\defeq \cs{\Z}\ct{N}{^a} + \ctcn{\Q}{^A}\ctcn{E}{^a_A}\quad \text{with}\quad\cs{\Z}\defeq -\cts{\ell}{_\mu}\ct{\Q}{^\mu} \geq 0\label{eq:Z}
	\end{equation}
is the \emph{asymptotic radiant super-Poynting} \cite{Fernandez-Alvarez_Senovilla20}. Observe that $ \cs{\W} $ is invariant under the choice of $ \cts{\ell}{_{\alpha}} $, whereas $ \cs{\Z} $ and $ \ctcn{\Q}{^A} $ depend on the choice of foliation --one can consider this decomposition on a single cut $ \Sc $ only, and then these quantities depend on the choice of that cut. From general formulae \cite{Fernandez-Alvarez_Senovilla-dS}, the relation between these quantities and the lightlike projections of $ \ct{d}{_{\alpha\beta\gamma}^\delta} $ is
	\begin{align}
		\cs{\W}& = 2\ctcnn{C}{_{AB}}\ctcnn{C}{^{AB}}=2\ctcnn{D}{_{AB}}\ctcnn{D}{^{AB}}\geq 0 \label{eq:W-d},\\
		\cs{\Z}&=4\ctcnn{C}{_A}\ctcnn{C}{^A}=4\ctcnn{D}{_A}\ctcnn{D}{^A}\geq 0 \label{eq:Z-d},\\
		\ctcn{\Q}{^A}&= 4\sqrt{2}\ctcnn{C}{_P}\ctcnn{C}{^{AP}}\spacef.
	\end{align}
Then, \cref{eq:DA-news,eq:DAB-news,eq:CA-news,eq:CAB-news} bring forth the connection between the asymptotic radiant supermomentum and the news tensor
	\begin{align}
		\cs{\W}&= 2 \ct{\dot{N}}{^{RT}}\ct{\dot{N}}{_{RT}} \geq 0\quad ,\label{eq:W-news}\\
		\cs{\Z}&\eqc 2 \cdc{_{R}}\ct{N}{^R_T} \cdc{_{M}}\ct{N}{^{MT}}  \geq 0\quad ,\label{eq:Z-news}\\
		\ctcn{\Q}{^A} &\eqc -4 \ct{\dot{N}}{^{MA}} \cdc{_{E}}\ct{N}{^E_{M}}\quad .\label{eq:QA-news}	
	\end{align}
	\subsection{Radiation condition}
		In \cite{Fernandez-Alvarez_Senovilla20} a new criterion to determine the presence of radiation at $ \scri $ escaping from the space-time was presented. The criterion holds in the $ \Lambda > 0$ case \cite{Fernandez-Alvarez_Senovilla20b} too. It translates into the following results --we refer the reader to \cite{Fernandez-Alvarez_Senovilla20} for the proofs--
		\begin{thm}[Radiation condition on a cut]\label{thm:noradCut}
		 There is no gravitational radiation on a given cut $ \Sc \subset \scri $ if and only if the radiant super-Poynting $ \cts{\Q}{^a} $ vanishes on that cut:\\
				 \begin{equation*}
		   \ct{N}{_{AB}}= 0 \quad\liff\quad \cts{\Q}{^a}\eqc 0 \quad (\liff\quad \cs{\Z}= 0).
				 \end{equation*}
		\end{thm}
		\begin{remark}\label{rmk:noradCut-eqstatement}
			Equivalently: there is no gravitational radiation on a given cut $ \Sc \subset \scri $ if and only if the radiant supermomentum is orthogonal to $\Sc$ everywhere and not collinear with $N^\alpha$.
		\end{remark}
		\begin{remark}\label{rmk:subtlety-1}
			The topology of $ \scri $ plays a key role in the proof presented in \cite{Fernandez-Alvarez_Senovilla20}. If the cuts do not have $ \mathbb{S}^2$-topology, 
				\begin{equation*}
					\cdc{_{M}}\ct{N}{_{B}^M}\eqc 0\hspace{3mm}  \Longleftrightarrow \hspace{3mm} \cts{\Q}{^a}\eqc 0\spacef,
				\end{equation*}
			still holds even if $ \ct{N}{_{AB}}\neq 0 $ may happen.
			 In any case, this does not pose a problem when considering open portions of $ \scri $, instead of single cuts --see \cref{rmk:subtlety-2}.
		\end{remark}
		\begin{thm}[No radiation on $ \Delta$] \label{thm:noradDelta} 
		There is no gravitational radiation on the open portion $ \Delta \subset \scri $ with the same topology of $ \scri $ if and only if the radiant supermomentum $ \ct{\Q}{^\alpha} $ vanishes on $\Delta$:
			 \begin{equation*}
				 \ct{N}{_{ab}} \eqsopen  0 \quad\liff\quad \ct{\Q}{^\alpha}\eqsopen  0 .
			 \end{equation*}
		\end{thm}	
		\begin{remark}
		 The following two re-statements are equivalent to that of \cref{thm:noradDelta}:
	 		\begin{itemize}
			 	\item  No gravitational radiation on $\Delta\subset \scri \iff \ct{\Q}{^\alpha}$ is orthogonal to all surfaces within $\Delta$ and not collinear with the generators.
			 	\item No gravitational radiation on $\Delta\subset \scri \iff \ct{N}{^\alpha}\evalat{\Delta}$ is a principal null vector of $\ct{d}{_{\alpha\beta\gamma}^\delta}\evalat{\Delta}$.
		 	\end{itemize}
		 The first point follows from \cref{rmk:noradCut-eqstatement}, particularising to any possible cut within $ \Delta $ --hence, implying that $ \ct{\Q}{^\alpha}\eqsopen 0 $. The second statement is a consequence of the general properties of radiant supermomenta \cite{Fernandez-Alvarez_Senovilla-dS}. Notice that the absence of radiation according to our criterion implies that $\ct{N}{^\alpha}$ is a multiple PND of the rescaled Weyl tensor and, therefore, according to the discussion following \cref{eq:d-projected-N} in the previous section, the leading term \eqref{eq:asymptotic-leading-term-weyl} in the peeling property vanishes. Nevertheless, it can still happen that the leading term  \eqref{eq:asymptotic-leading-term-weyl} vanishes and there is gravitational radiation at $\scri$, if $\ct{N}{^\alpha}$ is a non-multiple PND of the rescaled Weyl tensor. In that case the leading term in the peeling behaviour will be the type III Weyl tensor candidate of \cref{eq:rrw-e} times $\lambda^2$ and the radiant supermomentum is non-zero and points along $\ct{N}{^\alpha}$ at $\scri$. Another equivalent formulation of \cref{thm:noradDelta} is then
		 	\begin{itemize}
			 	\item  No gravitational radiation on $\Delta\subset \scri \iff $ $ \ctru{(N)}{d}{_{\alpha\beta\gamma}^\delta}\eqsopen 0 $ and $  \ctru{(III)}{e}{_{\alpha\beta\gamma}^\delta}\evalat{\Delta}$ has Petrov type N or zero for all space-time null geodesics $ \gamma $ reaching $ \Delta $.
		 	\end{itemize}
		 This implication follows from \cref{eq:xi3} or \cref{eq:rrw-e}. Hence, one has to distinguish two regimes of asymptotic gravitational radiation:
		 	\begin{enumerate}
		 	\item Space-times with non-vanishing leading order, $  \ctru{(N)}{d}{_{\alpha\beta\gamma}^\delta}\neqsopen 0$ --which holds for every space-time null geodesic $ \gamma $ arriving at $ \Delta $ by construction. These can have a Petrov type III, N or zero $ \lambda^2 $-term $ \ctru{(III)}{e}{_{\alpha\beta\gamma}^\delta}\evalat{\Delta} $. Observe that since in this regime $ \ct{N}{^\alpha} $ is not a repeated PND of $ \ct{d}{_{\alpha\beta\gamma}^\delta}\evalat{\Delta} $, there are always space-time null geodesics $ \gamma $ such that $ \ctru{(III)}{e}{_{\alpha\beta\gamma}^\delta}\evalat{\Delta} $ is of type N.
		 	\item Space-times with $ \ctru{(N)}{d}{_{\alpha\beta\gamma}^\delta}\eqsopen 0  $ but with a Petrov type III $ \lambda^2 $-term $ \ctru{(III)}{e}{_{\alpha\beta\gamma}^\delta}\evalat{\Delta} $ --which holds for every space-time null geodesics $ \gamma $ arriving at $ \Delta $. This is a curious case, as the flux of energy across $ \Delta $ carried by gravitational waves is constant in time, i.e., $ \ctd{N}{_{ab}}\eqsopen0 $ but $ \ct{N}{_{ab}}\neqsopen0 $. Moreover, the radiant supermomentum points along $ \ct{N}{^\alpha} $ and, therefore, it is orthogonal to every possible cut within $ \Delta $.
		 	\end{enumerate}
		 Their differences are an interesting object of study.
		\end{remark}
		
		\begin{remark}\label{rmk:subtlety-2}
			 Regarding \cref{rmk:subtlety-1}, it may be the case that even if one foliates $ \Delta $ by topological non-spheres, a different choice of foliation gives topological-$ \mathbb{S}^2 $ cuts. Hence, \cref{thm:noradCut} applies to those new cuts within $ \Delta $. It may be also the case that a foliation by topological-$ \mathbb{S}^2 $ cuts of a given $ \Delta $ is not possible --as it happens in the C-metric \cite{Ashtekar-Dray81}--, hence the situation described in \cref{rmk:subtlety-1} has to be considered. Observe that any foliation with $\cdc{_C} \ct{N}{_A ^C}=0$ is very special as the cuts are precisely defined by orthogonality to both $\ct{N}{^\alpha}$ and $\ct{Q}{^\alpha}$ if this is feasible (examples of such a situation can be exhibited in the C-metric and the Robinson-Trautman type N space-times, see the companion \cite{Fernandez-Alvarez_Senovilla-dS}). Hence, any other cut not belonging to the special foliation will have a non-vanishing radiant super-Poynting $ \cts{\Q}{^a} $, and therefore a non-zero $\cdc{_C} \ct{N}{_A ^C}$ and non-vanishing news, {\em unless} $\ctd{N}{_{AB}}=0$. However, our \cref{thm:noradDelta} requires the whole supermomentum $ \ct{\Q}{^\alpha} $, and not just $ \cts{\Q}{^a} $, to vanish, which involves $ \ctd{N}{_{AB}} $ as well --see \cref{eq:W-news}. Therefore, even if $ \cdc{_{C}}\ct{N}{_{A}^C} \eqc  0 $ on every cut $ \Sc\subset\Delta $ of such special foliation, our criterion still detects correctly the gravitational radiation.
		\end{remark}
		
	\subsection{Balance law}\label{ssec:structurebalance}
		It is possible to write a balance law describing the outgoing superenergy flux and the news tensor. Begin by  considering a connected portion $ \Delta\in\scri $, with $ \mathbb{R}\times\mathbb{S}^2 $ topology. Let it be bounded by two (non-intersecting) cuts, $ \Sc_{1} $ and $ \Sc_{2} $, the latter to the future of the former, and orthogonal lightlike vector fields (other than $ \ct{N}{^\alpha} $) $ \ctrd{1}{\ell}{^\alpha} $ and $ \ctrd{2}{\ell}{^\alpha} $ as in \cref{eq:ell-cut}, respectively. Consider any lightlike field $ \cts{\ell}{_\alpha}=\ct{\omega}{_{\alpha}^a}\cts{\ell}{_{a}} $ in $ \Delta $ with the properties of \cref{eq:ell-F-foliation}, such that 
			\begin{equation}
				\cts{\ell}{_\alpha}\eqSv{1}\ctrd{1}{\ell}{_\alpha}\spacef,\quad\cts{\ell}{_\alpha}\eqSv{2}\ctrd{2}{\ell}{_\alpha}\spacef.
			\end{equation}
		\Cref{eq:Q-divergence} decomposes as 
			\begin{equation}\label{eq:balanced-differential}
				\cts{\ell}{^\mu}\cd{_{\mu}}\cs{\W}+\cs{\W}\ct{\Psi}{^m_{m}}\eqsopen-\cds{_{m}}\cts{\Q}{^m}\spacef,
			\end{equation}
		where \cref{eq:inducedDevscri,eq:induced-connection-psi} were used. Using the quantities and notation introduced in \cref{ssec:basic-geometry}, integration of \cref{eq:balanced-differential} leads to a Gauss-law formula
			\begin{equation}\label{eq:balance-law-a}
				\int_{\Delta}\prn{\cts{\ell}{^\mu}\cd{_{\mu}}\cs{\W}+\cs{\W}\ct{\psi}{^m_{m}}}\epsilon=\csb{\Phi}{\Sc_2}-\csb{\Phi}{\Sc_1}\spacef,
			\end{equation}
		where $ \csb{\Phi}{\Sc} $ is the \emph{radiant superenergy density flux}, defined as
			\begin{equation}\label{eq:flux-superenergy}
				\csb{\Phi}{\Sc}\defeq\int_{\Sc}\cs{\Z}\csC{\epsilon}\geq 0\spacef,\quad \csb{\Phi}{\Sc}= 0\iff \ct{N}{_{AB}}\eqc 0\spacef.
			\end{equation}
		\Cref{eq:balance-law-a} shows that \emph{the change of the asymptotic radiant superenergy density $ \cs{\W} $ along any outgoing lightlike direction $ \cts{\ell}{^\alpha} $ in a volume $ \Delta $ is balanced by the flux of radiant superenergy density on the boundary of $ \Delta $} --constituted by the two cuts $ \Sc_{1,2} $.	Let us remark that this formula is valid in the presence of arbitrary matter fields --with the general assumption \ref{it:energytensorassumption} on page \pageref{it:energytensorassumption}. In other words, \cref{eq:balance-law-a} contains purely geometric terms. The choice of $ \cts{\ell}{^\alpha} $ does not change \cref{eq:balance-law-a}, as the difference between one choice and another can be checked to be a total divergence that integrates out \cite{Fernandez-Alvarez_Senovilla20}. Moreover, \cref{eq:balanced-differential} is gauge invariant. After some manipulation of the integrand and using \cref{eq:Z-news}, the radiant superenergy density flux reads
			\begin{equation}\label{eq:flux-news}
					\csb{\Phi}{\Sc}=\int_{\Sc}\ct{N}{_{RS}}\prn{2\cs{K}\ct{N}{^{RS}}-\cdc{_{M}}\cdc{^M}\ct{N}{^{RS}}}\csC{\epsilon}\spacef,
			\end{equation}
		where $ \cs{K} $ denotes the Gaussian curvature of the cuts. Observe that, although it does not manifest itself explicitly so, the integral on the right-hand side of \cref{eq:flux-news} is positive. It shows that the flux of radiant superenergy is indeed associated to the presence of gravitational waves and sourced, ultimately, by the news tensor --as one could already expect from \cref{eq:W-news}. The first term on the right-hand side of \cref{eq:flux-news} reminds us of the energy-momentum loss due to gravitational waves of \cref{eq:energy-loss-gw}. Without loss of generality , one can consider a foliation containing $ \Sc $ and select the function $ \cs{F} $ that appears in \cref{eq:energy-loss-gw} such that it fulfills \cref{eq:foliation-adapted} at least on $ \Sc $. If one does so, it is always possible to choose the conformal gauge in order to set
			\begin{equation}
				\cs{K}\frac{\dot{F}}{\alpha}=\text{constant}\spacef,\quad \text{at}\spacef\Sc_{1,2}\spacef.
			\end{equation}
		Under such gauge choice, \cref{eq:flux-news} reads
			\begin{equation}
					\csb{\Phi}{\Sc}=-\brkt{16\pi\cs{K}\frac{\dot{F}}{\alpha}\frac{\df \csrddu{{\tau}}{G}{\E}{\Sc_{C}}}{\df C}+\int_{\Sc}\ct{N}{_{RS}}\cdc{_{M}}\cdc{^M}\ct{N}{^{RS}}\csC{\epsilon}}\spacef,
			\end{equation}
		and  \cref{eq:balance-law-a} can be rewritten as
			\begin{equation}\label{eq:balance-law-b}
				\int_{\Delta}\prn{\cts{\ell}{^\mu}\cd{_{\mu}}\cs{\W}+\cs{\W}\ct{\psi}{^m_{m}}}\epsilon=-\brkt{16\pi\cs{K}\frac{\dot{F}}{\alpha}\frac{\df \csrddu{{\tau}}{G}{\E}{\Sc_{C}}}{\df C}+\int_{\Sc}\ct{N}{_{RS}}\cdc{_{M}}\cdc{^M}\ct{N}{^{RS}}\csC{\epsilon}}\spacef\bevalat{{\Sc_{1}}}^{\Sc_{2}}\spacef.
			\end{equation}
		The interpretation of this formulae is essentially the same as \cref{eq:balance-law-a}. Even so, let us point out that for fixed $ \Sc_{1,2} $, the change in the radiant superenergy density in the volume $ \Delta $ depends only on the initial and final evaluation of the news tensor $ \ct{N}{_{ab}} $, i.e., on $ \ct{N}{_{ab}}\evalat{\Sc_{1,2}} $.  In a way, the integral on the left-hand side of \cref{eq:balance-law-a,eq:balance-law-b} measures the failure of the system to recover its initial state. From another point of view, consider a gravitational system that is initially in equilibrium in the sense of having
			\begin{equation}\label{eq:balance-initial-condition}
				\ct{N}{_{AB}}\evalat{\Sc_{1}}=0\spacef.
			\end{equation}
		Then, the rate of change in the Bondi-Trautman energy at a later retarded time, i.e., on $ \Sc_{2} $, can be expressed as the change of $ \cs{\W} $ in the volume $ \Delta $ plus an additional term whose interpretation is not clear to us and that vanishes if and only if\footnote{Double implication holds true whenever $ \Sc $ has $ \mathbb{S}^2 $-topology.} so does $ \ct{N}{_{AB}} $
			\begin{equation}\label{eq:balance-law-c}
				\frac{\df \csrddu{{\tau}}{G}{\E}{\Sc_{2}}}{\df C}=-\frac{\alpha}{K\dot{F}16\pi}\brkt{\int_{\Delta}\prn{\cts{\ell}{^\mu}\cd{_{\mu}}\cs{\W}+\cs{\W}\ct{\psi}{^m_{m}}}\epsilon+\int_{\Sc_2}\ct{N}{_{RS}}\cdc{_{M}}\cdc{^M}\ct{N}{^{RS}}\csC{\epsilon}}\spacef.
			\end{equation}
		As a final remark, notice that when \cref{eq:balance-initial-condition} holds, \cref{eq:flux-superenergy} implies
			\begin{equation}
				\int_{\Delta}\prn{\cts{\ell}{^\mu}\cd{_{\mu}}\cs{\W}+\cs{\W}\ct{\psi}{^m_{m}}}\epsilon=0\spacef\iff\spacef\ct{N}{_{AB}}\evalat{\Sc_{2}}=0\spacef,
			\end{equation}
		and that $ \ct{N}{_{AB}}\evalat{\Sc_{1,2}} =0$ is a reasonable initial and final condition for any physical system that at first is in equilibrium, then undergoes a change that takes it out of equilibrium and finally settles down.

	\subsection{Alignment of supermomenta and the peeling property of the BR-tensor}
		The tools presented in \cref{ssec:peeling}, namely the asymptotic propagation of fields along null geodesics, can be applied to the physical Bel-Robinson tensor 
			\begin{equation}
				\pt{\T}{_{\alpha\beta\gamma\delta}}\defeq   \pt{C}{_{\alpha\mu\gamma}^\nu}\pt{C}{_{\delta\nu\beta}^\mu} + \ptru{*}{C}{_{\alpha\mu\gamma}^\nu}\ptru{*}{C}{_{\delta\nu\beta}^\mu}\spacef.
			\end{equation}
		 Consider this tensor field at point $ p(\lambda) $ and parallel propagate it along the curve $ \gamma $ defined as in \cref{ssec:peeling} to $ p_{0} $. This process defines a new tensor at $ p_{0} $ which we denote by $ \ppt{\T}{_{\alpha\beta\gamma\delta}} $. Application of $ \ct{L}{_{\beta}^\alpha} $ gives
		 	\begin{equation}\label{eq:asymptotic-propagation-br}
				 \lct{\underaccent{*}{\hat{\T}}}{_{\alpha\beta\gamma\delta}}\defeq  \Omega^4\prn{\lambda_{1}}\prn{ \lct{\underaccent{*}{\hat{C}}}{_{\alpha\mu\gamma}^\nu}\lct{\underaccent{*}{\hat{C}}}{_{\delta\nu\beta}^\mu}+ \lct{\underaccent{*}{\csru{*}{\hat{C}}}}{_{\alpha\mu\gamma}^\nu}\lct{\underaccent{*}{\csru{*}{\hat{C}}}}{_{\delta\nu\beta}^\mu}}\spacef,
			\end{equation}
		where $ \lct{\underaccent{*}{\hat{C}}}{_{\alpha\mu\gamma}^\nu} $ and $ \lct{\underaccent{*}{\csru{*}{\hat{C}}}}{_{\delta\nu\beta}^\mu} $ are the asymptotic propagated physical Weyl tensor \eqref{eq:asymptotic-propagation-weyl-tensor} and its Hodge dual, respectively. In order to arrive at \cref{eq:asymptotic-propagation-br} one has to use \cref{eq:operatorL-eta,eq:operatorL-scalar,eq:propagator-properties}. Then, it is possible to derive the \emph{peeling} property of the Bel-Robinson tensor,	
			\begin{thm}[Peeling of the Bel-Robinson tensor]\label{thm:peeling-br}
				Let $ \prn{\cs{M},\ct{g}{_{\alpha\beta}}} $ be a conformal completion of a physical space-time with $ \Lambda=0 $ as presented on page  \pageref{it:energytensorassumption} and let $ \gamma $ be a lightlike geodesic with affine parameter $ \lambda $ and tangent vector field $ \ct{\ell}{^\alpha} $ as in \cref{eq:null-geodesic-normalisation}. Also, let one end point $ p_{0} $ ($ \lambda=\lambda_{0}=0 $) of $ \gamma $ be at $ \scri^+ $ and the other one, $ p_{1} $ ($ \lambda=\lambda_{1}=-1 $), in $ \ps{M} $. Then, the asymptotic behaviour of the physical Bel-Robinson tensor $ \pt{\T}{_{\alpha\beta\gamma\delta}} $ along $ \gamma $ follows by application of \cref{it:asymptotic-propagation-first-step,it:asymptotic-propagation-second-step,it:asymptotic-propagation-third-step,} on page \pageref{it:asymptotic-propagation-first-step} and reads
					\begin{align}
					  	\frac{1}{\Omega\prn{\lambda_{1}}^4}\lct{\underaccent{*}{\hat{\T}}}{_{\alpha\beta\gamma\delta}}&=	\lambda^2 \ctru{(N)}{\D}{_{\alpha\beta\gamma\delta}}+\lambda^3 \ctru{(1)}{\X}{_{\alpha\beta\gamma\delta}}+\lambda^4 \ctru{(2)}{\X}{_{\alpha\beta\gamma\delta}}+\lambda^4\ctru{(III)}{\E}{_{\alpha\beta\gamma\delta}}+\lambda^5 \ctru{(3)}{\X}{_{\alpha\beta\gamma\delta}}\nonumber\\
					  &+\lambda^6 \ctru{(4)}{\X}{_{\alpha\beta\gamma\delta}}+\lambda^6\ctru{(II/D)}{\F}{_{\alpha\beta\gamma\delta}}+\lambda^7 \ctru{(5)}{\X}{_{\alpha\beta\gamma\delta}}+\lambda^8 \ctru{(6)}{\X}{_{\alpha\beta\gamma\delta}}+\lambda^8\ctru{(I)}{\G}{_{\alpha\beta\gamma\delta}}\nonumber\\
					  &+\lambda^9 \ctru{(7)}{\X}{_{\alpha\beta\gamma\delta}}+\lambda^{10} \ctru{(8)}{\X}{_{\alpha\beta\gamma\delta}}+\lambda^{10}\ctru{(I)}{\cH}{_{\alpha\beta\gamma\delta}}+\mathcal{O}\prn{\lambda^{11}}\spacef,\label{eq:peeling-br}
					\end{align}
				near $ \lambda=\lambda_{0} $, where
					\begin{align}
						\ctru{(N)}{\D}{_{\alpha\beta\gamma\delta}}\defeq&\ctru{(N)}{d}{_{\alpha\mu\gamma}^\nu}\ctru{(N)}{d}{_{\delta\nu\beta}^\mu} + \ctru{(N)*}{d}{_{\alpha\mu\gamma}^\nu}\ctru{(N)*}{d}{_{\delta\nu\beta}^\mu}\spacef,\label{eq:rrBR-D}\\
						\ctru{(III)}{\E}{_{\alpha\beta\gamma\delta}}\defeq&\ctru{(III)}{e}{_{\alpha\mu\gamma}^\nu}\ctru{(III)}{e}{_{\delta\nu\beta}^\mu} + \ctru{(III)*}{e}{_{\alpha\mu\gamma}^\nu}\ctru{(III)*}{e}{_{\delta\nu\beta}^\mu}\spacef,\label{eq:rrBR-E}\\
						\ctru{(II/D)}{\F}{_{\alpha\beta\gamma\delta}}\defeq&\ctru{(II/D)}{f}{_{\alpha\mu\gamma}^\nu}\ctru{(II/D)}{f}{_{\delta\nu\beta}^\mu} + \ctru{(II/D)*}{f}{_{\alpha\mu\gamma}^\nu}\ctru{(II/D)*}{f}{_{\delta\nu\beta}^\mu}\spacef,\label{eq:rrBR-F}\\
						\ctru{(I)}{\G}{_{\alpha\beta\gamma\delta}}\defeq&\ctru{(I)}{g}{_{\alpha\mu\gamma}^\nu}\ctru{(I)}{g}{_{\delta\nu\beta}^\mu} + \ctru{(I)*}{g}{_{\alpha\mu\gamma}^\nu}\ctru{(I)*}{g}{_{\delta\nu\beta}^\mu}\spacef,\label{eq:rrBR-G}\spacef
					\end{align}
				are basic superenergy tensors labelled with the Petrov type of the Weyl-tensor candidate they are built with, respectively; the Weyl-tensor candidates are the ones of \cref{thm:peeling-weyl} described in \cref{tab:rrw-tensors}. The tensor fields $ \ctru{(a)}{\X}{_{\alpha\beta\gamma\delta}}$ with $ a=1,2,3,4,5,6 $ are symmetric and traceless, and contain cross terms:
					\begin{align}
						\ctru{(1)}{\X}{_{\alpha\beta\gamma\delta}}\defeq&\ctru{(N)}{d}{_{\alpha\mu\gamma}^\nu}\ctru{(III)}{e}{_{\delta\nu\beta}^\mu} + \ctru{(N)*}{d}{_{\alpha\mu\gamma}^\nu}\ctru{(N)*}{e}{_{\delta\nu\beta}^\mu}+\ctru{(III)}{e}{_{\alpha\mu\gamma}^\nu}\ctru{(N)}{d}{_{\delta\nu\beta}^\mu} + \ctru{(III)*}{e}{_{\alpha\mu\gamma}^\nu}\ctru{(N)*}{d}{_{\delta\nu\beta}^\mu}\spacef,\\
						\ctru{(2)}{\X}{_{\alpha\beta\gamma\delta}}\defeq&\ctru{(N)}{d}{_{\alpha\mu\gamma}^\nu}\ctru{(II/D)}{f}{_{\delta\nu\beta}^\mu} + \ctru{(N)*}{d}{_{\alpha\mu\gamma}^\nu}\ctru{(II/D)*}{f}{_{\delta\nu\beta}^\mu}+\ctru{(II/D)}{f}{_{\alpha\mu\gamma}^\nu}\ctru{(N)}{d}{_{\delta\nu\beta}^\mu} + \ctru{(II/D)*}{f}{_{\alpha\mu\gamma}^\nu}\ctru{(N)*}{d}{_{\delta\nu\beta}^\mu}\spacef,\\
						\ctru{(3)}{\X}{_{\alpha\beta\gamma\delta}}\defeq&\ctru{(N)}{d}{_{\alpha\mu\gamma}^\nu}\ctru{(I)}{g}{_{\delta\nu\beta}^\mu} + \ctru{(N)*}{d}{_{\alpha\mu\gamma}^\nu}\ctru{(I)*}{g}{_{\delta\nu\beta}^\mu}+\ctru{(I)}{g}{_{\alpha\mu\gamma}^\nu}\ctru{(N)}{d}{_{\delta\nu\beta}^\mu} + \ctru{(I)*}{g}{_{\alpha\mu\gamma}^\nu}\ctru{(N)*}{d}{_{\delta\nu\beta}^\mu}\nonumber\\
						+&\ctru{(II/D)}{f}{_{\alpha\mu\gamma}^\nu}\ctru{(III)}{e}{_{\delta\nu\beta}^\mu} + \ctru{(II/D)*}{f}{_{\alpha\mu\gamma}^\nu}\ctru{(N)*}{e}{_{\delta\nu\beta}^\mu}+\ctru{(III)}{e}{_{\alpha\mu\gamma}^\nu}\ctru{(II/D)}{f}{_{\delta\nu\beta}^\mu} + \ctru{(III)*}{e}{_{\alpha\mu\gamma}^\nu}\ctru{(II/D)*}{f}{_{\delta\nu\beta}^\mu}\spacef,\\
						\ctru{(4)}{\X}{_{\alpha\beta\gamma\delta}}\defeq&\ctru{(N)}{d}{_{\alpha\mu\gamma}^\nu}\ctru{(I)}{h}{_{\delta\nu\beta}^\mu} + \ctru{(N)*}{d}{_{\alpha\mu\gamma}^\nu}\ctru{(I)*}{h}{_{\delta\nu\beta}^\mu}+\ctru{(I)}{h}{_{\alpha\mu\gamma}^\nu}\ctru{(N)}{d}{_{\delta\nu\beta}^\mu} + \ctru{(I)*}{h}{_{\alpha\mu\gamma}^\nu}\ctru{(N)*}{d}{_{\delta\nu\beta}^\mu}\nonumber\\
						+&\ctru{(I)}{g}{_{\alpha\mu\gamma}^\nu}\ctru{(III)}{e}{_{\delta\nu\beta}^\mu} + \ctru{(I)*}{g}{_{\alpha\mu\gamma}^\nu}\ctru{(N)*}{e}{_{\delta\nu\beta}^\mu}+\ctru{(III)}{e}{_{\alpha\mu\gamma}^\nu}\ctru{(I)}{g}{_{\delta\nu\beta}^\mu} + \ctru{(III)*}{e}{_{\alpha\mu\gamma}^\nu}\ctru{(I)*}{g}{_{\delta\nu\beta}^\mu}\spacef.
					\end{align}
			\end{thm}
			\begin{proof}
			 Application of \cref{it:asymptotic-propagation-first-step,it:asymptotic-propagation-second-step} leads to \cref{eq:asymptotic-propagation-br}. Then, a direct calculation of the Taylor series around $ \lambda_{0} $ yields \cref{eq:peeling-br}.
			\end{proof}
		The interest of the above result lies in the following remarkable property of supermomenta:
			\begin{corollary}
				Let conditions of \cref{thm:peeling-br} hold and $ \pt{\P}{_{\alpha}} $ be the supermomentum associated with a causal vector field $ \pt{u}{^\alpha} $, constructed with the physical Bel-Robinson tensor $ \pt{\T}{_{\alpha\beta\gamma\delta}} $. Then, the asymptotic behaviour of the supermomentum along $ \gamma $  follows by application of \cref{it:asymptotic-propagation-first-step,it:asymptotic-propagation-second-step,it:asymptotic-propagation-third-step,} on page \pageref{it:asymptotic-propagation-first-step} and reads
								\begin{equation}\label{eq:sm-convergence}
									\lct{\underaccent{*}{\hat{\P}}}{_{\alpha}}=\Omega^6\prn{\lambda_{1}}\prn{\ct{\ell}{_{\mu}}\lct{\hat{\underaccent{*}{u}}}{^{\mu}}}^3\cs{\W}\lambda^2\ct{\ell}{_{\alpha}}+\mathcal{O}\prn{\lambda^3}\spacef,
								\end{equation}
							where $ \cs{\W} $ is the asymptotic radiant superenergy \eqref{eq:radiant-superenergy} and $ \lct{\hat{\underaccent{*}{u}}}{^{\mu}}\defeq \ct{g}{^{\nu\mu}}\ct{L}{_{\nu}^\rho}\ppt{u}{_{\rho}}\prn{\lambda_{0}} $.
						\end{corollary}
						\begin{proof}
							\Cref{it:asymptotic-propagation-first-step} together with \cref{eq:propagator-properties} provides us with
								\begin{equation}
									\ppt{\P}{_{\alpha}}=\Omega^6\prn{\lambda}\ppt{u}{^\beta}\ppt{u}{^\gamma}\ppt{u}{^\delta}\ppt{\T}{_{\alpha\beta\gamma\delta}}\spacef.
								\end{equation}
							Next, one applies \cref{it:asymptotic-propagation-second-step} and uses \cref{eq:operatorL-scalar},
								\begin{equation}
							 		\lct{\underaccent{*}{\hat{\P}}}{_{\alpha}}=\frac{\Omega^6\prn{\lambda}}{\Xi^6\prn{\lambda}}\lct{\underaccent{*}{\hat{u}}}{^{\beta}}	\lct{\underaccent{*}{\hat{u}}}{^{\gamma}}	\lct{\underaccent{*}{\hat{u}}}{^{\delta}}\lct{\underaccent{*}{\hat{\T}}}{_{\alpha\beta\gamma\delta}}\spacef.
								\end{equation}
							This last expression has a Taylor expansion around $ \lambda_{0} $ that reads
								\begin{equation}
									\lct{\underaccent{*}{\hat{\P}}}{_{\alpha}}=\Omega^{10}\prn{\lambda_{1}}\lct{\underaccent{*}{\hat{u}}}{^{\beta}}	\lct{\underaccent{*}{\hat{u}}}{^{\gamma}}	\lct{\underaccent{*}{\hat{u}}}{^{\delta}}\ctru{(N)}{\D}{_{\alpha\beta\gamma\delta}}\lambda^2+\mathcal{O}\prn{\lambda^3}\spacef.
								\end{equation}
							where \cref{eq:peeling-br} was used. Finally, using \cref{eq:rrw-d,eq:asymptotic-leading-term-weyl} together with \cref{eq:radiant-superenergy}, the result follows.
					\end{proof}
			\begin{remark}
				Observe that \cref{eq:sm-convergence} is well behaved at $ \lambda=\lambda_{0} $ if and only if $  \ct{\ell}{_{\mu}}\lct{\hat{u}}{^{\mu}} $ does not diverge there. The equation, when regular, shows that at leading order only the $ \ct{N}{^\alpha} $ component of $ \pt{u}{^\alpha} $ contributes to the physical supermomentum transported along a null geodesic reaching $ \scri $. This is in natural agreement with \eqref{thm:noradDelta}, which bases the determination of outgoing gravitational radiation precisely on the asymptotic radiant supermomentum \cref{eq:radiant-super-momentum}, i.e., a radiant supermomentum for the `observer' $ \ct{N}{^\alpha} $.
			\end{remark}
			\begin{remark}
				Notice that $ \pt{u}{^\alpha} $ has to be causal, and in particular can be lightlike. Hence the result applies to physical \emph{radiant} supermomenta too. In other words, {\em all} supermomenta, radiant or not, tend to align towards infinity with one single direction given by the asymptotic radiant supermomentum. This supports our criterion that a vanishing radiant supermomentum characterizes absence of gravitational radiation arriving at infinity.
			\end{remark}

%% file: section-6/section-6.tex
	\begin{itemize}
	\item The computational effort in obtaining the radiant asymptotic supermomentum is small in comparison to other characterisations which need of a suitable choice of conformal frame --e.g.,  the computation of the news tensor $ \ct{N}{_{ab}} $ as the time derivative of the asymptotic shear in a Bondi gauge, or the determination of $ \ct{\rho}{_{ab}} $.
	\item The classical criterion by means of the news tensor field $ \ct{N}{_{ab}} $ has been shown to be equivalent to the tidal-based criterion. This proves that  the tidal techniques are reliable and suited to the problem.
	\item It is possible to adapt our geometrical derivation of the peeling behaviour, and the general asymptotic propagation of fields, to the $ \Lambda\neq 0 $ scenarios. Also, it would be interesting to see if the endomorphism $ \ct{L}{_{\alpha}^\beta} $ can be derived for curves other than null geodesics.
	\end{itemize}

%% file: appendices/appendices.tex
	\section{Conformal-gauge transformations}\label{app:conformal-gauge-transformations}
		Some gauge-transformation formulae are presented. Only those which apply specifically to $ \Lambda=0 $ are shown; other expressions are listed in \cite{Fernandez-Alvarez_Senovilla-dS}.\\
		
		Quantities of $ \prn{\scri,\ms{_{ab}}} $:
		\begin{align}
		\cts{N}{^a}&\eqs\frac{1}{\omega}\cts{N}{^a}\spacef,\\
		\msg{_{ab}}&\eqs\omega^2\ms{_{ab}}\spacef,\label{eq:conformal-rescaling-scri-afs}\\
		\ctg{\epsilon}{_{abc}}&\eqs\omega^3\ct{\epsilon}{_{abc}}\spacef,\\
		\ctsg{\Gamma}{^a_{bc}} &\eqs \cts{\Gamma}{^a_{bc}} +\cts{C}{^a_{bc}}\spacef,\cts{C}{^a_{bc}}=\frac{1}{\omega}\prn{2\delta{_{
					(b}^a}\cts{\omega}{_{c)}}-\ms{_{bc}}\cts{\omega}{^a}}\quad\\
		\ctsg{R}{_{ab}}&\eqs\cts{R}{_{ab}}- \frac{1}{\omega}\cds{_a}\cts{\omega}{_{b}}-\frac{1}{\omega}\ms{_{ab}}\cds{_{m}}\cts{\omega}{^m}+\frac{2}{\omega^2}\cts{\omega}{_a}\cts{\omega}{_b}\spacef,\label{eq:gauge-afsRicciTensorScriApp}\\
		\csSg{R}&\eqs\frac{1}{\omega^2}\csS{R}-\frac{2}{\omega^3}\ms{^{mp}}\cds{_m}\cts{\omega}{_p} +\frac{2}{\omega^4}\cts{\omega}{_{m}}\cts{\omega}{^m}\spacef,\label{eq:gauge-afsRicciScalarScriApp}\\
		\ctsg{S}{_{ab}}&\eqs \cts{S}{_{ab}}+\frac{2}{\omega^2}\cts{\omega}{_a}\cts{\omega}{_b}-\frac{1}{\omega}\cds{_a}\cts{\omega}{_b}-\frac{1}{2\omega^2}\cts{\omega}{_s}\cts{\omega}{^s}\ms{_{ab}}\spacef,
		\end{align}
	where $ \ct{g}{^{\alpha\mu}}\omega_{\mu}=\ct{e}{^\alpha_{a}}\cts{\omega}{^a} $.\\
	
		Quantities associated to a cut $ \prn{\Sc,\mc{_{AB}}} $:
				\begin{align}
				\mcg{_{AB}}&\eqc\omega^2\mc{_{AB}}\spacef,\label{eq:gauge-afs-mc}\\
				\ctcg{\epsilon}{_{AB}}&\eqc \omega^2\ctc{\epsilon}{_{AB}}\spacef,\\
				\ctcg{\Gamma}{^A_{BC}} &\eqc \ctc{\Gamma}{^A_{BC}} +\ctc{C}{^A_{BC}}\spacef,\ctc{C}{^A_{BC}}\eqc\frac{1}{\omega}\mc{^{AT}}\prn{2\mc{_{T(B}}\ctc{\omega}{_{A)}}-\mc{_{AB}}\ctc{\omega}{_T}}\quad\\
				\ctcg{R}{_{AB}}&\eqc\ctc{R}{_{AB}}+\frac{1}{\omega^2}\mc{_{AB}}\ctc{\omega}{_{M}}\ctc{\omega}{^M}-\frac{1}{\omega}\mc{_{AB}}\cdc{_{M}}\ctc{\omega}{^M}\spacef,\label{eq:gauge-afsRicciTensorCutApp}\\
				\ctcg{R}{}&\eqc\frac{1}{\omega^2}\ctc{R}{}+\frac{2}{\omega^4}\ctc{\omega}{_{M}}\ctc{\omega}{^M}-\frac{2}{\omega^3}\cdc{_M}\ctc{\omega}{^M} \spacef,\label{eq:gauge-afsRicciScalarCutApp}\\
				\ctcg{S}{_{AB}}&\eqc \ctc{S}{_{AB}} +\frac{2}{\omega^2}\ctc{\omega}{_A}\ctc{\omega}{_B}-\frac{1}{\omega}\cdc{_A}\ctc{\omega}{_B}-\frac{1}{2\omega^2}\ctc{\omega}{_P}\ctc{\omega}{^P}\mc{_{AB}}\spacef,\label{eq:gauge-afsSAB}
				\end{align}
				
	The lightlike projections of the rescaled Weyl tensor on $ \scri $, calculated with respect to $ \ct{N}{^a} $, have the following gauge transformations:
				\begin{align}
				\ctn{\tilde{D}}{^{ab}}&\eqs \frac{1}{\omega^5} \ctn{D}{^{ab}}\spacef,	&\ctn{\tilde{C}}{^{ab}}&\eqs \frac{1}{\omega^5} \ctn{C}{^{ab}}\spacef,\\
				\ctn{\tilde{D}}{_{ab}}&\eqs \frac{1}{\omega} \ctn{D}{_{ab}}\spacef,	&\ctn{\tilde{C}}{_{ab}}&\eqs \frac{1}{\omega} \ctn{C}{_{ab}}\spacef,\\
				\ctcnn{\tilde{D}}{_{A}}&\eqs\frac{1}{\omega^2}\ctcnn{D}{_{A}}\spacef,&\ctcnn{\tilde{C}}{_{A}}&\eqs\frac{1}{\omega^2}\ctcnn{C}{_{A}} \spacef.
				\end{align}